\documentclass[
 reprint,
 amsmath,amssymb,
 aps,
]{revtex4-1}
\usepackage{lineno}
\usepackage{graphicx}
\usepackage{dcolumn}
\usepackage{mathrsfs}
\usepackage{amsmath}
\usepackage{bm}
\usepackage{color}
\usepackage{ulem}
\usepackage{float}
\usepackage{CJK}
\usepackage{color}
\usepackage[
bookmarks=true,
colorlinks,
linkcolor=blue,
urlcolor=blue,
citecolor=blue,
plainpages=false,
pdfpagelabels,
final,
breaklinks=true
]{hyperref}

\definecolor{purple}{RGB}{128,0,128}

\begin{document}
\preprint{APS/123-QED}

\title{Impacts of symmetry energy slope on the oscillation frequencies of neutron stars with short-range correlation and admixed dark matter}
\author{Bin Hong$^{1}$}
\email{hongbin@tongji.edu.cn}
\author{ZhongZhou Ren$^{1}$}
\email{zren@tongji.edu.cn}
\author{Chen Wu$^{ 2}$}
\author{XueLing Mu$^{ 3}$}
\affiliation{$^{1}$ School of Physics Science and Engineering, Tongji University, Shanghai 200092, China
}
\affiliation{ $^{2}$ Shanghai Advanced Research Institute, Chinese Academy of Sciences, Shanghai 201210, China}
\affiliation{
$^{3}$ College of Electronic Information and Electrical Engineering, Chengdu University, Chendu 610106, China
}



\date{\today}

\begin{abstract}
  Oscillation modes of compact stars, in general, can serve as a fingerprint in determining the equation of state (EOS) of dense matter. In this study, we examine the impact of symmetry energy slope ($L$) on the oscillation frequencies of neutron stars (NSs) with nucleon-nucleon short range correlation (SRC) and admixed dark matter (DM) for the first time within the relativistic mean-field theory. By adjusting the $L$, we revise the EOS and coupling parameters in light of the SRC and DM effects, and construct the new sets. The results reveal that NSs containing SRC and DM inside are more likely to satisfy the observational constraints, and we find that smaller $L$ exhibits larger fundamental non-radial and radial frequencies, and that the effect on Large Separation (LG) is also mainly concentrated in the low-mass region. Moreover, we update the linear relationship between the non-radial frequency and mean density, and we further give empirical relations between non-radial and radial frequencies and tidal deformability at different $L$ for 1.4$M_{\odot}$ and 2$M_{\odot}$. These findings will enable us to more effectively confine the NS EOSs, in turn, also provide a strategy to place constraints on the $L$.
\end{abstract}

\maketitle


\section{Introduction}
As a class of very dense objects in astronomy, the description of matter at high density inside NSs has attracted a lot of attention from nuclear physics \citep{RevModPhys.89.015007}, particle physics, and astrophysics \citep{Jiang_2020,Luo_2022}. However, given the non-perturbative nature of nuclear forces, we cannot derive the EOS directly from quantum chromodynamics (QCD). As a result, it is more common to construct NS EOSs from the microscopic first principles, such as $\chi EFT$ \citep{RevModPhys.81.1773,RevModPhys.85.197,HOLT201335,PhysRevLett.110.032504,PhysRevLett.116.062501}, or from the self-consistent phenomenological models, typical approaches like the Skyrme-Hartree-Fock \citep{STONE2007587,PhysRevC.68.034324,PhysRevC.85.035201} and Gogny-Hartree-Fock \citep{PhysRevC.83.065809,PhysRevC.96.065806}, or from parameterization models \citep{PhysRevD.79.124032,Luo_2022}. These models are also strongly constrained by multi-messenger observations, including the recent series of discoveries of twice-solar-mass ($2 M_{\odot}$) NSs \citep{RN583,Fonseca_2016,Arzoumanian_2018,Antoniadis1233232,RN584,Fonseca_2021}, the tidal deformability extracted from the binary NS merger event GW170817 \citep{PhysRevX.9.011001,PhysRevLett.123.141101}, the simultaneous mass and radius measurements of the isolated PSR J0030$+$0451 by NICER (Neutron Star Interior Composition Explorer) \citep{Miller_2019,Riley_2019} and the gravitational wave event GW190814 from the coalescence of a stellar-mass black hole and a mysterious compact star \citep{Abbott_2020}. To some extent, these observations undoubtedly disprove certain EOSs, making it imperative to establish plausible ones that can describe the low density properties while meeting the observational constraints.

Nuclear microscopic scale nucleon-nucleon SRC \citep{ARRINGTON2012898,ATTI20151,RevModPhys.89.045002} and cosmic macroscopic size DM have proven to be relatively challenging issues. Several theoretical works have demonstrated that, in contrast to the free Fermi gas, the SRC originating from the strongly repulsive core of nuclear force and its tensor part leads to an appreciable depletion below the Fermi surface, with some nucleons occupying regions above the Fermi surface and giving rise to a high-momentum tail, as has been verified by experiments like $(e,e'p)$ \citep{PhysRevLett.93.182501} and $(e,e'NN)$ \citep{PhysRevLett.81.2213,STARINK200033}. Furthermore, the SRC effects also play an important role in nuclear physics \citep{PhysRevC.73.035805,PhysRevC.90.064312,PhysRevC.92.011601,PhysRevC.91.025803,PhysRevC.96.054603,PhysRevC.100.054325}, such as allowing us to have a better understanding of the EMC effect \citep{PhysRevLett.106.052301,PhysRevC.85.047301} and to explain the neutrino oscillation measurements \citep{PhysRevLett.111.022501,PhysRevLett.111.022502} as well as density-dependent behavior of nuclear symmetry energy \citep{PhysRevC.91.025803}. When it is extended to study NSs, tidal deformability \citep{PhysRevC.101.065202,Hong_2022}, mass-radius relations \citep{LI2019118,LU2021122200,PhysRevD.105.023008} and cooling efficiency will all be affected \citep{souza2020shortrange}. For DM, many observations such as gravitational lensing, galaxy rotation curves, velocity dispersions, galaxy clusters and cosmic microwave background have predicted its existence (for a review, see \citep{BERTONE2005279,RevModPhys.90.045002,BUENABAD20221}). Although there are many candidates for DM, the origin and properties remain a mystery, moreover, DM does not interact directly with normal matter, but it has a more pronounced gravitational effect on dense objects like NSs \citep{PhysRevD.83.083512}, and play a important role in determining NS mass-radius relation \citep{Quddus_2020,PhysRevD.105.023008} and tidal deformability \citep{PhysRevD.104.063028,10.1093/mnras/staa1435}. Despite this, there is still little ongoing research to incorporate nucleon-nucleon SRC into the admixed DM NSs and to further examine their effects.

The symmetry energy and its slope play a crucial role in determining the equation of state of pure neutron matter, which involves determining the neutron skin thickness in neutron-rich matter, the dynamics of heavy-ion collisions, and also affects the structure and properties of NSs, helping us understand isospin asymmetry physics under extreme conditions. Given that $L$ characterizes the density dependence of the symmetry energy, it not only determines the behavior of high-density asymmetry nuclear matter but also directly affects the supernova, and even the production of heavy elements in nucleosynthesis. Moreover, its value remains large uncertain and still requires theoretical and experimental studies. Therefore, we choose the symmetry energy slope $L$ to study its effect on the properties of NSs under the background of short-range correlations and dark matter admixed. In the work \citep{PhysRevC.95.034324} the authors used two different methods to deduce its range around $L=64.29\pm11.84$ MeV and $L=53.85\pm10.29$ MeV, respectively, while the latest study employs Bayesian analysis to estimate $L$ to be around $L=70^{+21}_{-18}$ MeV \citep{PhysRevD.104.063032}. Recently, a more accurate model-independent experiment PREX-2 reported very thick $^{208}$Pb neutron skin thickness ($R_{\text {skin }}^{208}$) \citep{PhysRevLett.126.172502}, and based on this result with taking into account the strong correlation between $R_{\text {skin }}^{208}$ and $L$, the work \citep{PhysRevLett.126.172503} extracted a $L$ up to $(106\pm37)$ MeV, a value that seriously challenges our understanding of neutron-rich matter near the nuclear saturation density and will further aggravate the uncertainty in $L$. Additionally, as the oscillation frequency is highly dependent on the NS internal structure, it is possible to understand the internal physics by analyzing GW signals from compact star oscillations in addition to signals from binary NS mergers. The presence of any non-axisymmetric perturbation in NSs caused by dynamical instabilities, such as starquakes, magnetic reconfiguration, and rotating \citep{Jones}, is known to cause non-radial oscillations and thus emission of gravitational waves. Furthermore, depending on the restoring force, they can be classified as fundamental $f$-modes, gravity $g$-modes, pure time $w$-modes, pressure $p$-modes, and rotational $r$-modes \citep{10.1093/mnras/255.1.119,Lau_2010,RN84}. It is well known that non-radial oscillations of NSs are also accompanied by radial oscillations \citep{PhysRevD.75.084038}, and although radial oscillations do not produce gravitational radiation directly, they can couple and further amplify it \citep{PhysRevD.73.084010,PhysRevD.75.084038}. Many excellent works have been carried out to investigate the radial and non-radial oscillations, for example, by studying the difference in the oscillation frequency caused by the different components inside a NS, it is possible to identify whether it is a hadronic, hybrid or quark star \citep{1992AA,PhysRevD.82.063006,PhysRevC.96.065803,Ranea_Sandoval_2018,Pereira_2018,PhysRevD.103.103003,PhysRevD.103.123009,PhysRevD.103.063015}. Moreover, numerical relativistic simulations show that binary NSs merger to form a hypermassive NS may experience strong radial oscillations and radiate GWs at kHz \citep{PhysRevD.90.023002,PhysRevLett.113.091104}, and the identification of quasi-periodic oscillations in giant flares caused by NS torsional oscillations can also help us to effectively constrain the NS crust properties \citep{PhysRevLett.108.201101,10.1093/mnrasl/sls006,10.1093/mnras/sty1755,10.1093/mnras/stz2385}.

Given the fact that the $L$ is still theoretically uncertain, and that the NS oscillation frequency is expected to detect in the near future with the upgrade of the observational devices \citep{PhysRevD.103.044024,Punturo_2010,PhysRevD.91.082001}, so it is necessary to establish a link between them, which will not only assist us in establishing a bridge between theory and observation to better constrain the EOSs, but also to infer the theoretical value of $L$ from frequency. In light of this, we introduce the nucleon-nucleon SRC in the interior of admixed DM neutron stars and rebuilt the EOSs that not only satisfy the properties of nuclear matter at saturation density, but also satisfy the constraints from multi-messenger observation. Based on this, we examine the impact of $L$ on the NS radial and non-radial oscillations, build their relationship and predict their frequency ranges.

This article is organized as follows: Section II provides a basic introduction to theoretical models of SRC and DM in NSs. Section III discusses in detail the parameter construction that meets the constraints from saturation nuclear behaviors and multi-messenger observations. Section IV is devoted to investigate the implication of $L$ on the NS oscillation frequency, and Section V provides a brief summary and remark. Finally, Appendix shows a comprehensive derivation of relativistic nucleon coupling parameters by incorporating the SRC effect.

\section{Relativistic mean field theory with Short-Range correlation and Dark Matter Admixed}
In this paper, we adopt relativistic mean field theory (RMF) with $\sigma\omega\rho$ model. Based on this framework, we absorb the SRC and introduce the DM to give the SRC-DM-revised RMF. In the mean-field approximation, the energy density and pressure are expressed as follows:
\begin{eqnarray}
\nonumber
&&\mathcal{E}=\sum_{N}\langle\psi_{N}^{+} i \dot{\psi}_{N}\rangle+\sum_{l}\langle\psi_{l}^{+} i \dot{\psi}_{l}\rangle+\frac{1}{2} m_{\sigma}^{2} \sigma_{0}^{2}-\frac{1}{2} m_{\omega}^{2} \omega_{0}^{2}\\
\nonumber
&&-\frac{1}{2} m_{\varrho}^{2} \varrho_{0}^{2}-\Lambda_{\omega}\left(g_{\varrho N} \varrho_{0}\right)^{2}\left(g_{\omega N} \omega_{0}\right)^{2}+\frac{1}{3}g_{2} \sigma_{0}^{3}+\frac{1}{4} g_{3} \sigma_{0}^{4}\\
\end{eqnarray}
and
\begin{eqnarray}
\nonumber
&&P=\frac{1}{3} \sum_{N}\langle\psi_{N}^{+}(-i \alpha \cdot \nabla) \psi_{N}\rangle+\frac{1}{3} \sum_{l}\langle\psi_{l}^{+}(-i \alpha \cdot\nabla) \psi_{l}\rangle \\
\nonumber
\\
\nonumber
&&-\frac{1}{2} m_{\sigma}^{2} \sigma_{0}^{2}+\frac{1}{2} m_{\omega}^{2} \omega_{0}^{2}+\frac{1}{2} m_{\varrho}^{2} \varrho_{0}^{2}+\Lambda_{\omega}\left(g_{\varrho N} \varrho_{0}\right)^{2}\left(g_{\omega N} \omega_{0}\right)^{2}\\
&&-\frac{1}{3} g_{2} \sigma_{0}^{3}-\frac{1}{4} g_{3} \sigma_{0}^{4}\\
\nonumber
\end{eqnarray}
where $N$ and $l$ denote the nucleon $N(n,p)$ and lepton $l(e,\mu)$ degrees of freedom, and $m_{\sigma}$,$m_{\omega}$,$m_{\varrho}$ correspond to the mass of $\sigma,\omega,\rho$. Coupling parameters $g_{\sigma N},g_{\omega N},g_{\varrho N},g_{2},g_{3},\Lambda_{\omega}$ need
to be further revised by incorporating SRC and DM and the detailed analytical derivation have been attached in the Appendix. For DM component, here we adopt the lightest neutralino as the Fermi DM candidate with a mass of $m_{\chi}$=200 GeV. Moreover, DM and nucleons do not interact directly but through coupled Higgs fields, with the following Lagrangian form \citep{PhysRevD.96.083004}:
\begin{eqnarray}
\nonumber
\mathcal{L}_{\mathrm{DM}}=&& \bar{\chi}\left[i \gamma^{\mu} \partial_{\mu}-M_{\chi}+y h\right] \chi+\frac{1}{2} \partial_{\mu} h \partial^{\mu} h \\
&&-\frac{1}{2} M_{h}^{2} h^{2}+\sum_{N} f \frac{m_{N}}{v} \bar{\psi}_{N} h \psi_{N},\\
\nonumber
\end{eqnarray}
where $\chi$, $h$ denote the DM and Higgs field, and $y$ is the coupling strength between DM and Higgs field which usually takes values between 0.001 and 0.1, here we adopt the generally used value of 0.07 \citep{PhysRevD.64.015001,PhysRevD.96.083004}. $M_{h}$ refers to the mass of the Higgs boson, which is 125 GeV. $f \frac{m_{N}}{v} $ is the effective Yukawa coupling strength between nucleon and Higgs field, where $v$ is the Higgs vacuum expectation value with $v=246$ GeV and $f$ is the Higgs-nucleon formation factor. Based on the lattice computations, we choose the optimal value of $f$ as 0.3, which is not only consistent with theoretical result \citep{PhysRevD.88.055025} but also satisfies the experimental constraints of PandaX-II \citep{PhysRevLett.117.121303}, PandaX-4T \citep{PhysRevLett.127.261802} for the DM-nucleon scattering cross section.

Finally, we assume that the DM number density is about 1000 times lower than the number density of nucleons, and if we take the typical nuclear saturation number density $n_{0}=0.16$ fm$^{-3}$, the calculated DM Fermi energy $k_{F}^{\mathrm{DM}}$ is about 0.033 GeV. Applying the mean field approximation, the effective mass of nucleons and DM and its scalar density are given as,
\begin{eqnarray}
&&M_{N}^{\star}=m_{N}-g_{\sigma} \sigma_{0}-\frac{f m_{N}}{v} h_{0}, \\
&&M_{\chi}^{\star}=m_{\chi}-y h_{0},\\
&&\rho_{s}^{\mathrm{DM}}=\frac{ M_{\chi}^{*}}{ \pi^{2}} \int_{0}^{k_{F}^{\mathrm{DM}}} \frac{k^{2} \mathrm{d}k}{\left(k^{2}+M_{\chi}^{* 2}\right)^{1 / 2}},
\end{eqnarray}
while the energy density and pressure of dark matter are
\begin{eqnarray}
\mathcal{E}_{\mathrm{DM}}=\frac{1}{\pi^{2}} \int_{0}^{k_{F}^{\mathrm{DM}}} k^{2} \mathrm{d} k \sqrt{k^{2}+\left(M_{\chi}^{\star}\right)^{2}}+\frac{1}{2} M_{h}^{2} h_{0}^{2},
\end{eqnarray}
and
\begin{eqnarray}
P_{\mathrm{DM}}=\frac{1}{3 \pi^{2}} \int_{0}^{k_{F}^{\mathrm{DM}}} \frac{k^{4} \mathrm{d} k}{\sqrt{k^{2}+\left(M_{\chi}^{\star}\right)^{2}}}-\frac{1}{2} M_{h}^{2} h_{0}^{2}.
\end{eqnarray}

Both $\langle\psi_{N}^{+} i \dot{\psi_{N}}\rangle$ and $\langle\psi_{N}^{+}(-i \alpha \cdot \nabla) \psi_{N}\rangle$ in the Eqs. (1) and (2) depend directly on the momentum distribution of the nucleon in the mean-field approximation, and therefore also on the SRC. Recent theoretical and experimental studies indicate that the short-range correlation produces a single nucleon momentum distribution that approximately satisfies the $\sim1/k^{4}$ distribution near the Fermi surface \citep{PhysRevC.96.054603,PhysRevC.91.025803,PhysRevC.92.011601}. The $1/k^4$ functional form has been shown to provide a good fit to experimental data on nucleon momentum distributions, and it is insensitive to the specific details of the nucleon-nucleon interaction and the nuclear wave function, making it a versatile tool in the study of nucleon momentum distributions. The form can be applied in a range of contexts, including electron scattering experiments \citep{PhysRevLett.93.182501,PhysRevLett.81.2213,STARINK200033}. Although the parameterized model mentioned above can well characterize the high-momentum tail caused by nucleon short-range correlations and is consistent with experimental results, there is still one thing to point out that the correlation percentage between nucleons and the form of high-momentum distribution caused by SRCs still depend on the model, and different models also have their own advantages in characterizing SRC, for example, the Green's function method \cite{doi:10.1142/6821} provides a powerful framework for describing nucleon SRCs in nuclear systems, and can help to shed light on the physics underlying these phenomena.  In contrast, the $1/k^{4}$ form has the advantage of portraying a high-momentum tail on the one hand in its simple form, and on the other hand, it can qualitatively give similar high-momentum tail calculated using microscopic self-consistent Green's function (SCGF) theory \citep{PhysRevC.89.044303,Rios_2020}. Moreover, because of its simplicity, the $1/k^{4}$ form has also been widely used in understanding nuclear matter properties \citep{PhysRevC.96.054603,PhysRevC.90.064312,PhysRevC.92.011601,PhysRevC.91.025803,PhysRevC.96.054603,PhysRevC.104.034603} and NS prosperties \citep{PhysRevC.101.065202,Hong_2022,LI2019118,LU2021122200,PhysRevD.105.023008}.

For asymmetry nuclear matter, based on the microscopic theoretical model predicting an approximate linear correlation between isospin-asymmetry and single-nucleon distribution, we can express the nucleon momentum distribution in the following form:
\begin{equation}
f(k)_{p}=\left\{
\begin{array}{lll}
C_{1}^{(p)}, & k \leq k_{F}^{(p)} \\
\frac{C_{2}^{(p)}(1-\beta)}{k^{4}}, & k_{F}^{(p)}<k \leq \lambda k_{F}^{(p)} \\
0, & k>\lambda k_{F}^{(p)}\\
\end{array}\right.
\end{equation}

\begin{equation}
f(k)_{n}=\left\{
\begin{array}{ll}C_{1}^{(n)}, & k \leq k_{F}^{(n)} \\
 \frac{C_{2}^{(n)}(1+\beta)}{k^{4}} , & k_{F}^{(n)}<k \leq \lambda k_{F}^{(n)} \\
 0, & k>\lambda k_{F}^{(n)}
 \end{array}\right.
\end{equation}
where $\beta$ is the isospin-asymmetry, and $k_{F}$ and $\lambda$ are the Fermi momentum and high-momentum cutoff, respectively. According to experiments with proton-knockout reactions utilizing megaelectron-energy electron beams, around $20\%\sim25\%$ of the nucleons have high momentum distribution \citep{R.SUBEDIR}. Furthermore, the deuteron momentum distribution suggests that the cutoff $\lambda$ should be around $2.75\pm0.25$ \citep{PhysRevC.92.045205,PhysRevC.91.025803}. In the next discussion, we take the cutoff to be 2.75, while taking high-momentum ratio to be typically 25\%. Combining the normalization conditions, we can fix the values of $C_{1}^{(n,p)}$, $C_{2}^{(n,p)}$, and they are $C_{1}^{(p)}=1-0.25(1-\beta)$, $C_{2}^{(p)}=0.25k_{F}^{(p)4}/(3-3/\lambda)$, $C_{1}^{(n)}=1-0.25(1+\beta)$ and $C_{2}^{(n)}=0.25k_{F}^{(n)4}/(3-3/\lambda)$ respectively. After incorporating results into Eqs. (1) and (2), the SRC-revised energy density and pressure are derived as follows:
 \begin{widetext}
\begin{eqnarray}
\nonumber
&&\mathcal{E}_{\mathrm{SRC}}=\sum_{N} \frac{1}{\pi^{2}} \int_{0}^{k_{N}} \mathrm{~d} k k^{2} \sqrt{k^{2}+m^{* 2}}+\sum_{l} \frac{1}{\pi^{2}} \int_{0}^{k_{l}} \mathrm{d} k k^{2} \sqrt{k^{2}+m_{l}^{2}}+\sum_{N} \frac{1}{3 \pi^{2}} k_{N}^{3}\left(g_{\omega N} \omega_{0}+g_{\varrho N} I_{3 N} \varrho_{0}\right)\\
\nonumber
&&-\frac{0.25(1-\beta)}{\pi^{2}} \int_{0}^{k_{F}^{(n)}} \mathrm{d} k k^{2}\left(g_{\omega n} \omega_{0}+g_{\varrho n} I_{3 n} \varrho_{0}+\sqrt{k^{2}+m^{* 2}}\right)
-\frac{0.25(1+\beta)}{\pi^{2}} \int_{0}^{k_{F}^{(p)}} \mathrm{d} k k^{2}\left(g_{\omega p} \omega_{0}+g_{\varrho p} I_{3 p} \varrho_{0}+\sqrt{k^{2}+m^{* 2}}\right)\\
\nonumber
&&+\frac{C_{2}^{(n)}(1-\beta)}{\pi^{2}} \int_{k_{F}^{(n)}}^{\lambda k_{F}^{(n)}} \mathrm{d} k k^{2} \frac{\left(g_{\omega n} \omega_{0}+g_{\varrho n} I_{3 n} \varrho_{0}+\sqrt{k^{2}+m^{* 2}}\right)}{k^{4}}+\frac{C_{2}^{(p)}(1+\beta)}{\pi^{2}} \int_{k_{F}^{(p)}}^{\lambda k_{F}^{(p)}} \mathrm{d} k k^{2} \frac{\left(g_{\omega p} \omega_{0}+g_{\varrho p} I_{3 p} \varrho_{0}+\sqrt{k^{2}+m^{* 2}}\right)}{k^{4}}\\
\nonumber
&&+\frac{1}{2} m_{\sigma}^{2} \sigma_{0}^{2}-\frac{1}{2} m_{\omega}^{2} \omega_{0}^{2}-\frac{1}{2} m_{\varrho}^{2} \varrho_{0}^{2}-\Lambda_{\omega}\left(g_{\varrho N} \varrho_{0}\right)^{2}\left(g_{\omega N} \omega_{0}\right)^{2}+\frac{1}{3}g_{2} \sigma_{0}^{3}+\frac{1}{4} g_{3} \sigma_{0}^{4}\\
\end{eqnarray}
and
\begin{eqnarray}
\nonumber
&&P_{\mathrm{SRC}}=\frac{1}{3} \sum_{N} \frac{1}{\pi^{2}} \int_{0}^{k_{N}} \mathrm{~d} k \frac{k^{4}}{\sqrt{k^{2}+m^{* 2}}}+\frac{1}{3} \sum_{l} \frac{1}{\pi^{2}} \int_{0}^{k_{l}} \mathrm{d} k \frac{k^{4}}{\sqrt{k^{2}+m_{l}^{2}}}-\frac{1}{2} m_{\sigma}^{2} \sigma_{0}^{2}+\frac{1}{2} m_{\omega}^{2} \omega_{0}^{2}+\frac{1}{2} m_{\varrho}^{2} \varrho_{0}^{2}+\Lambda_{\omega}\left(g_{\varrho N} \varrho_{0}\right)^{2}\left(g_{\omega N} \omega_{0}\right)^{2}\\
\nonumber
&&-\frac{1}{3} g_{2} \sigma_{0}^{3}-\frac{1}{4} g_{3} \sigma_{0}^{4}-\frac{1}{3} \frac{0.25(1-\beta)}{\pi^{2}} \int_{0}^{k_{F}^{(n)}} \mathrm{d} k \frac{k^{4}}{\sqrt{k^{2}+m^{* 2}}}+\frac{C_{2}^{(n)}(1-\beta)}{3\pi^{2}} \int_{k_{F}^{(n)}}^{\lambda k_{F}^{(n)}} \mathrm{d} k \frac{k^{4}}{k^{4}\sqrt{k^{2}+m^{* 2}}}\\
&&-\frac{1}{3} \frac{0.25(1+\beta)}{\pi^{2}} \int_{0}^{k_{F}^{(p)}} \mathrm{d} k \frac{k^{4}}{\sqrt{k^{2}+m^{* 2}}}+\frac{C_{2}^{(p)}(1+\beta)}{3\pi^{2}} \int_{k_{F}^{(p)}}^{\lambda k_{F}^{(p)}} \mathrm{d} k \frac{k^{4}}{k^{4}\sqrt{k^{2}+m^{* 2}}}
\end{eqnarray}
 \end{widetext}
Then the total energy density and pressure, revised jointly by admixing DM as well as SRC, are expressed as
\begin{eqnarray}
\mathcal{E}=\mathcal{E}_{\mathrm{DM}}+\mathcal{E}_{\mathrm{SRC}},~~~ P=P_{\mathrm{DM}}+P_{\mathrm{SRC}}.
\end{eqnarray}

As a part of standard NS calculation scheme, in the NS outer crust, where the density is around $6.3 \times 10^{-12} \mathrm{fm}^{-3} \leqslant n \leqslant 2.46 \times 10^{-4}\mathrm{fm}^{-3}$, we employ the Baym-Pethick-Sutherland (BPS) EOS \citep{1971ApJ...170..299B}. For NS inner crust area with a density of $2.46 \times 10^{-4} \mathrm{fm}^{-3} \leqslant n \leqslant n_{t}$, we adopt the polytropic parametrized EOSs of $P=a+b \varepsilon^{4 / 3}$\citep{PhysRevC.93.014619,PhysRevC.79.035802,Carriere_2003}, where $a$ and $b$ are related to the core-crust transition $n_{t}$. In this study, we examine the possible weak interaction between DM and ordinary baryonic matter, which occurs via the exchange of Higgs bosons. When solving the hydrostatic Tolman-Oppenheimer-Volkoff (TOV) equation, they resemble the degrees of freedom of hyperons inside NSs, hence we can regard the mixed system of DM and baryonic matter as a single-fluid system. Adopting the static spherically symmetric space-time background given by
\begin{equation}
d s^{2}=-e^{\nu(r)} d t^{2}+e^{\lambda(r)} d r^{2}+r^{2}\left(d \theta^{2}+\sin ^{2} \theta d \phi^{2}\right).
\end{equation}
with solving the Einstein equation, we can derive the Tolman-Oppenheimer-Volkoff (TOV) equation \citep{Oppenheimer1939374}
\begin{equation}
\frac{\mathrm{d} p}{\mathrm{d} r}=-\frac{(p+\epsilon)\left(M+4 \pi r^{3} p\right)}{r(r-2 M)}, \\
\end{equation}
\begin{equation}
\mathrm{d} M=4 \pi r^{2} \epsilon \mathrm{d} r,
\end{equation}
in which the corresponding metric functions express as
\begin{equation}
e^{\lambda(r)}=(1-2m / r)^{-1},
\end{equation}
\begin{equation}
\nu(r)=2\int_{r}^{\infty} d r^{\prime} \frac{e^{\lambda\left(r^{\prime}\right)}}{r^{\prime 2}}\left(m+4 \pi r^{\prime 3} p\right).
\end{equation}
Then with the beta equilibrium and charge conservation condition, the NS mass and radius can be determined.

\begin{figure*}[ht!]
\centering
\includegraphics[width=6in]{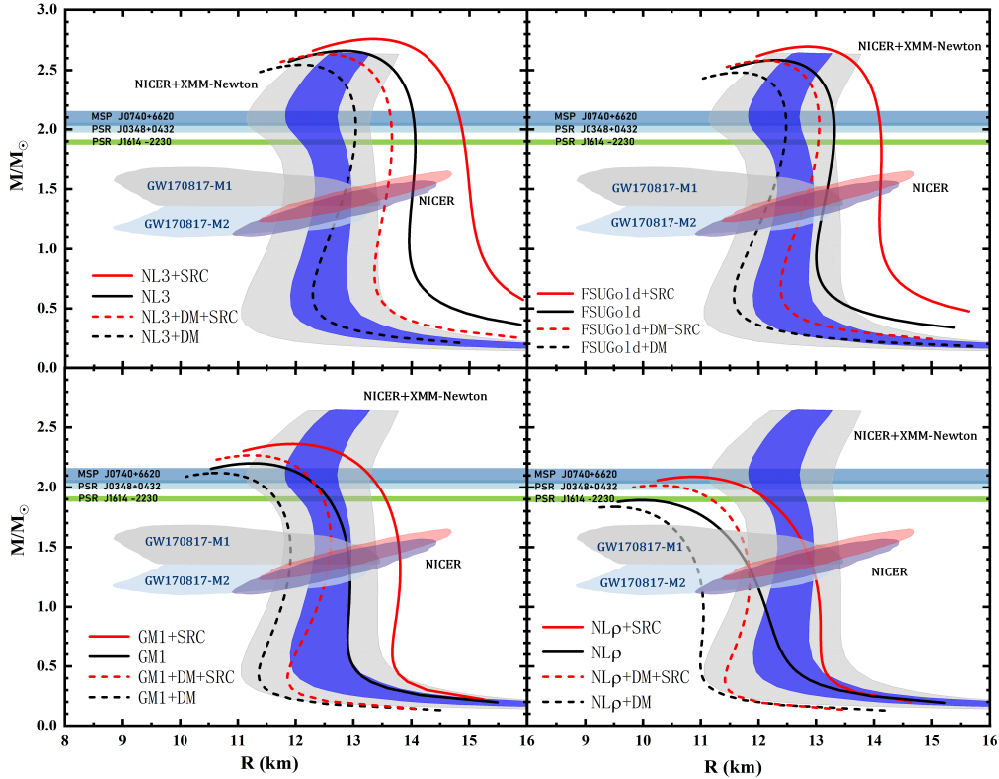}
\caption{The NS mass-radius relationships are calculated from four widely used relativistic parameter sets, each of which considers four cases, corresponding to the original version (black solid line), the SRC-revised version (red solid line), the DM-revised version (black dashed line), and both the SRC and DM-revised version (red dashed line). The three horizontal light-colored bars indicate the three massive NSs MSP J0740+6620, PSR J0348+0432 and PSR J1614-2230, the middle elliptical shaded region is the constraints given by GW170817 and NICER, the vertical long shaded region is the mass-radius constraint given by NICER in conjunction with XMM-Newton. \label{fig:1}}
\end{figure*}

\section{The construction of parameters}
To better visualize the implications of SRC and DM on NSs, we choose four extensively used parameter sets to characterize the NS mass-radius relation, namely NL3 \citep{PhysRevC.55.540}, FSUGold \citep{PhysRevLett.95.122501}, GM1 \citep{PhysRevLett.67.2414} and NL$\rho$ \citep{PhysRevC.65.045201}, as these sets can provide a good description of the properties of ground state nuclei and can be extrapolated to study standard NS matter. Moreover, as we mentioned before after the introduction of SRC in the mean field, in addition to the revisions of energy density and pressure, the coupling parameters between nucleons also need to be revised accordingly. The revised results are shown in Fig.1, where the black solid line indicates the mass-radius curve obtained from the original version, while the red solid line indicates the result of SRC effect, which makes the EOSs stiffer due to the additional correlation effect, being consistent with the conclusions from \citep{PhysRevC.101.065202,LU2021122200}. The black dashed line indicates the case of admixed DM, and it can be found that, compared to the original parameters, the introduction of DM substantially softens the EOS. To further compare with the astronomical observations, the constraints from different groups are attached to the figure, where the three horizontal light-colored bars indicate the three massive NSs \citep{RN583,Arzoumanian_2018,Antoniadis1233232,RN584,Fonseca_2021}, the middle elliptical shaded region indicates the constraints given by GW170817 \citep{PhysRevLett.119.161101,PhysRevLett.121.161101,PhysRevLett.120.172703,PhysRevLett.120.261103,PhysRevLett.121.091102} and NICER \citep{Miller_2019,Riley_2019,Miller_2021}, the vertical long shaded region indicates the mass-radius constraint given by NICER in conjunction with XMM-Newton through Bayesian analysis \citep{Riley_2021,Miller_2021}, in which the blue and gray region represent the 75\% and 90\% confidence intervals, respectively. Unfortunately, considering the SRC alone would largely increase the radius of NSs and thus make it easier to deviate from the astronomical constraints, as in the case of parameter NL3 and FSUGold, while considering admixed DM alone would make the EOSs excessively soft, as in the case of parameter NL$\rho$. If the effects of both SRC and DM are considered inside the NSs, as shown by the red dotted line, their complementary mechanism makes the EOS curves easier to satisfy the observational constraints. Although we cannot draw an absolute conclusion from this due to the model dependence of the parameters, all four parameter sets support our conclusion of having both SRC and DM effects in the NS interior. Moreover, the SRC and admixed DM exhibit similar effects on the mass-radius curves for all four sets, and in view of this, here we have chosen only GM1 for the next discussion. The values are shown in Table I, where the first column is the original version for GM1 and the second column is the SRC and DM revised version (GM1+SD). By adjusting the range of $L$ while maintaining the other saturation properties, we investigate the impact of $L$ on the modified version. Such an approach keep the isoscalar nature unchanged and can satisfy the properties at the saturation density, and has been used in many works \citep{PhysRevC.101.065202,PhysRevC.98.065804}. We extract $L$ located around 50$\sim$90 MeV within a relatively credible interval based on previous experimental and theoretical studies, and its corresponding modified versions (GM1+SD+L) are shown in the second half of the Table I.

\begin{table*}
\caption{\label{tab:table1}
 The original relativistic parameter set GM1 and its associated revised version, GM1+SD denotes the set revised by incorporating SRC and DM, and GM1+SD+L denotes the new parameters based on the GM1+SD by adjusting the isovector $L$ while keeping the isoscalar properties (saturation density $n_{0}$, incompressibility coefficient $K$, nucleon effective mass $m^{*}/m$, binding energy per nucleon $-B/A$ ) unchanged.
}
\begin{ruledtabular}
\begin{tabular}{cccccccccccc}
             & GM1& GM1+SD &  GM1+SD+L50 & GM1+SD+L60 &GM1+SD+L70 & GM1+SD+L80 &GM1+SD+L90  \\ \hline
$m_{N}(\textrm{MeV})$ &939&939&939&939&939&939&939\\
$m_{\sigma}(\textrm{MeV})$ &512&512&512&512&512&512&512\\
$m_{\omega}(\textrm{MeV})$ &783&783&783&783&783&783&783\\
$m_{\varrho}(\textrm{MeV})$ &770&770&770&770&770&770&770\\
$n_{0}(\textrm{fm}^{-3})$&0.153&0.153&0.153&0.153&0.153&0.153&0.153\\
$K(\textrm{MeV})$ &300&300&300&300&300&300&300\\
$m^{*}/m$ &0.70&0.70&0.70&0.70&0.70&0.70&0.70\\
$-B/A(\textrm{MeV})$ &16.3&16.3&16.3&16.3&16.3&16.3&16.3\\
$E_{\textrm{sym}}(\textrm{MeV})$ &32.5&32.5&32.5&32.5&32.5&32.5&32.5\\
$L(\textrm{MeV})$ &94&94&50&60&70&80&90\\
 \hline
$g_{\sigma}$ &8.910&9.321&9.321&9.321&9.321&9.321&9.321\\
$g_{\omega}$  &10.610&9.645&9.645&9.645&9.645&9.645&9.645\\
$g_{\varrho}$  &8.196&13.566&14.934&14.587&14.263&13.960&13.675\\
$b$  &0.0029&0.0029&0.004&0.004&0.004&0.004&0.004\\
$c$  &-0.001&-0.001&-0.003&-0.003&-0.003&-0.003&-0.003\\
$\Lambda_{\omega}$ &0&0.002&0.011&0.009&0.007&0.005&0.003\\
\end{tabular}
\end{ruledtabular}
\end{table*}

As a next step, we need to decide a key question: whether the sets constructed by different $L$ are able to describe empirical properties at low densities while also fulfilling constraints from multi-messenger observations?  To this end, we have calculated the binding energy per nucleon (E/A) for symmetry nuclear matter (SNM) and pure neutron matter (PNM), as shown in Fig.2. The different color curves in the upper part indicate different $L$ for PNM, in which below the saturation number density $n_{0}$ (corresponding to the black vertical dashed line), the smaller $L$ gives a relatively large E/A, while the opposite is true above $n_{0}$, and these results are in agreement with \citep{PhysRevC.104.015802}. Furthermore, in order to constrain the PNM behavior at subsaturation density, we adopt the $\chi EFT$  given in \citep{PhysRevLett.110.032504}, which presents next-to-next-to-next-to-leading order (N$^{3}$LO) in the chiral expansion based on potentials developed by Epelbaum, Gl$\ddot{o}$ckle and Mei{\ss}ner (EGM). As shown in the light green area (the lower right corner corresponds to the enlarged version), our revised parameters are largely in compliance with the constraints. Furthermore, it can be seen that $\chi EFT$ is more supportive of smaller $L$. The blue line shows the E/A for SNM, which, unlike PNM, is clearly independent of a specific $L$ and has a minimum value at $n_{0}$, corresponding to the most stable ground state, of about -16 MeV.

\begin{figure}[ht!]
\centering
\includegraphics[width=3.3in]{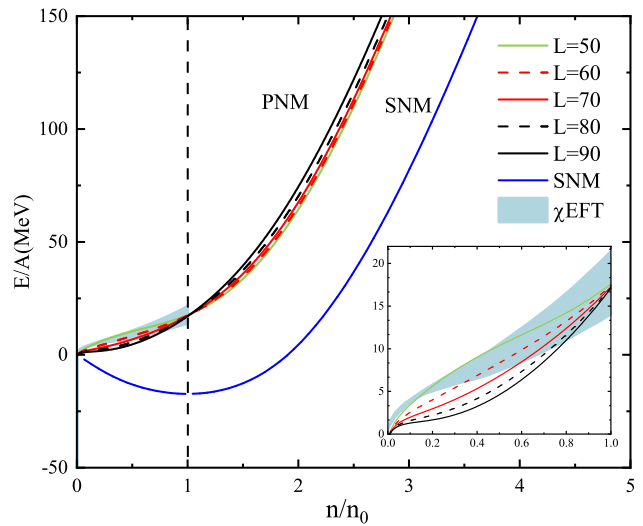}
\caption{The energy per nucleon (E/A) as a function of number density in pure neutron matter (PNM) and symmetry nuclear matter (SNM). The light green area gives constraints from $\chi EFT$. \label{fig:2}}
\end{figure}

\begin{figure}[ht!]
\centering
\includegraphics[width=3.3in]{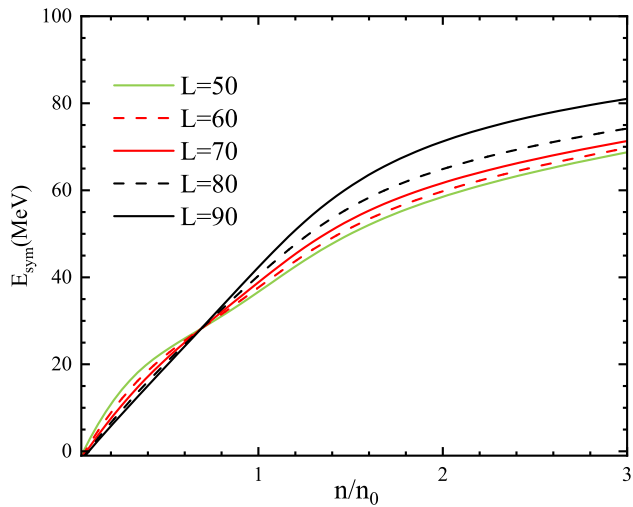}
\caption{The symmetry energy as a function of baryon number density for different $L$s.  \label{fig:4}}
\end{figure}

Fig.3 shows the symmetry energy $E_{\mathrm{sym}}$ as a function of density for different $L$. As can be seen, $E_{\mathrm{sym}}$ with a smaller $L$ have a lower (higher) value than those with a larger $L$ at $n > n_{0}$ ($n < n_{0}$), similar behavior is observed in \citep{10.1093/ptep/ptaa016,PhysRevC.104.015802}. Fig.4 gives the pressure of SNM, and we includes the experimental constraints obtained from collective flow data in heavy-ion collisions which are often used to constrain the EOS of SNM \citep{Huang_2020,PhysRevC.84.065810}. The enclosed area represents experimental data according to \citep{doi:10.1126/science.1078070}, and our result is in line with this constraint.
\begin{figure}[ht!]
\centering
\includegraphics[width=3.5in]{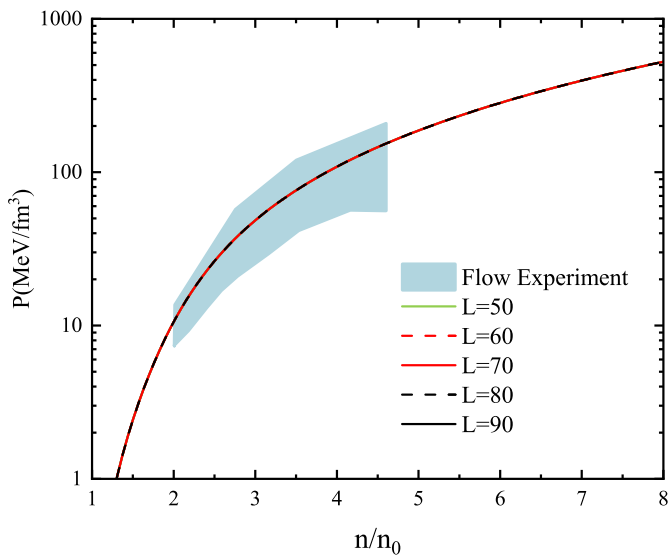}
\caption{Pressure as a function of baryon number density for different $L$s in SNM.  \label{fig:4}}
\end{figure}

Apart from that, reasonable parameters also need to satisfy the constraints from multi-messenger observations, including being able to generate massive NSs, meeting the mass-radius constraint derived from NICER+XMM-Newton, and satisfying the tidal deformability range given by GW170817. Fig.5 depicts the mass-radiu relation, with different colors representing different $L$s and with shaded regions showing the same constraints as in Fig.1. The influence trend of the symmetry energy slope $L$ on the mass-radius curve can be observed to be similar, where a larger $L$ can support a harder EOS and hence give a larger radius. In addition, the range of  $L$ we have selected, from the lowest 50 MeV to the highest 90 MeV, can yield to the observational constraints in both low-mass and massive NS regions. Specially, at low-mass of $1.4 M_{\odot}$, the radius varies from 11.5 km ($L$ = 50) to 12.5 km ($L$ = 90), and at massive mass of $2 M_{\odot}$, the radius varies from 11.87 km ($L$ = 50) to 12.16 km ($L$ = 90). In addition to the above, the tidal deformability-mass relations at different $L$, along with respective tidal deformability of two NSs in GW170817 are given in Fig.6, where the vertical line in the left panel indicates the tidal constraint inferred from GW170817, the solid dot in the right panel represents the $1.4 M_{\odot}$ NSs, and the gray dashed lines indicate the confidence intervals of 50\% and 90\%, respectively. As can be seen in Fig.6, a smaller $L$ gives a relatively smaller tidal deformability, which can be inspired by Fig.4, where a smaller $L$ yields a smaller radius, so a smaller radius makes it harder for a star to distort in the tidal field, resulting in a smaller tidal value. More importantly, we find that the tidal deformability obtained from the five sets after absorbing SRC and DM all fall within the range given by the GW170817 event.

\begin{figure}[ht!]
\centering
\includegraphics[width=3.5in]{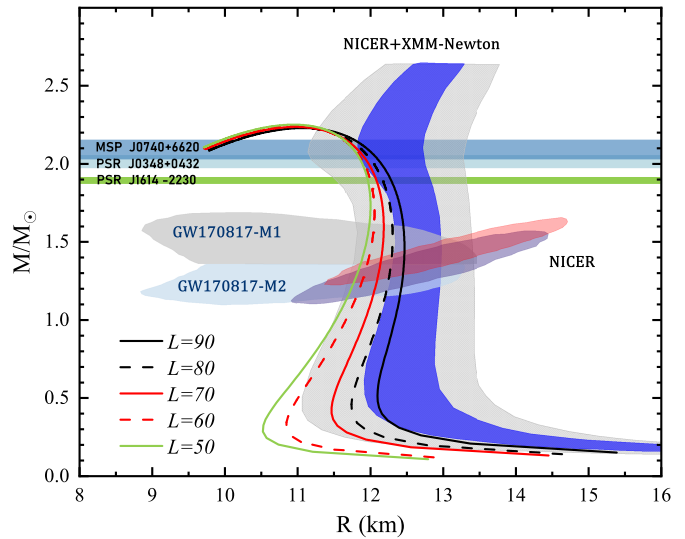}
\caption{The mass-radius relations for different $L$s, with the horizontal lines indicating the three massive NSs MSP J0740+6620, PSR J0348+0432 and PSR J1614-2230. The shaded areas indicate the constraints from GW170817 as well as NICER and XMM-Newton, respectively.  \label{fig:5}}
\end{figure}

\begin{figure*}[ht!]
\centering
\includegraphics[width=6.5in]{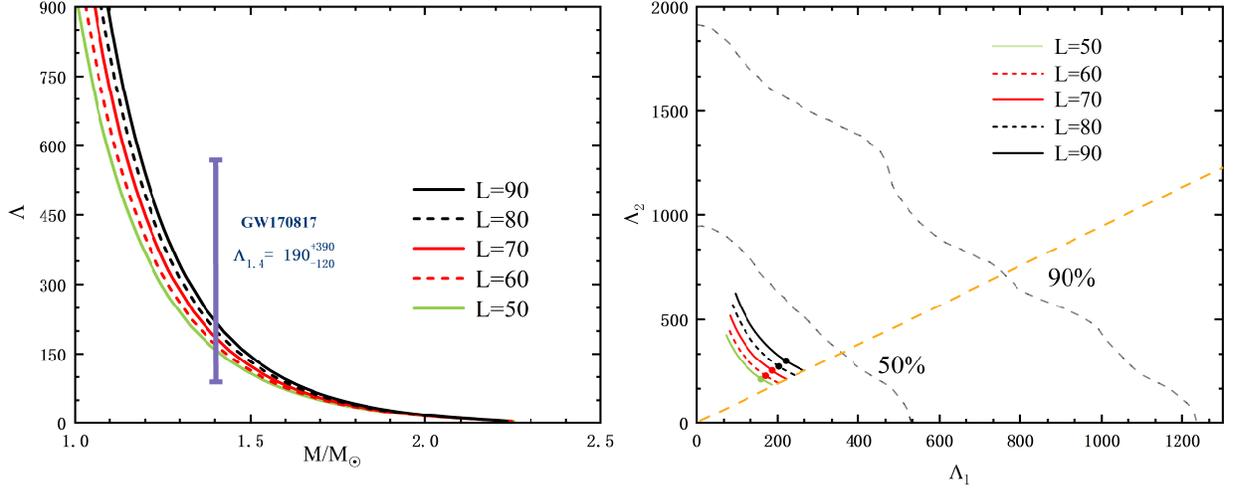}
\caption{Left: Relationship between tidal deformability and mass, with different colors indicating different $L$s. Vertical line shows constraint from GW170817. Right: The relationship between $\Lambda_{1}$ and $\Lambda_{2}$ in GW170817 event calculated with different $L$, with solid dots indicating the $1.4 M_{\odot}$ NSs. The gray dash-lines indicate the 50\% and 90\% confidence intervals. \label{fig:6}}
\end{figure*}

As a whole, after taking into account the SRC and admixed DM in NS interior, the five parameter sets constructed by varying $L$ based on the GM1 can not only characterize the behavior of nuclear matter at low densities (Figs.2-4), but also meet the requirements inferred from recent multi-messenger observations (Figs.5-6). Next, using these parameter sets, we investigate the effect of $L$ on the currently undetectable NS oscillation frequencies.

\section{Non-radial oscillation and radial oscillation}
\subsection{Non-radial oscillation}
Thorne and Campollataro \citep{1967ApJ...149..591T} are credited with being the ones who first suggested discussing non-radial modes within the context of general relativity. In our work, we consider a non-rotating NS with an ideal fluid interior, using the Cowling approximation \citep{10.1093/mnras/101.8.367,1983ApJ...268..837M,10.1093/mnras/stx3067}, which ignores the space-time metric perturbation but retains density perturbations caused by fluid oscillations \citep{RN84}. Recent research indicates that the difference between the $f$-mode calculated using the Cowling approximation approach and the complete linearized equations of general relativity is less than 20\%, the $p$-mode error is about 10\% \citep{10.1093/mnras/289.1.117}, and the $g$-mode error is only a few percent \citep{PhysRevD.65.024010}. This shows the practicality of Cowling approximation \citep{V_squez_Flores_2014}, and based on this, the fluid perturbations consist of a spherical harmonic function $Y_{l m}(\theta, \phi)$ and a time-dependent part e$^{\mathrm{i} \omega t}$, resulting in the following Lagrangian fluid displacements associated with infinitesimal oscillatory perturbations:
\begin{eqnarray}
\nonumber
\zeta^{\mathrm{i}}=&& {\left[\mathrm{e}^{-\Lambda(r)} W(r),-V(r) \partial_{\theta},-V(r) \sin ^{-2} \theta \partial_{\phi}\right] r^{-2} } \\
&& \times \mathrm{Y}_{l m}(\theta, \phi) \mathrm{e}^{\mathrm{i} \omega t},
\end{eqnarray}
where $W(r)$ and $V(r)$ satisfy
\begin{eqnarray}
\nonumber
&&\frac{\mathrm{d} W(r)}{\mathrm{d} r}=\frac{\mathrm{d} \varepsilon}{\mathrm{d} p}\left[\omega^{2} r^{2} \mathrm{e}^{\Lambda(r)-2 \phi(r)} V(r)+\frac{\mathrm{d} \Phi(r)}{\mathrm{d} r} W(r)\right] \\
\nonumber
&&-l(l+1) \mathrm{e}^{\Lambda(r)} V(r), \\
\\
\nonumber
&&\frac{\mathrm{d} V(r)}{\mathrm{d} r}=2 \frac{\mathrm{d} \Phi(r)}{\mathrm{d} r} V(r)-\frac{1}{r^{2}} \mathrm{e}^{\Lambda(r)} W(r),
\end{eqnarray}
with
\begin{eqnarray}
\frac{\mathrm{d} \Phi(r)}{\mathrm{d} r}=\frac{-1}{\varepsilon(r)+p(r)} \frac{\mathrm{d} p}{\mathrm{~d} r}.
\end{eqnarray}
the above equations can be viewed as the eigenvalue equations of $\omega$ for suitable boundary conditions. Inside NS($r=0$), $W(r)$ and $V(r)$ exhibit the following approximate behaviors:
\begin{eqnarray}
W(r)=A r^{l+1}, \quad V(r)=-\frac{A}{l} r^{l},
\end{eqnarray}
with A being an arbitrary constant. The outer boundary condition is the pressure will disappear at the NS surface
\begin{eqnarray}
\omega^{2} \mathrm{e}^{\Lambda(R)-2 \Phi(R)} V(R)+\left.\frac{1}{R^{2}} \frac{\mathrm{d} \Phi(r)}{\mathrm{d} r}\right|_{r=R} W(R)=0 .
\end{eqnarray}

In this paper, we determine the typical non-radial bar mode instability for quadrupole oscillations ($l$ = 2) \citep{PhysRevD.87.084010} using the above revised parameter sets, and show the $f$-mode frequencies under different $L$s in Fig.7. If we take a closer look at the region roughly between $1.4 M_{\odot}$ and $2 M_{\odot}$, a smaller $L$ is more likely to excite $f$-mode oscillations, giving a larger frequency. The curve starts to bend after approaching maximum mass, after which it corresponds to the unstable NS. In the stable region, the frequency increases with mass, for $1.4 M_{\odot}$, different $L$s give frequencies in the range of 2.10 kHz-2.19 kHz, and for $2 M_{\odot}$, in the range of 2.35 kHz-2.39 kHz. If future gravitational wave detectors like the Cosmic Explorer or Einstein Telescope \citep{PhysRevD.103.044024,Punturo_2010,PhysRevD.91.082001} are able to pick up the frequency in this range, we may have a good reason to speculate about the approximate mass of this NS.
\begin{figure}[ht!]
\centering
\includegraphics[width=3.5in]{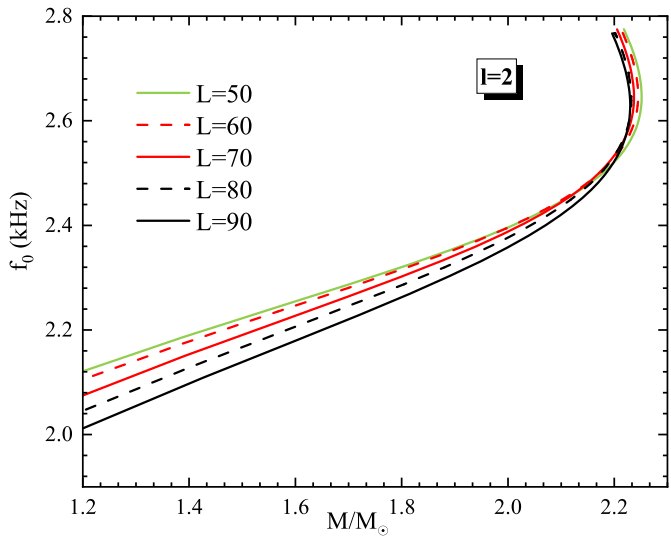}
\caption{The $f$-mode frequency of non-radial oscillations as a function of NS mass. We determine the typical non-radial bar mode instability for quadrupole oscillations ($l$ = 2)  \label{fig:7}}
\end{figure}

\begin{figure*}[ht!]
\centering
\includegraphics[width=6.5in]{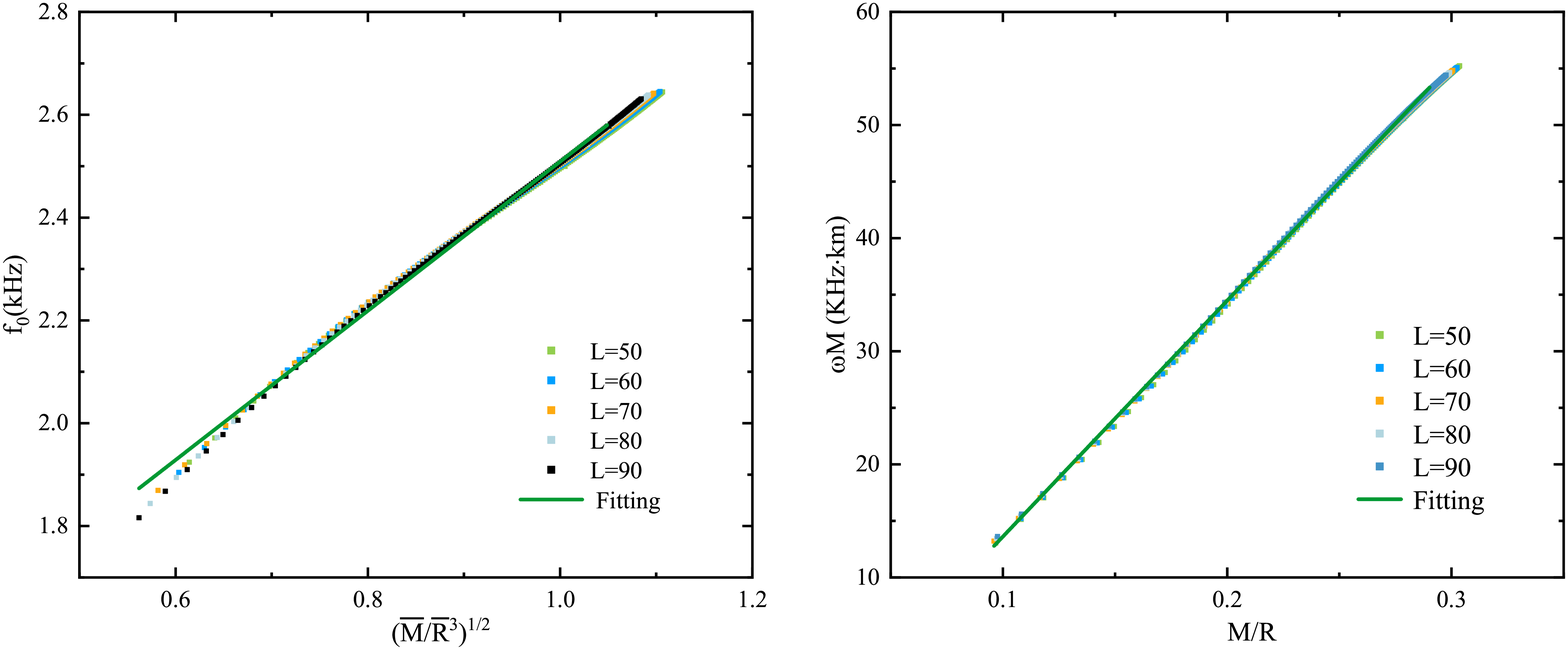}
\caption{Left: Linear relationship between $f$-mode frequency and $\sqrt{\frac{\bar{M}}{\bar{R}^{3}}}$ with different $L$.  Right: The relationship between $\omega M$ and compactness $M/R$. \label{fig:8}}
\end{figure*}

In GW asteroseismology, we can relate the oscillation frequency and damping timescale to NS bulk properties like mass, radius, and tidal deformability, and then establish an empirical relationship between them \citep{PhysRevD.88.044052,PhysRevD.70.124015,PhysRevD.87.084010,PhysRevLett.107.101102,PhysRevC.103.035810,PhysRevC.99.045806}. The original study was proposed by Andersson and Kokkotas \citep{PhysRevLett.77.4134,10.1046/j.1365-8711.1998.01840.x}, who suggested that the stellar dynamical timescale is related to its mean density and established a empirical relationship as
\begin{eqnarray}
f(\mathrm{kHz})=a+b \sqrt{\frac{\bar{M}}{\bar{R}^{3}}},
\end{eqnarray}
where the dimensionless parameters $\bar{M}=M/(1.4M_{\odot})$ and $\bar{R}=R/(10\mathrm{km})$. A further evaluation of this relationship was carried out by taking into account the rotation effect \citep{PhysRevD.88.044052} and exotic matter \citep{PhysRevD.70.124015} in the NS. However, these studies are not supported by the current results of multi-messenger astronomy and need further adjustment. Our goal is to re-calibrate this empirical relationship by considering SRC and admixed DM inside NSs, with satisfying the behaviours near saturation density and multi-messenger observations. The specific results are shown on the left in Fig.8, where the different scatter points indicate the results given by different $L$s and the solid green line is the result of linear fitting. We present the fitting result in Table 2, as a comparison, we also present the results by other researchers. It is found that incorporating SRC and DM within NSs still maintains a good linear relationship, and the discrepancies are very small compared to other models, further illustrating the model-independent nature. Another similar scenario is that of a excellent linear correlation between $\omega M$ and $M/R$, which has been used to analyze $g$-mode, $p$-mode, and $f$-mode, and the fitting is shown in the right of Fig.8 with $\omega M=208.687(M/R)-7.257$, which shows a stronger linear correlation than the case on the left. These empirical relations can be used to easily decode NS internal information once the $f$-mode frequencies can be detected. On the one hand, the $f$-mode frequency is capable of not only providing information on the NS density, but also inferring the compactness ($M/R$) which further correlates to NS tidal deformability. On the other hand, as the NICER detector are upgraded to simultaneously give accurate masses and radii in the foreseeable future, we can in turn be able to predict its $f$-mode frequency well. As mentioned, since the compactness controls the tidal deformability directly, we can also relate the $f$-mode frequency to the tidal deformability by extrapolating the relationship between $f$-mode and compactness. For this purpose we plot $\Lambda_{1.4}$ and $\Lambda_{2.0}$ with the corresponding $f$-mode frequencies in Fig.9, where the dots in different shades indicate different $L$s. It can be found that the larger $L$ is, the larger corresponding radius is (see Fig.5 for details), and the NS is more likely to undergo a deformation resulting in a larger tidal deformability (see Fig.6), but a larger radius is more conducive to NS stability and thus more difficult to vibrate(see Fig.7), so the larger $L$, the larger tidal deformability and the smaller $f$-mode frequency. Furthermore, the frequency and tidal deformability show a good linear correlation, and we have fitted with $f_{0}=-0.001\Lambda_{1.4}+2.386$ and $f_{0}=-0.001\Lambda_{2.0}+2.719$, and the range of $\Lambda_{1.4}$ and $\Lambda_{2.0}$ extracted from different $L$ both satisfy the constraints from GW170817 and theoretical calculations \citep{PhysRevLett.121.091102}. This connection not only explains the influence of $L$ on frequency and tidal deformability, but it also predicts the observations. Together with the empirical relation shown in Fig.8, the constraints on the EOSs will become increasingly rigorous.

\begin{figure*}[ht!]
\centering
\includegraphics[width=6.5in]{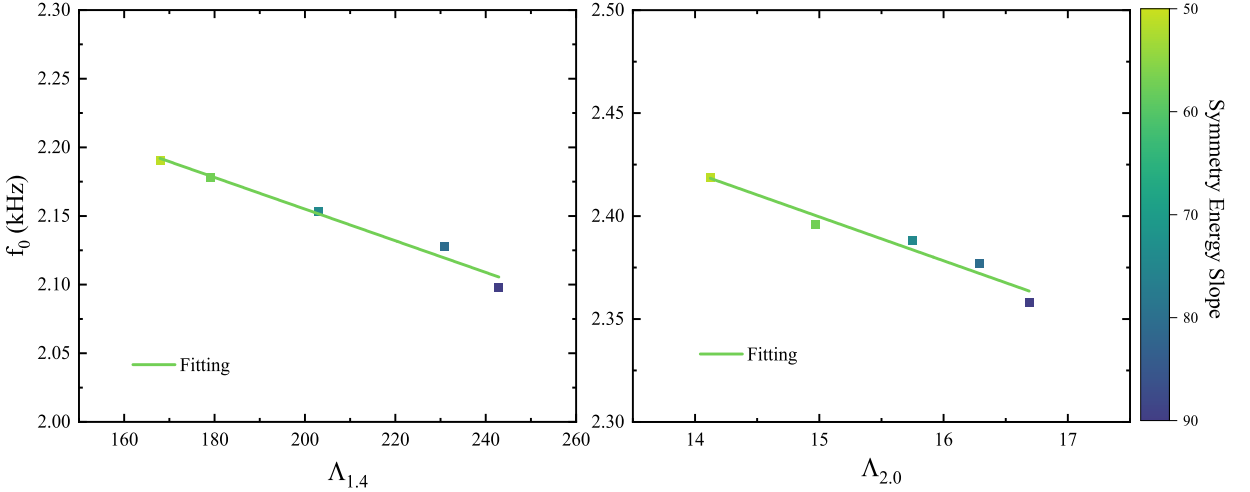}
\caption{Left and right show the linear relationship between the $f$-mode frequency and $\Lambda_{1.4}$ and $\Lambda_{2.0}$ respectively, where the different scatter points indicate the results given by different $L$s and the solid green line indicates the linear fitting. \label{fig:9}}
\end{figure*}

\begin{table}
\caption{\label{tab:table2}
 The numerical fitting results data from Fig.8. The variables $a$ and $b$ stand for the intercept and slope in Eq.(20), respectively, and the first row, which highlights the data of this work. The results from other groups are also provided for comparison in the final four lines.
}
\begin{ruledtabular}
\begin{tabular}{lccccccccccc}
             CASES &  a(kHz)&&  b(kHz) \\ \hline
\textbf{SRC and admixed DM} &\textbf{1.0581}&&\textbf{1.4507}\\
Andersson and Kokkotas &0.78&&1.6350\\
\citep{PhysRevLett.77.4134}& &&\\
Omar Benhar et al.&0.79&&1.5000\\
\citep{PhysRevD.70.124015}& &&\\
Daniela D. Doneva et al.&1.5620&&1.1510\\
\citep{PhysRevD.88.044052}& &&\\
Bikram Keshari Pradhan et al.&1.0750&&1.4120\\
\citep{PhysRevC.103.035810}& &&\\
\end{tabular}
\end{ruledtabular}
\end{table}

\begin{figure*}[ht!]
\centering
\includegraphics[width=6.5in]{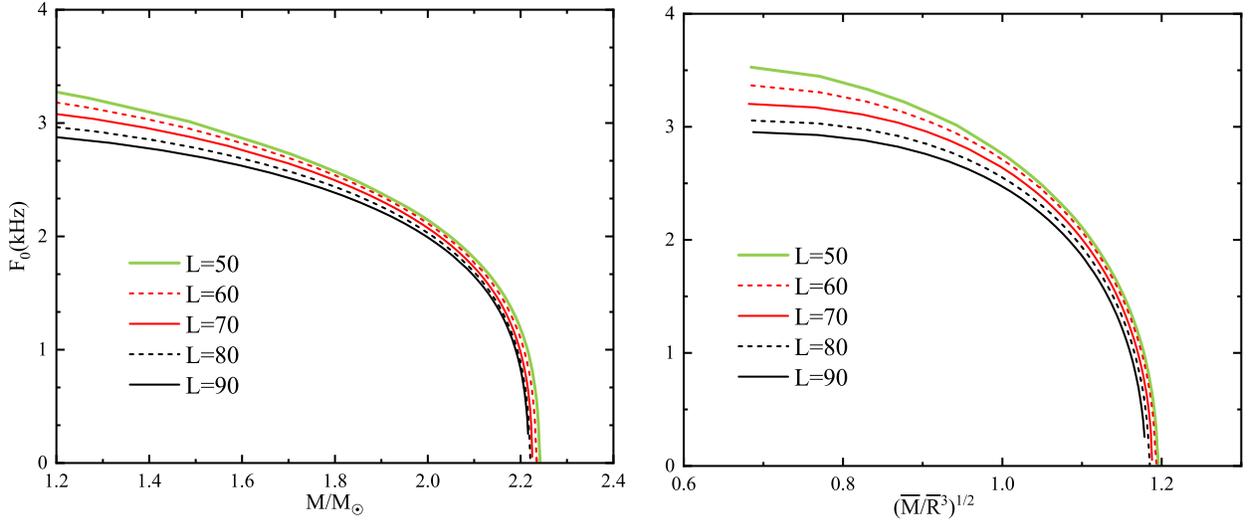}
\caption{Left: The radial $F_{0}$-mode frequency as a function of NS mass for different $L$s.  Right: Relationship between radial $F_{0}$-mode frequency and $\sqrt{\frac{\bar{M}}{\bar{R}^{3}}}$ for different $L$s. \label{fig:10}}
\end{figure*}

\begin{figure*}[ht!]
\centering
\includegraphics[width=6.5in]{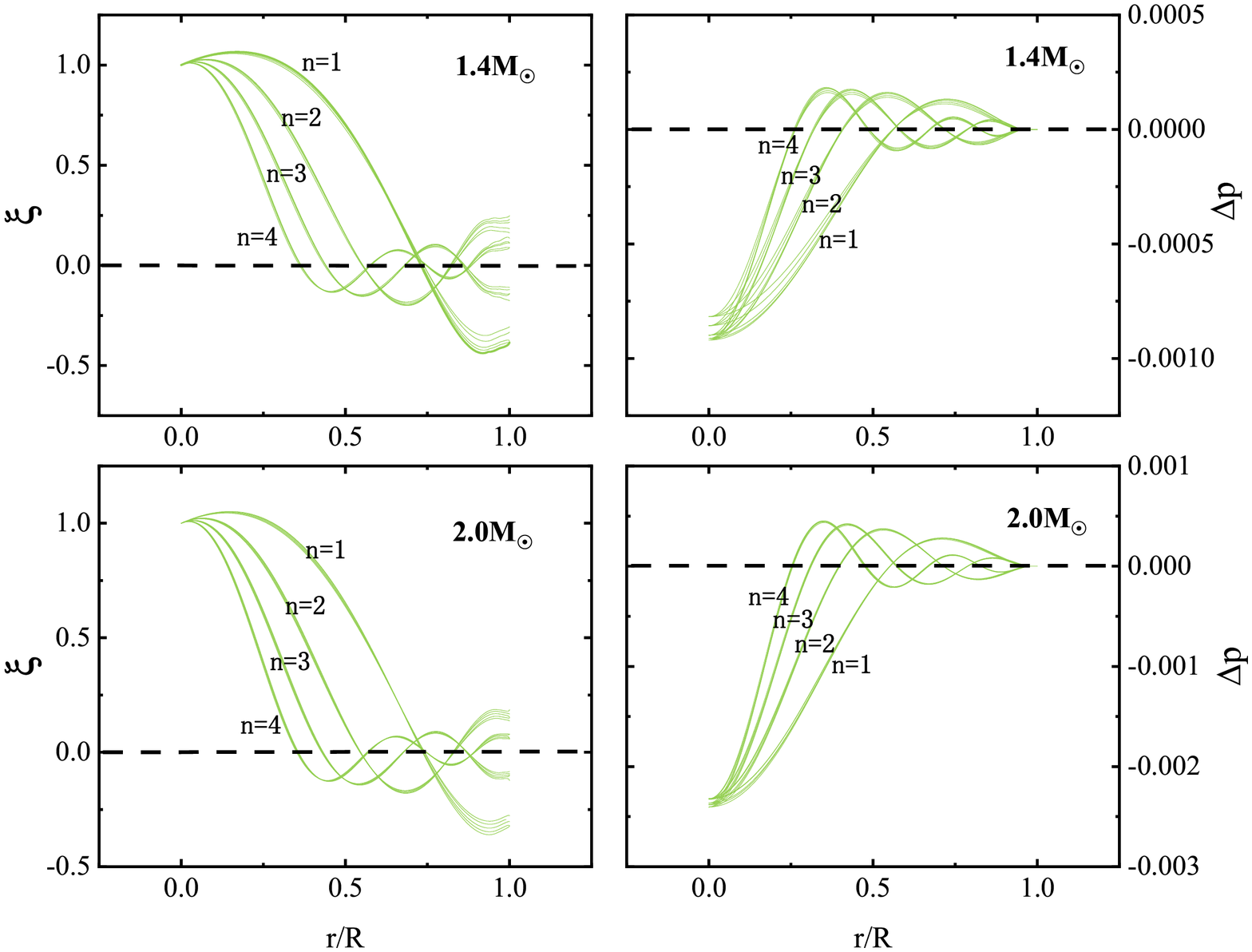}
\caption{ The perturbations $\xi$ and $P$ of a NS with 1.4$M_{\odot}$ (upper panel) or 2$M_{\odot}$ (down panel) as a function of the dimensionless radial parameter $r/R$ for different overtone numbers $n$=1,2,3,4. Different symmetry energy slopes are denoted by thin lines in each overtone group. \label{fig:11}}
\end{figure*}

\subsection{Radial oscillation}
It has been shown that radial oscillations of NSs, accompanied by non-radial oscillations\citep{PhysRevD.75.084038}, are a special class of non-radial oscillations, which in astroseismology correspond to spherical harmonic functions with zero orbital angular momentum. A seminal paper by Chandrasekhar \citep{PhysRevLett.12.114} was the first to discuss radial perturbations of compact stars and in the past 60 years, different work has been done to study the NSs' radial oscillations under the general relativity \citep{1983ApJS...53...93G,1977ApJ...217..799C,PhysRevD.73.084010,PhysRevD.75.084038,PhysRevD.101.103003,PhysRevD.101.063025,PhysRevD.103.103003}. Some studies have pointed out that radial oscillation can be used to distinguish between traditional NSs and strange stars \citep{1992AA} as well as hybrid stars \citep{Pereira_2018,PhysRevD.103.103003}, and can also often be used to describe the NS stability. Despite the fact that radial oscillations cannot directly produce gravitational radiation, they can couple and amplify gravitational wave signals \citep{PhysRevD.73.084010,PhysRevD.75.084038}. As described in Ref.\citep{1997A&A...325..217G}, the differential equations that characterize infinitesimal radial oscillations can be expressed as:
\begin{eqnarray}
\frac{\mathrm{d} \xi}{\mathrm{d} r}=-\frac{1}{r}\left(3 \xi+\frac{\Delta p}{\Gamma p}\right)-\frac{\mathrm{d} p}{\mathrm{~d} r} \frac{\xi}{(p+\varepsilon)},
\end{eqnarray}
\begin{eqnarray}
\nonumber
\frac{\mathrm{d} \Delta p}{\mathrm{~d} r}=&& \xi\left\{\omega^2 \mathrm{e}^{\lambda-v}(p+\varepsilon) r-4 \frac{\mathrm{d} p}{\mathrm{~d} r}\right\}+\xi\left\{\left(\frac{\mathrm{d} p}{\mathrm{~d} r}\right)^2 \frac{r}{(p+\varepsilon)}\right.\\
\nonumber
&&\left.-8 \pi \mathrm{e}^\lambda(p+\varepsilon) p r\right\}+\Delta p\left\{\frac{\mathrm{d} p}{\mathrm{~d} r} \frac{1}{(p+\varepsilon)}\right.\\
&&\left.-4 \pi(p+\varepsilon) r \mathrm{e}^\lambda\right\}
\end{eqnarray}
where $\omega$ is the radial eigenfrequency, $\xi$($\equiv\triangle r/r$) and $\triangle p$ characterize the radial displacement and Lagrangian perturbation of pressure respectively, both of which rely on the harmonic time form of $e^{i\omega t}$. $\Gamma$ stands for the relativistic adiabatic index:
\begin{eqnarray}
\Gamma=\left(1+\frac{\varepsilon}{p}\right) \frac{dp}{d\varepsilon}
\end{eqnarray}
Numerically solving above equations requires two boundary conditions, one is that in the NS interior ($r$ = 0) the term associated with $1/r$ must be finite \citep{1997A&A...325..217G}, so we have
\begin{equation}
(\Delta p)_{\text {center }}=-3(\xi \Gamma p)_{\text {center }}
\end{equation}
with $\xi(0)=1$ for the normalization condition and the second one is that the pressure perturbation near NS surface should become negligible, i.e. $\Delta p\rightarrow 0$ for $r\rightarrow R$. Taking a given NS EOS as an input, the above oscillation equations are Sturm-Liouville boundary value problems \citep{1992AA} and using the numerical shooting method we can get a series of so-called eigenvalue frequencies $\omega^{2}_{1}<\omega^{2}_{2}<...<\omega^{2}_{n}<...$, with $n$ being the number of nodes.

\begin{table*}
\caption{\label{tab:table3}
Eleven sets of high-mode frequencies $\nu_{n}$ under different $L$s. The two sets of data in each column represent the 1.4$M_{\odot}$ and 2$M_{\odot}$ counterparts respectively
}
\begin{ruledtabular}
\begin{tabular}{cccccccccccc}
            \textrm{ Overtone number} &&  $L$=50 MeV&&    $L$=60 MeV  &&    $L$=70 MeV &&    $L$=80 MeV&&    $L$=90 MeV  \\ \hline
$n$=1 &&3.095~(2.142)&&3.031~(2.113)&&2.952~(2.074)&&2.861~(2.031)&&2.777~(1.991)\\
$n$=2 &&7.885~(7.086)&&7.705~(7.031)&&7.498~(6.952)&&7.284~(6.842)&&7.090~(6.735)\\
$n$=3 &&12.141~(11.129)&&11.839~(11.029)&&11.507~(10.895)&&11.183~(10.715)&&10.888~(10.545)\\
$n$=4 &&16.288~(15.025)&&15.871~(14.877)&&15.417~(14.687)&&14.991~(14.441)&&14.594~(14.213)\\
$n$=5 &&20.395~(18.868)&&19.866~(18.671)&&19.294~(18.425)&&18.765~(18.114)&&18.256~(17.824)\\
$n$=6 &&24.482~(22.683)&&23.844~(22.438)&&23.153~(22.137)&&22.522~(21.768)&&21.906~(21.415)\\
$n$=7 &&28.558~(26.484)&&27.814~(26.192)&&26.999~(25.835)&&26.261~(25.401)&&25.516~(24.994)\\
$n$=8 &&32.626~(30.275)&&31.777~(29.936)&&30.829~(29.526)&&29.978~(29.031)&&29.066~(28.565)\\
$n$=9 &&36.688~(34.061)&&35.734~(33.676)&&34.638~(33.212)&&33.661~(32.657)&&32.521~(32.131)\\
$n$=10&&40.746~(37.842)&&39.684~(37.412)&&38.417~(36.894)&&37.392~(36.280)&&35.815~(35.694)\\
$n$=11&&44.798~(41.620)&&43.628~(41.145)&&42.158~(40.575)&&40.872~(39.901)&&38.895~(39.253)\\
\end{tabular}
\end{ruledtabular}
\end{table*}

Our next step is to examine how $L$ influences radial oscillations in terms of fundamental radial frequencies ($F_{0}$) and consecutive frequency differences (so-called Large Separation)$\triangle \nu_{n} =(\omega_{n}-\omega_{n-1})/2\pi$ \citep{refId0,PhysRevD.96.083013}. The fundamental radial frequencies $F_{0}$($n=0$) as a function of NS mass are given on the left of Fig.10. The smaller $L$ exhibit larger radial frequency, similar to the result from non-radial one, and the effect of $L$ on the frequency is more pronounced in low mass regions, where the frequency drops from 3.11 kHz ($L=50$ MeV) to 2.76 kHz ($L=90$ MeV) for 1.4$M_{\odot}$ and from 2.16 kHz to 1.99 kHz for 2$M_{\odot}$. Furthermore, unlike non-radial oscillations, the radial frequency decreases more rapidly at massive NS regions, and tends to zero when the mass indefinitely approaches the maximum stable mass (MSM). This is because when a NS reaches its MSM, any small radial perturbation does not cause the star to oscillate. Due to this fact, the linear relationship between the radial frequency and the mean density is no longer satisfied as it was for the non-radial one (see the right of Fig.10). Moreover, we can apply radial frequency equivalently to determine the star equilibrium condition, where a stable NS condition requires $\partial M / \partial n_{c}\geq0$ in the mass-radius curve, in which $\partial M / \partial n_{c}=0$ gives the MSM, corresponding to the point in figure where the frequency is zero. This is important because the MSM obtained from the mass-radius curve can only be given by a theoretical model calculation, whereas the radial oscillation frequency allows the MSM to be determined from an observed perspective. Combined with the results from non-radial oscillations together, we can infer whether the NS has a MSM, as radial oscillations are less likely to occur near the MSM (see Fig.10), while non-radial oscillations are more likely to produce high frequencies (see Fig.7), which means that if a small radial frequency be detected along with a high non-radial frequency, the NS is closer to its MSM. This will be possible in the near future with the upgrade of high-sensitivity GW detectors, and from this point of view, NS astroseismology could provide at least one new way to determine the NS MSM.

\begin{figure*}[ht!]
\centering
\includegraphics[width=6.5in]{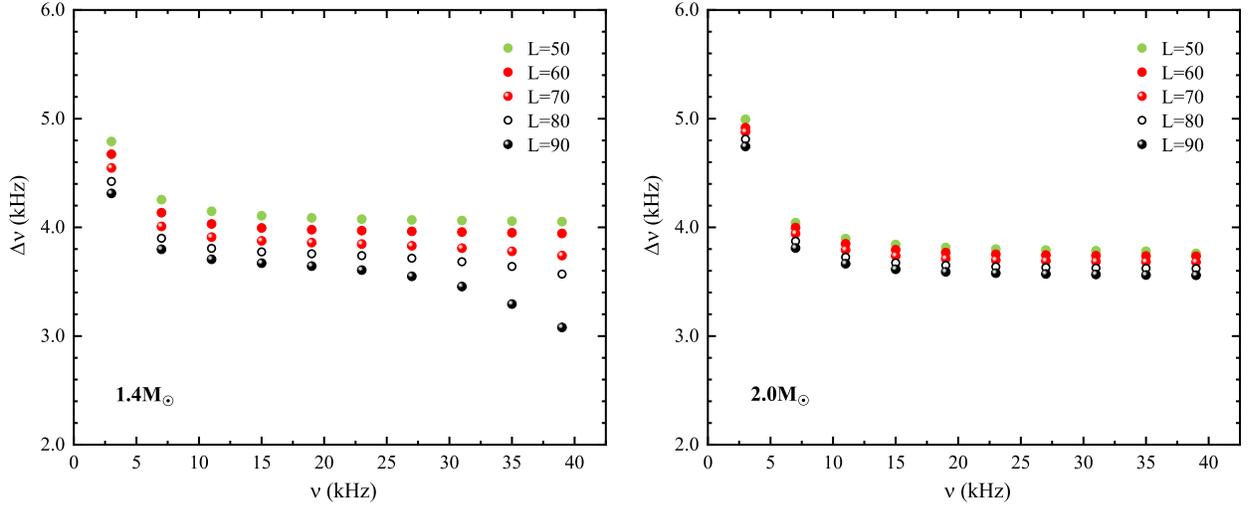}
\caption{Consecutive frequency differences (so-called Large Separation) $\triangle \nu =(\omega_{n}-\omega_{n-1})/2\pi$ (in kHz) as a function of $\nu$. Left and right panel consider 1.4$M_{\odot}$ and 2$M_{\odot}$ respectively, in which different color dots represent different $L$s. \label{fig:12}}
\end{figure*}

\begin{figure*}[ht!]
\centering
\includegraphics[width=6.5in]{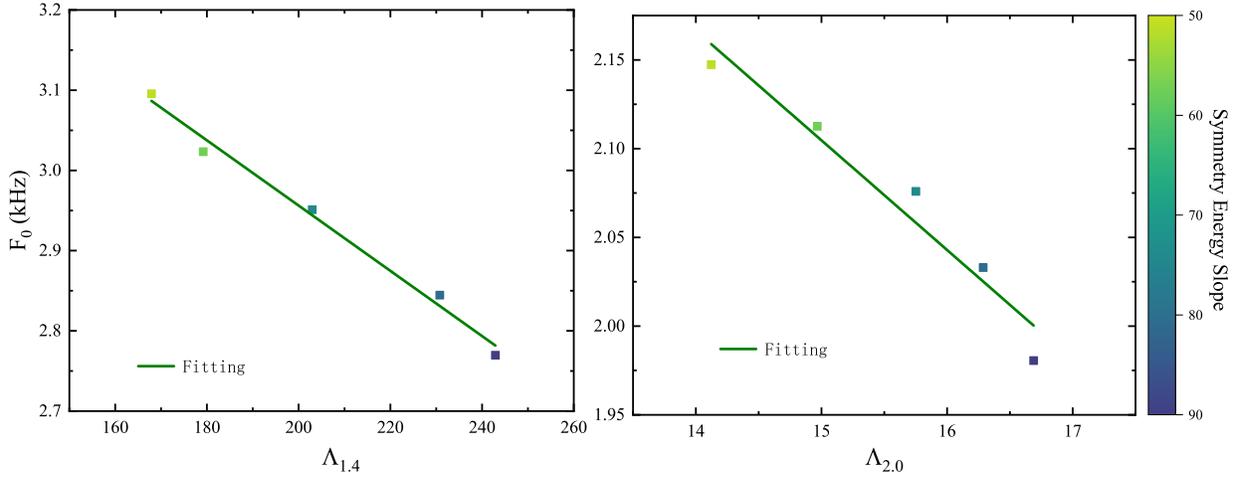}
\caption{Left and right show the linear relationship between the radial $F_{0}$-mode frequency and $\Lambda_{1.4}$ and $\Lambda_{2.0}$ respectively, where the different scatter points indicate the results given by different $L$s and the solid green line is the linear fitting. \label{fig:13}}
\end{figure*}
For high-mode oscillations with overtone numbers $n>0$, we examined the implications of $L$ for their eigenfunctions. Fig.11 shows the first four high-mode eigenfunctions, with $\xi$ shown in the left part and $P$ in the right, and 1.4$M_{\odot}$ and 2$M_{\odot}$ correspond to upper and down panel, respectively. The intersection of eigenfunction curve with dash-line represents the overtone number, for example, $n$=1 indicates that there is only one node. Furthermore, as the overtone numbers increase, the curve fluctuations become more prominent and reflect a higher frequency. In the NS interior, the different symmetry energy slopes (corresponding to different lines under each overtone curve) have little impact on the high-mode oscillation. It is essential to emphasize that the $F_{0}$-mode still remains the most promising frequency band for detection since the larger overtone numbers $n$ produce higher the frequencies, which would far exceed the sensitivity of current detection devices, making it still difficult to detect.

The consecutive frequency differences or Large Separation (LS) can also be extracted from the high-mode eigenfunctions, which is a measurable quantity that has been used to identify the DM component\citep{PhysRevD.96.083013}. To further examine the possible implications from $L$, we calculated 11 consecutive high-mode oscillation frequencies listed in Table 3, and plot the results in Fig.12, with the left and right parts showing 1.4$M_{\odot}$ and 2$M_{\odot}$ cases, respectively. It can be seen that the LS in both cases is roughly around 4 kHz to 5 kHz, which is consistent with the study of strange quark stars admixed DM \citep{PhysRevD.96.083013}. Moreover, LS decreases as $L$ increases, where for 1.4$M_{\odot}$ the decrease in LG is about 0.8 kHz or more as $L$ grows from 50 MeV to 90 MeV, while for 2$M_{\odot}$ LG decreases by approximately 0.4 kHz, which implies that the effect of $L$ on LG is more pronounced for low-mass NS than for massive NS, as also implied from the previous discussion on this point (see Figs.5-6). Finally, as discussed in section 4.1, the non-radial oscillations give a very strong linear correlation between frequency and tidal deformability for different $L$s, and as an analogy we also give results for the radial ones, apart from the different frequency values, 1.4$M_{\odot}$ and 2$M_{\odot}$ also satisfy a fairly good linear relationship, with the fitted results being $F_{0}=-0.004\Lambda_{1.4}+3.769 $ and $F_{0}=-0.062\Lambda_{2.0}+3.032$ shown in Fig.13 respectively. These results together with the non-radial oscillations will be more helpful in constraining EOSs for NSs.

\section{Summary}
In this study, we absorb the nucleon-nucleon SRC and admixed DM effect in NS interior, adopting the widely accepted SRC model and assuming the DM component to be one thousandth of the conventional nucleon number density. As a result of incorporating SRC and DM, we re-derive the pressure and energy density at the relativistic mean field theory and revise the corresponding nucleon coupling parameters. It is found that the simultaneous introduction of SRC and DM enables NS to better satisfy the observational constraints, and on this basis a typical set of parameters GM1 is selected to evaluate the impact of $L$ on the oscillation frequency of NSs with incorporating SRC and DM. For the purpose of verifying the feasibility of sets constructed from different $L$s, we calculated its behavior at low and high densities and discovered that it not only fulfills the properties near the saturation nuclear matter density well, but also meets the astronomical constraints. The following are the important physical findings of this study.

1. For non-radial oscillations, absorbing SRC and DM, the smaller $L$ is more likely to excite $f$-mode for stable NSs, thus giving larger frequencies.

2. After absorbing SRC and DM, we updated the empirical relationship between frequency and mean density with $f=1.0581+1.4507\sqrt{\bar{M}/\bar{R}^{3}}$, and the relationship between $\omega M$ and the compactness parameter with $\omega M=208.687(M/R)-7.257$, these relations can be used to decode NS internal information from the $f$-mode frequencies.

3. Our results indicate that a larger $L$ results in a larger tidal deformability and a smaller non-radial frequency. Furthermore, the frequency and tidal deformability show a good linear correlation, corresponding to $f_{0}=-0.001\Lambda_{1.4}+2.386$ and $f_{0}=-0.001\Lambda_{2.0}+2.719$ for 1.4$M_{\odot}$ and 2$M_{\odot}$.

4. For fundamental radial oscillations, the smaller $L$ exhibit larger radial frequency, and the effect of $L$ on the frequency is more pronounced in low mass regions, where the frequency drops from 3.11 kHz to 2.76 kHz for 1.4$M_{\odot}$ and from 2.16 kHz to 1.99 kHz for 2$M_{\odot}$ as $L$ changes from 50 MeV to 90 MeV.  Moreover, with the features of non-radial frequencies, we can provide a methodology for estimating the maximum stable mass of NS.

5. Finally, we evaluate the effect of $L$ on the high-mode eigenfunction as well as LG, and find that the $L$ has little impact on the high-mode oscillation and the effect on LG is more pronounced for low-mass NS than for massive NS. We also found that the frequency of radial oscillations also satisfies a strong linear relationship with tidal deformability, for which we fitted this relationship at 1.4$M_{\odot}$ and 2$M_{\odot}$, corresponding to $F_{0}=-0.004\Lambda_{1.4}+3.769 $ and $F_{0}=-0.062\Lambda_{2.0}+3.032$, respectively.

The above findings establish a link between $L$ and NS oscillation frequency, which allows us to understand how the $L$ affects the frequency and on other NS properties, with the continuous improvement of observational equipments, it is anticipated that this band frequency will be observed in the future, which in turn can be used to constrain $L$.


\acknowledgments
Bin thanks the anonymous referee for his or her kind guidance. This work is supported by the National Natural Science Foundation of China ( Grants No.12035011, No.11975167, No.11961141003, No.11905103, No.11761161001, No.11375086), and by the National Key R\&D Program of China (Contracts No. 2018YFA0404403 and No.2016YFE0129300).

\section*{Data Availability Statement}
This manuscript has no associated data or the data will not be deposited. [Authors’ comment: All results are shown in figures or table and are clearly given. The experimental data we cited can be obtained from references.]

\section*{appendix}
In this study within SRC-revised RMFT, in addition to the energy density and pressure that need to be revised, the six nucleon coupling parameters of $g_{\sigma N},~g_{\omega N},~g_{\rho N},~b,~c,\Lambda_{\omega}$ also need to be revised, which rely on the saturation number density $n_{0}$, binding energy per nucleon $B/A$, incompressibility coefficient $K$ and nucleon effective mass $m^{*}$. Next we will re-derive their connection by considering SRC effect. Given that the sum of binding energy per nucleon $B/A$ and nucleon mass $m_{B}$ determine energy per nucleon $E/A$, i.e. $E/A=B/A+m_{B}$, thus we can yield
\begin{eqnarray}
\nonumber
&&\frac{B}{A}+m_{B}=(\frac{g_{\omega}}{m_{\omega}})^{2}n_{0}+C_{1}\sqrt{k_{F}^{2}+m^{* 2}}\\
\nonumber
&&+\frac{C_{2}\lambda^{3}\sqrt{(\lambda k_{F})^{2}+m^{* 2}} }{(\lambda k_{F})^{4}}-\frac{C_{2}\sqrt{ k_{F}^{2}+m^{* 2}} }{k_{F}^{4}}\\
&&+\frac{1}{k_{F}^{2}}\frac{\partial C_{2}}{\partial k_{F}} \int_{k_{F}}^{\lambda k_{F}} \frac{k^{2}\sqrt{k^{2}+m^{* 2}}}{k^{4}} \mathrm{d} k,
\end{eqnarray}
where $m^{*}$ stands for the nucleon effective mass, and since the isospin-asymmetry $\beta=0$ for saturation density, we get $C_{1}^{(n)}=C_{1}^{(p)}=C_{1}$, $C_{2}^{(n)}=C_{2}^{(p)}=C_{2}$, $k_{F}^{(n)}=k_{F}^{(p)}=k_{F}$. As described in RMFT, the isoscalar-scalar meson $\sigma$ meet
\begin{eqnarray}
\nonumber
&&\left(\frac{m_{\sigma}}{g_{\sigma}}\right)^{2}\left(m_{B}-m^{*}\right)+m_{B}\left(m_{B}-m^{*}\right)^{2} b+c\left(m_{B}-m^{*}\right)^{3}\\
\nonumber
&&=2\Bigg[\frac{C_{1}}{\pi^{2}} \int_{0}^{k_{F}} \frac{m^{*} k^{2}}{\sqrt{k^{2}+m^{* 2}}} \mathrm{~d} k\\
&&+\frac{C_{2}}{\pi^{2}} \int_{k_{F}}^{\lambda k_{F}} \frac{m^{*} k^{2}}{k^{4} \sqrt{k^{2}+m^{* 2}}} \mathrm{~d} k\Bigg],
\end{eqnarray}
and energy density satisfies as
\begin{eqnarray}
\nonumber
&&\frac{1}{2}\left(m_{B}-m^{*}\right)^{2}\left(\frac{m_{\sigma}}{g_{\sigma}}\right)^{2}+\frac{1}{3} m_{B}\left(m_{B}-m^{*}\right)^{3} b\\
\nonumber
&&+\frac{1}{4}\left(m_{B}-m^{*}\right)^{4}c+\frac{1}{2}\left(\frac{g_{\omega}}{m_{\omega}}\right)^{2} n_{0}^{2}\\
\nonumber
&&=\left(B / A+m_{B}\right) n_{0}-2\times\Bigg[\frac{C_{1}}{\pi^{2}} \int_{0}^{k_{F}} k^{2}\sqrt{k^{2}+m^{* 2}}\mathrm{d} k\\
&&+\frac{C_{2}}{\pi^{2}} \int_{k_{F}}^{\lambda k_{F}} \frac{k^{2}\sqrt{k^{2}+m^{* 2}}}{k^{4}} \mathrm{d} k  \Bigg].
\end{eqnarray}

At saturation density, the incompressibility coefficient reads as
\begin{eqnarray}
\nonumber
&&K=(\frac{g_{\omega}}{m_{\omega}})^{2}\cdot\frac{6k_{F}^{3}}{\pi^{2}} +\Bigg(AA+BB+CC\Bigg)^{'k}3 k_{F}\\
\nonumber
&&-   \Bigg(AA+BB+CC\Bigg)^{'m*}  \frac{6 k_{F}\left\langle\bar{\psi}\psi\right\rangle^{' k}  }{\Bigg[(\frac{m_{\sigma}}{g_{\sigma}})^{2}+2\left\langle\bar{\psi}\psi\right\rangle^{' m*}  - \frac{d^{2}U}{d\sigma_{0}^{2}}\frac{1}{g_{\sigma}^{2}} \Bigg]},\\
\end{eqnarray}
where prime refers to the partial derivative of momentum $k$ and nucleon effective mass $m^{*}$, and $AA$,$BB$,$CC$ and $ \left\langle\bar{\psi} \psi\right\rangle$ read as
\begin{eqnarray}
AA&&=C_{1} \sqrt{k_{F}^{2}+m^{* 2}},\\
BB&&=\frac{C_{2}\lambda^{3}\sqrt{(\lambda k_{F})^{2}+m^{* 2}} }{(\lambda k_{F})^{4}}-\frac{C_{2}\sqrt{ k_{F}^{2}+m^{* 2}} }{ k_{F}^{4}},\\
CC&&=\frac{C'_{2}}{k_{F}^{2}} \int_{k_{F}}^{\lambda k_{F}} \frac{k^{2}\sqrt{k^{2}+m^{* 2}}}{k^{4}} \mathrm{d} k,\\
\nonumber
  \left\langle\bar{\psi} \psi\right\rangle &&=\frac{2C_{1}}{\pi^{2}} \int_{0}^{k_{F}} \frac{m^{*} k^{2}}{\sqrt{k^{2}+m^{* 2}}} \mathrm{~d} k+\frac{2C_{2}}{\pi^{2}} \int_{k_{F}}^{\lambda k_{F}} \frac{m^{*} k^{2}}{k^{4} \sqrt{k^{2}+m^{* 2}}} \mathrm{~d} k.\\
\end{eqnarray}

Moreover, we can correlate the isovector coupling parameters $g_{\varrho}$ and $\Lambda_{\omega}$ with symmetry energy $E_{\text{sym}}$ and its slope $L$, for symmetry energy at saturation density, it can be divided into two parts and reads as
\begin{eqnarray}
\nonumber
&&E_{\mathrm{sym}}(n_{0})=\frac{1}{2} \frac{\partial^{2}(E / A)}{\partial \beta^{2}}\Bigg|_{\beta=0}=\frac{1}{2}\left[\frac{\partial^{2}(\epsilon / n)}{\partial \beta^{2}}\right]_{\beta=0}\\
&&=E_{1\mathrm{sym}}(n_{0})+E_{2\mathrm{sym}}(n_{0}),
\end{eqnarray}
where two parts correspond to
\begin{eqnarray}
E_{1\mathrm{sym}}(n_{0})=\frac{k_{0}^{3}}{12 \pi^{2}} \frac{g_{\varrho}^{2}}{m_{\varrho}^{2}+2\Lambda_{\omega}\left(g_{\omega} \omega_{0}\right)^{2} g_{\varrho}^{2}}
\end{eqnarray}
and
\begin{eqnarray}
\nonumber
&&E_{2\mathrm{sym}}(n_{0})=\frac{k_{0}^{2}}{8\sqrt{k_{0}^{2}+m^{* 2}}}\\
\nonumber
&&+\frac{1}{4}\sqrt{k_{0}^{2}+m^{* 2}}+ \Bigg[ (\frac{C'_{2}}{2k_{0}^{2}})'\cdot\frac{k_{0}}{3}-\frac{ C'_{2}}{  k_{0}^{2}}\Bigg]   \int_{k_{0}}^{\lambda k_{0}} \mathrm{~d} k \frac{k^{2}\sqrt{k^{2}+m^{* 2}}}{k^{4}}\\
\nonumber
&&+\bigg(\frac{C'_{2}k_{0}}{3}-C_{2}\bigg)\bigg(\frac{\lambda^{3} \sqrt{\left(\lambda k_{0}\right)^{2}+m^{* 2}}}{(\lambda k_{0})^{4}}- \frac{\sqrt{k_{0}^{2}+m^{* 2}}}{ k_{0}^{4}}\bigg)\\
&&+\frac{k_{0}C_{2}}{6}\bigg(\frac{\lambda^{3} \sqrt{\left(\lambda k_{0}\right)^{2}+m^{* 2}}}{(\lambda k_{0})^{4}}- \frac{\sqrt{k_{0}^{2}+m^{* 2}}}{ k_{0}^{4}}\bigg)'.
\end{eqnarray}
While, for symmetry energy slope $L$, it also can be divided into two parts accordingly
\begin{eqnarray}
L(n_{0})=&3n_{0}\frac{\partial E_{\mathrm{sym}}}{\partial n}\Bigg|_{n=n_{0}}=L_{1}(n_{0})+L_{2}(n_{0}),
\end{eqnarray}
where $L_{1,2}$ represent as
\begin{eqnarray}
&L_{1}(n_{0})=3E_{3\mathrm{sym}}  \Bigg[ 1-32\frac{g_{\omega}^{2}}{m_{\omega}^{2}} \times  g_{\omega}\omega \Lambda_{\omega} \times E_{3\mathrm{sym}}     \Bigg]
\end{eqnarray}
and
\begin{eqnarray}
\nonumber
&&L_{2}(n_{0})=k(E_{2\mathrm{sym}})^{'k}\\
\nonumber
&&-\frac{2k (E_{2\mathrm{sym}})^{'m*}\langle\bar{\psi} \psi\rangle^{\prime k} }{\Bigg[ (\frac{m_{\sigma}}{g_{\sigma}})^{2}+2\langle\bar{\psi} \psi\rangle^{\prime m*}+2bm(m-m^{*})+3c(m-m^{*})^{2} \Bigg]}.\\
\end{eqnarray}
At this point, it is clear that the six nucleon coupling parameters under the SRC-revised RMFT are successfully associated with six saturation nuclear parameters, and can be quantitatively solved.

\bibliography{reference}

\begin{thebibliography}{137}%
\makeatletter
\providecommand \@ifxundefined [1]{%
 \@ifx{#1\undefined}
}%
\providecommand \@ifnum [1]{%
 \ifnum #1\expandafter \@firstoftwo
 \else \expandafter \@secondoftwo
 \fi
}%
\providecommand \@ifx [1]{%
 \ifx #1\expandafter \@firstoftwo
 \else \expandafter \@secondoftwo
 \fi
}%
\providecommand \natexlab [1]{#1}%
\providecommand \enquote  [1]{``#1''}%
\providecommand \bibnamefont  [1]{#1}%
\providecommand \bibfnamefont [1]{#1}%
\providecommand \citenamefont [1]{#1}%
\providecommand \href@noop [0]{\@secondoftwo}%
\providecommand \href [0]{\begingroup \@sanitize@url \@href}%
\providecommand \@href[1]{\@@startlink{#1}\@@href}%
\providecommand \@@href[1]{\endgroup#1\@@endlink}%
\providecommand \@sanitize@url [0]{\catcode `\\12\catcode `\$12\catcode
  `\&12\catcode `\#12\catcode `\^12\catcode `\_12\catcode `\%12\relax}%
\providecommand \@@startlink[1]{}%
\providecommand \@@endlink[0]{}%
\providecommand \url  [0]{\begingroup\@sanitize@url \@url }%
\providecommand \@url [1]{\endgroup\@href {#1}{\urlprefix }}%
\providecommand \urlprefix  [0]{URL }%
\providecommand \Eprint [0]{\href }%
\providecommand \doibase [0]{http://dx.doi.org/}%
\providecommand \selectlanguage [0]{\@gobble}%
\providecommand \bibinfo  [0]{\@secondoftwo}%
\providecommand \bibfield  [0]{\@secondoftwo}%
\providecommand \translation [1]{[#1]}%
\providecommand \BibitemOpen [0]{}%
\providecommand \bibitemStop [0]{}%
\providecommand \bibitemNoStop [0]{.\EOS\space}%
\providecommand \EOS [0]{\spacefactor3000\relax}%
\providecommand \BibitemShut  [1]{\csname bibitem#1\endcsname}%
\let\auto@bib@innerbib\@empty
\bibitem [{\citenamefont {Oertel}\ \emph {et~al.}(2017)\citenamefont {Oertel},
  \citenamefont {Hempel}, \citenamefont {Kl\"ahn},\ and\ \citenamefont
  {Typel}}]{RevModPhys.89.015007}%
  \BibitemOpen
  \bibfield  {author} {\bibinfo {author} {\bibfnamefont {M.}~\bibnamefont
  {Oertel}}, \bibinfo {author} {\bibfnamefont {M.}~\bibnamefont {Hempel}},
  \bibinfo {author} {\bibfnamefont {T.}~\bibnamefont {Kl\"ahn}}, \ and\
  \bibinfo {author} {\bibfnamefont {S.}~\bibnamefont {Typel}},\ }\href
  {\doibase 10.1103/RevModPhys.89.015007} {\bibfield  {journal} {\bibinfo
  {journal} {Rev. Mod. Phys.}\ }\textbf {\bibinfo {volume} {89}},\ \bibinfo
  {pages} {015007} (\bibinfo {year} {2017})}\BibitemShut {NoStop}%
\bibitem [{\citenamefont {Jiang}\ \emph {et~al.}(2020)\citenamefont {Jiang},
  \citenamefont {Tang}, \citenamefont {Wang} \emph {et~al.}}]{Jiang_2020}%
  \BibitemOpen
  \bibfield  {author} {\bibinfo {author} {\bibfnamefont {J.-L.}\ \bibnamefont
  {Jiang}}, \bibinfo {author} {\bibfnamefont {S.-P.}\ \bibnamefont {Tang}},
  \bibinfo {author} {\bibfnamefont {Y.-Z.}\ \bibnamefont {Wang}},  \emph
  {et~al.},\ }\href {\doibase 10.3847/1538-4357/ab77cf} {\bibfield  {journal}
  {\bibinfo  {journal} {The Astrophysical Journal}\ }\textbf {\bibinfo {volume}
  {892}},\ \bibinfo {pages} {55} (\bibinfo {year} {2020})}\BibitemShut
  {NoStop}%
\bibitem [{\citenamefont {Luo}\ \emph {et~al.}(2022)\citenamefont {Luo},
  \citenamefont {Tang}, \citenamefont {Jiang}, \citenamefont {Gao},\ and\
  \citenamefont {Wei}}]{Luo_2022}%
  \BibitemOpen
  \bibfield  {author} {\bibinfo {author} {\bibfnamefont {C.-N.}\ \bibnamefont
  {Luo}}, \bibinfo {author} {\bibfnamefont {S.-P.}\ \bibnamefont {Tang}},
  \bibinfo {author} {\bibfnamefont {J.-L.}\ \bibnamefont {Jiang}}, \bibinfo
  {author} {\bibfnamefont {W.-H.}\ \bibnamefont {Gao}}, \ and\ \bibinfo
  {author} {\bibfnamefont {D.-M.}\ \bibnamefont {Wei}},\ }\href {\doibase
  10.3847/1538-4357/ac6175} {\bibfield  {journal} {\bibinfo  {journal} {The
  Astrophysical Journal}\ }\textbf {\bibinfo {volume} {930}},\ \bibinfo {pages}
  {4} (\bibinfo {year} {2022})}\BibitemShut {NoStop}%
\bibitem [{\citenamefont {Epelbaum}\ \emph {et~al.}(2009)\citenamefont
  {Epelbaum}, \citenamefont {Hammer},\ and\ \citenamefont
  {Mei\ss{}ner}}]{RevModPhys.81.1773}%
  \BibitemOpen
  \bibfield  {author} {\bibinfo {author} {\bibfnamefont {E.}~\bibnamefont
  {Epelbaum}}, \bibinfo {author} {\bibfnamefont {H.-W.}\ \bibnamefont
  {Hammer}}, \ and\ \bibinfo {author} {\bibfnamefont {U.-G.}\ \bibnamefont
  {Mei\ss{}ner}},\ }\href {\doibase 10.1103/RevModPhys.81.1773} {\bibfield
  {journal} {\bibinfo  {journal} {Rev. Mod. Phys.}\ }\textbf {\bibinfo {volume}
  {81}},\ \bibinfo {pages} {1773} (\bibinfo {year} {2009})}\BibitemShut
  {NoStop}%
\bibitem [{\citenamefont {Hammer}\ \emph {et~al.}(2013)\citenamefont {Hammer},
  \citenamefont {Nogga},\ and\ \citenamefont {Schwenk}}]{RevModPhys.85.197}%
  \BibitemOpen
  \bibfield  {author} {\bibinfo {author} {\bibfnamefont {H.-W.}\ \bibnamefont
  {Hammer}}, \bibinfo {author} {\bibfnamefont {A.}~\bibnamefont {Nogga}}, \
  and\ \bibinfo {author} {\bibfnamefont {A.}~\bibnamefont {Schwenk}},\ }\href
  {\doibase 10.1103/RevModPhys.85.197} {\bibfield  {journal} {\bibinfo
  {journal} {Rev. Mod. Phys.}\ }\textbf {\bibinfo {volume} {85}},\ \bibinfo
  {pages} {197} (\bibinfo {year} {2013})}\BibitemShut {NoStop}%
\bibitem [{\citenamefont {Holt}\ \emph {et~al.}(2013)\citenamefont {Holt},
  \citenamefont {Kaiser},\ and\ \citenamefont {Weise}}]{HOLT201335}%
  \BibitemOpen
  \bibfield  {author} {\bibinfo {author} {\bibfnamefont {J.~W.}\ \bibnamefont
  {Holt}}, \bibinfo {author} {\bibfnamefont {N.}~\bibnamefont {Kaiser}}, \ and\
  \bibinfo {author} {\bibfnamefont {W.}~\bibnamefont {Weise}},\ }\href
  {\doibase https://doi.org/10.1016/j.ppnp.2013.08.001} {\bibfield  {journal}
  {\bibinfo  {journal} {Progress in Particle and Nuclear Physics}\ }\textbf
  {\bibinfo {volume} {73}},\ \bibinfo {pages} {35} (\bibinfo {year}
  {2013})}\BibitemShut {NoStop}%
\bibitem [{\citenamefont {Tews}\ \emph {et~al.}(2013)\citenamefont {Tews},
  \citenamefont {Kr\"uger}, \citenamefont {Hebeler},\ and\ \citenamefont
  {Schwenk}}]{PhysRevLett.110.032504}%
  \BibitemOpen
  \bibfield  {author} {\bibinfo {author} {\bibfnamefont {I.}~\bibnamefont
  {Tews}}, \bibinfo {author} {\bibfnamefont {T.}~\bibnamefont {Kr\"uger}},
  \bibinfo {author} {\bibfnamefont {K.}~\bibnamefont {Hebeler}}, \ and\
  \bibinfo {author} {\bibfnamefont {A.}~\bibnamefont {Schwenk}},\ }\href
  {\doibase 10.1103/PhysRevLett.110.032504} {\bibfield  {journal} {\bibinfo
  {journal} {Phys. Rev. Lett.}\ }\textbf {\bibinfo {volume} {110}},\ \bibinfo
  {pages} {032504} (\bibinfo {year} {2013})}\BibitemShut {NoStop}%
\bibitem [{\citenamefont {Lynn}\ \emph {et~al.}(2016)\citenamefont {Lynn},
  \citenamefont {Tews}, \citenamefont {Carlson}, \citenamefont {Gandolfi},
  \citenamefont {Gezerlis}, \citenamefont {Schmidt},\ and\ \citenamefont
  {Schwenk}}]{PhysRevLett.116.062501}%
  \BibitemOpen
  \bibfield  {author} {\bibinfo {author} {\bibfnamefont {J.~E.}\ \bibnamefont
  {Lynn}}, \bibinfo {author} {\bibfnamefont {I.}~\bibnamefont {Tews}}, \bibinfo
  {author} {\bibfnamefont {J.}~\bibnamefont {Carlson}}, \bibinfo {author}
  {\bibfnamefont {S.}~\bibnamefont {Gandolfi}}, \bibinfo {author}
  {\bibfnamefont {A.}~\bibnamefont {Gezerlis}}, \bibinfo {author}
  {\bibfnamefont {K.~E.}\ \bibnamefont {Schmidt}}, \ and\ \bibinfo {author}
  {\bibfnamefont {A.}~\bibnamefont {Schwenk}},\ }\href {\doibase
  10.1103/PhysRevLett.116.062501} {\bibfield  {journal} {\bibinfo  {journal}
  {Phys. Rev. Lett.}\ }\textbf {\bibinfo {volume} {116}},\ \bibinfo {pages}
  {062501} (\bibinfo {year} {2016})}\BibitemShut {NoStop}%
\bibitem [{\citenamefont {Stone}\ and\ \citenamefont
  {Reinhard}(2007)}]{STONE2007587}%
  \BibitemOpen
  \bibfield  {author} {\bibinfo {author} {\bibfnamefont {J.}~\bibnamefont
  {Stone}}\ and\ \bibinfo {author} {\bibfnamefont {P.-G.}\ \bibnamefont
  {Reinhard}},\ }\href {\doibase https://doi.org/10.1016/j.ppnp.2006.07.001}
  {\bibfield  {journal} {\bibinfo  {journal} {Progress in Particle and Nuclear
  Physics}\ }\textbf {\bibinfo {volume} {58}},\ \bibinfo {pages} {587}
  (\bibinfo {year} {2007})}\BibitemShut {NoStop}%
\bibitem [{\citenamefont {Rikovska~Stone}\ \emph {et~al.}(2003)\citenamefont
  {Rikovska~Stone}, \citenamefont {Miller}, \citenamefont {Koncewicz},
  \citenamefont {Stevenson},\ and\ \citenamefont
  {Strayer}}]{PhysRevC.68.034324}%
  \BibitemOpen
  \bibfield  {author} {\bibinfo {author} {\bibfnamefont {J.}~\bibnamefont
  {Rikovska~Stone}}, \bibinfo {author} {\bibfnamefont {J.~C.}\ \bibnamefont
  {Miller}}, \bibinfo {author} {\bibfnamefont {R.}~\bibnamefont {Koncewicz}},
  \bibinfo {author} {\bibfnamefont {P.~D.}\ \bibnamefont {Stevenson}}, \ and\
  \bibinfo {author} {\bibfnamefont {M.~R.}\ \bibnamefont {Strayer}},\ }\href
  {\doibase 10.1103/PhysRevC.68.034324} {\bibfield  {journal} {\bibinfo
  {journal} {Phys. Rev. C}\ }\textbf {\bibinfo {volume} {68}},\ \bibinfo
  {pages} {034324} (\bibinfo {year} {2003})}\BibitemShut {NoStop}%
\bibitem [{\citenamefont {Dutra}\ \emph {et~al.}(2012)\citenamefont {Dutra},
  \citenamefont {Louren\ifmmode~\mbox{\c{c}}\else \c{c}\fi{}o}, \citenamefont
  {S\'a~Martins}, \citenamefont {Delfino}, \citenamefont {Stone},\ and\
  \citenamefont {Stevenson}}]{PhysRevC.85.035201}%
  \BibitemOpen
  \bibfield  {author} {\bibinfo {author} {\bibfnamefont {M.}~\bibnamefont
  {Dutra}}, \bibinfo {author} {\bibfnamefont {O.}~\bibnamefont
  {Louren\ifmmode~\mbox{\c{c}}\else \c{c}\fi{}o}}, \bibinfo {author}
  {\bibfnamefont {J.~S.}\ \bibnamefont {S\'a~Martins}}, \bibinfo {author}
  {\bibfnamefont {A.}~\bibnamefont {Delfino}}, \bibinfo {author} {\bibfnamefont
  {J.~R.}\ \bibnamefont {Stone}}, \ and\ \bibinfo {author} {\bibfnamefont
  {P.~D.}\ \bibnamefont {Stevenson}},\ }\href {\doibase
  10.1103/PhysRevC.85.035201} {\bibfield  {journal} {\bibinfo  {journal} {Phys.
  Rev. C}\ }\textbf {\bibinfo {volume} {85}},\ \bibinfo {pages} {035201}
  (\bibinfo {year} {2012})}\BibitemShut {NoStop}%
\bibitem [{\citenamefont {Loan}\ \emph {et~al.}(2011)\citenamefont {Loan},
  \citenamefont {Tan}, \citenamefont {Khoa},\ and\ \citenamefont
  {Margueron}}]{PhysRevC.83.065809}%
  \BibitemOpen
  \bibfield  {author} {\bibinfo {author} {\bibfnamefont {D.~T.}\ \bibnamefont
  {Loan}}, \bibinfo {author} {\bibfnamefont {N.~H.}\ \bibnamefont {Tan}},
  \bibinfo {author} {\bibfnamefont {D.~T.}\ \bibnamefont {Khoa}}, \ and\
  \bibinfo {author} {\bibfnamefont {J.}~\bibnamefont {Margueron}},\ }\href
  {\doibase 10.1103/PhysRevC.83.065809} {\bibfield  {journal} {\bibinfo
  {journal} {Phys. Rev. C}\ }\textbf {\bibinfo {volume} {83}},\ \bibinfo
  {pages} {065809} (\bibinfo {year} {2011})}\BibitemShut {NoStop}%
\bibitem [{\citenamefont {Gonzalez-Boquera}\ \emph {et~al.}(2017)\citenamefont
  {Gonzalez-Boquera}, \citenamefont {Centelles}, \citenamefont {Vi\~nas},\ and\
  \citenamefont {Rios}}]{PhysRevC.96.065806}%
  \BibitemOpen
  \bibfield  {author} {\bibinfo {author} {\bibfnamefont {C.}~\bibnamefont
  {Gonzalez-Boquera}}, \bibinfo {author} {\bibfnamefont {M.}~\bibnamefont
  {Centelles}}, \bibinfo {author} {\bibfnamefont {X.}~\bibnamefont {Vi\~nas}},
  \ and\ \bibinfo {author} {\bibfnamefont {A.}~\bibnamefont {Rios}},\ }\href
  {\doibase 10.1103/PhysRevC.96.065806} {\bibfield  {journal} {\bibinfo
  {journal} {Phys. Rev. C}\ }\textbf {\bibinfo {volume} {96}},\ \bibinfo
  {pages} {065806} (\bibinfo {year} {2017})}\BibitemShut {NoStop}%
\bibitem [{\citenamefont {Read}\ \emph {et~al.}(2009)\citenamefont {Read},
  \citenamefont {Lackey}, \citenamefont {Owen},\ and\ \citenamefont
  {Friedman}}]{PhysRevD.79.124032}%
  \BibitemOpen
  \bibfield  {author} {\bibinfo {author} {\bibfnamefont {J.~S.}\ \bibnamefont
  {Read}}, \bibinfo {author} {\bibfnamefont {B.~D.}\ \bibnamefont {Lackey}},
  \bibinfo {author} {\bibfnamefont {B.~J.}\ \bibnamefont {Owen}}, \ and\
  \bibinfo {author} {\bibfnamefont {J.~L.}\ \bibnamefont {Friedman}},\ }\href
  {\doibase 10.1103/PhysRevD.79.124032} {\bibfield  {journal} {\bibinfo
  {journal} {Phys. Rev. D}\ }\textbf {\bibinfo {volume} {79}},\ \bibinfo
  {pages} {124032} (\bibinfo {year} {2009})}\BibitemShut {NoStop}%
\bibitem [{\citenamefont {Demorest}\ \emph {et~al.}(2010)\citenamefont
  {Demorest}, \citenamefont {Pennucci}, \citenamefont {Ransom} \emph
  {et~al.}}]{RN583}%
  \BibitemOpen
  \bibfield  {author} {\bibinfo {author} {\bibfnamefont {P.~B.}\ \bibnamefont
  {Demorest}}, \bibinfo {author} {\bibfnamefont {T.}~\bibnamefont {Pennucci}},
  \bibinfo {author} {\bibfnamefont {S.~M.}\ \bibnamefont {Ransom}},  \emph
  {et~al.},\ }\href {\doibase 10.1038/nature09466} {\bibfield  {journal}
  {\bibinfo  {journal} {Nature}\ }\textbf {\bibinfo {volume} {467}},\ \bibinfo
  {pages} {1081} (\bibinfo {year} {2010})}\BibitemShut {NoStop}%
\bibitem [{\citenamefont {Fonseca}\ \emph {et~al.}(2016)\citenamefont
  {Fonseca}, \citenamefont {Pennucci}, \citenamefont {Ellis} \emph
  {et~al.}}]{Fonseca_2016}%
  \BibitemOpen
  \bibfield  {author} {\bibinfo {author} {\bibfnamefont {E.}~\bibnamefont
  {Fonseca}}, \bibinfo {author} {\bibfnamefont {T.~T.}\ \bibnamefont
  {Pennucci}}, \bibinfo {author} {\bibfnamefont {J.~A.}\ \bibnamefont {Ellis}},
   \emph {et~al.},\ }\href {\doibase 10.3847/0004-637x/832/2/167} {\bibfield
  {journal} {\bibinfo  {journal} {The Astrophysical Journal}\ }\textbf
  {\bibinfo {volume} {832}},\ \bibinfo {pages} {167} (\bibinfo {year}
  {2016})}\BibitemShut {NoStop}%
\bibitem [{\citenamefont {Arzoumanian}\ \emph {et~al.}(2018)\citenamefont
  {Arzoumanian}, \citenamefont {Brazier}, \citenamefont {Burke-Spolaor} \emph
  {et~al.}}]{Arzoumanian_2018}%
  \BibitemOpen
  \bibfield  {author} {\bibinfo {author} {\bibfnamefont {Z.}~\bibnamefont
  {Arzoumanian}}, \bibinfo {author} {\bibfnamefont {A.}~\bibnamefont
  {Brazier}}, \bibinfo {author} {\bibfnamefont {S.}~\bibnamefont
  {Burke-Spolaor}},  \emph {et~al.},\ }\href {\doibase
  10.3847/1538-4365/aab5b0} {\bibfield  {journal} {\bibinfo  {journal} {The
  Astrophysical Journal Supplement Series}\ }\textbf {\bibinfo {volume}
  {235}},\ \bibinfo {pages} {37} (\bibinfo {year} {2018})}\BibitemShut
  {NoStop}%
\bibitem [{\citenamefont {Antoniadis}\ \emph {et~al.}(2013)\citenamefont
  {Antoniadis}, \citenamefont {Freire}, \citenamefont {Wex} \emph
  {et~al.}}]{Antoniadis1233232}%
  \BibitemOpen
  \bibfield  {author} {\bibinfo {author} {\bibfnamefont {J.}~\bibnamefont
  {Antoniadis}}, \bibinfo {author} {\bibfnamefont {P.~C.~C.}\ \bibnamefont
  {Freire}}, \bibinfo {author} {\bibfnamefont {N.}~\bibnamefont {Wex}},  \emph
  {et~al.},\ }\href {https://science.sciencemag.org/content/340/6131/1233232}
  {\bibfield  {journal} {\bibinfo  {journal} {Science}\ }\textbf {\bibinfo
  {volume} {340}} (\bibinfo {year} {2013})}\BibitemShut {NoStop}%
\bibitem [{\citenamefont {Cromartie}\ \emph {et~al.}(2020)\citenamefont
  {Cromartie}, \citenamefont {Fonseca}, \citenamefont {Ransom} \emph
  {et~al.}}]{RN584}%
  \BibitemOpen
  \bibfield  {author} {\bibinfo {author} {\bibfnamefont {H.~T.}\ \bibnamefont
  {Cromartie}}, \bibinfo {author} {\bibfnamefont {E.}~\bibnamefont {Fonseca}},
  \bibinfo {author} {\bibfnamefont {S.~M.}\ \bibnamefont {Ransom}},  \emph
  {et~al.},\ }\href {\doibase 10.1038/s41550-019-0880-2} {\bibfield  {journal}
  {\bibinfo  {journal} {Nature Astronomy}\ }\textbf {\bibinfo {volume} {4}},\
  \bibinfo {pages} {72} (\bibinfo {year} {2020})}\BibitemShut {NoStop}%
\bibitem [{\citenamefont {Fonseca}\ \emph {et~al.}(2021)\citenamefont
  {Fonseca}, \citenamefont {Cromartie},\ and\ \citenamefont
  {Pennucci}}]{Fonseca_2021}%
  \BibitemOpen
  \bibfield  {author} {\bibinfo {author} {\bibfnamefont {E.}~\bibnamefont
  {Fonseca}}, \bibinfo {author} {\bibfnamefont {H.~T.}\ \bibnamefont
  {Cromartie}}, \ and\ \bibinfo {author} {\bibfnamefont {T.~T.}\ \bibnamefont
  {Pennucci}},\ }\href {\doibase 10.3847/2041-8213/ac03b8} {\bibfield
  {journal} {\bibinfo  {journal} {The Astrophysical Journal Letters}\ }\textbf
  {\bibinfo {volume} {915}},\ \bibinfo {pages} {L12} (\bibinfo {year}
  {2021})}\BibitemShut {NoStop}%
\bibitem [{\citenamefont {Abbott}\ \emph {et~al.}(2019)\citenamefont {Abbott},
  \citenamefont {Abbott}, \citenamefont {Abbott} \emph
  {et~al.}}]{PhysRevX.9.011001}%
  \BibitemOpen
  \bibfield  {author} {\bibinfo {author} {\bibfnamefont {B.~P.}\ \bibnamefont
  {Abbott}}, \bibinfo {author} {\bibfnamefont {R.}~\bibnamefont {Abbott}},
  \bibinfo {author} {\bibfnamefont {T.~D.}\ \bibnamefont {Abbott}},  \emph
  {et~al.} (\bibinfo {collaboration} {LIGO Scientific Collaboration and Virgo
  Collaboration}),\ }\href {\doibase 10.1103/PhysRevX.9.011001} {\bibfield
  {journal} {\bibinfo  {journal} {Phys. Rev. X}\ }\textbf {\bibinfo {volume}
  {9}},\ \bibinfo {pages} {011001} (\bibinfo {year} {2019})}\BibitemShut
  {NoStop}%
\bibitem [{\citenamefont {Fasano}\ \emph {et~al.}(2019)\citenamefont {Fasano},
  \citenamefont {Abdelsalhin}, \citenamefont {Maselli},\ and\ \citenamefont
  {Ferrari}}]{PhysRevLett.123.141101}%
  \BibitemOpen
  \bibfield  {author} {\bibinfo {author} {\bibfnamefont {M.}~\bibnamefont
  {Fasano}}, \bibinfo {author} {\bibfnamefont {T.}~\bibnamefont {Abdelsalhin}},
  \bibinfo {author} {\bibfnamefont {A.}~\bibnamefont {Maselli}}, \ and\
  \bibinfo {author} {\bibfnamefont {V.}~\bibnamefont {Ferrari}},\ }\href
  {\doibase 10.1103/PhysRevLett.123.141101} {\bibfield  {journal} {\bibinfo
  {journal} {Phys. Rev. Lett.}\ }\textbf {\bibinfo {volume} {123}},\ \bibinfo
  {pages} {141101} (\bibinfo {year} {2019})}\BibitemShut {NoStop}%
\bibitem [{\citenamefont {Miller}\ \emph {et~al.}(2019)\citenamefont {Miller},
  \citenamefont {Lamb}, \citenamefont {Dittmann} \emph {et~al.}}]{Miller_2019}%
  \BibitemOpen
  \bibfield  {author} {\bibinfo {author} {\bibfnamefont {M.~C.}\ \bibnamefont
  {Miller}}, \bibinfo {author} {\bibfnamefont {F.~K.}\ \bibnamefont {Lamb}},
  \bibinfo {author} {\bibfnamefont {A.~J.}\ \bibnamefont {Dittmann}},  \emph
  {et~al.},\ }\href {\doibase 10.3847/2041-8213/ab50c5} {\bibfield  {journal}
  {\bibinfo  {journal} {The Astrophysical Journal}\ }\textbf {\bibinfo {volume}
  {887}},\ \bibinfo {pages} {L24} (\bibinfo {year} {2019})}\BibitemShut
  {NoStop}%
\bibitem [{\citenamefont {Riley}\ \emph {et~al.}(2019)\citenamefont {Riley},
  \citenamefont {Watts}, \citenamefont {Bogdanov} \emph {et~al.}}]{Riley_2019}%
  \BibitemOpen
  \bibfield  {author} {\bibinfo {author} {\bibfnamefont {T.~E.}\ \bibnamefont
  {Riley}}, \bibinfo {author} {\bibfnamefont {A.~L.}\ \bibnamefont {Watts}},
  \bibinfo {author} {\bibfnamefont {S.}~\bibnamefont {Bogdanov}},  \emph
  {et~al.},\ }\href {\doibase 10.3847/2041-8213/ab481c} {\bibfield  {journal}
  {\bibinfo  {journal} {The Astrophysical Journal}\ }\textbf {\bibinfo {volume}
  {887}},\ \bibinfo {pages} {L21} (\bibinfo {year} {2019})}\BibitemShut
  {NoStop}%
\bibitem [{\citenamefont {Abbott}\ \emph {et~al.}(2020)\citenamefont {Abbott},
  \citenamefont {Abbott}, \citenamefont {Abraham} \emph
  {et~al.}}]{Abbott_2020}%
  \BibitemOpen
  \bibfield  {author} {\bibinfo {author} {\bibfnamefont {R.}~\bibnamefont
  {Abbott}}, \bibinfo {author} {\bibfnamefont {T.~D.}\ \bibnamefont {Abbott}},
  \bibinfo {author} {\bibfnamefont {S.}~\bibnamefont {Abraham}},  \emph
  {et~al.},\ }\href {\doibase 10.3847/2041-8213/ab960f} {\bibfield  {journal}
  {\bibinfo  {journal} {The Astrophysical Journal}\ }\textbf {\bibinfo {volume}
  {896}},\ \bibinfo {pages} {L44} (\bibinfo {year} {2020})}\BibitemShut
  {NoStop}%
\bibitem [{\citenamefont {Arrington}\ \emph {et~al.}(2012)\citenamefont
  {Arrington}, \citenamefont {Higinbotham}, \citenamefont {Rosner},\ and\
  \citenamefont {Sargsian}}]{ARRINGTON2012898}%
  \BibitemOpen
  \bibfield  {author} {\bibinfo {author} {\bibfnamefont {J.}~\bibnamefont
  {Arrington}}, \bibinfo {author} {\bibfnamefont {D.}~\bibnamefont
  {Higinbotham}}, \bibinfo {author} {\bibfnamefont {G.}~\bibnamefont {Rosner}},
  \ and\ \bibinfo {author} {\bibfnamefont {M.}~\bibnamefont {Sargsian}},\
  }\href {\doibase https://doi.org/10.1016/j.ppnp.2012.04.002} {\bibfield
  {journal} {\bibinfo  {journal} {Progress in Particle and Nuclear Physics}\
  }\textbf {\bibinfo {volume} {67}},\ \bibinfo {pages} {898} (\bibinfo {year}
  {2012})}\BibitemShut {NoStop}%
\bibitem [{\citenamefont {degli Atti}(2015)}]{ATTI20151}%
  \BibitemOpen
  \bibfield  {author} {\bibinfo {author} {\bibfnamefont {C.~C.}\ \bibnamefont
  {degli Atti}},\ }\href {\doibase
  https://doi.org/10.1016/j.physrep.2015.06.002} {\bibfield  {journal}
  {\bibinfo  {journal} {Physics Reports}\ }\textbf {\bibinfo {volume} {590}},\
  \bibinfo {pages} {1} (\bibinfo {year} {2015})}\BibitemShut {NoStop}%
\bibitem [{\citenamefont {Hen}\ \emph {et~al.}(2017)\citenamefont {Hen},
  \citenamefont {Miller}, \citenamefont {Piasetzky},\ and\ \citenamefont
  {Weinstein}}]{RevModPhys.89.045002}%
  \BibitemOpen
  \bibfield  {author} {\bibinfo {author} {\bibfnamefont {O.}~\bibnamefont
  {Hen}}, \bibinfo {author} {\bibfnamefont {G.~A.}\ \bibnamefont {Miller}},
  \bibinfo {author} {\bibfnamefont {E.}~\bibnamefont {Piasetzky}}, \ and\
  \bibinfo {author} {\bibfnamefont {L.~B.}\ \bibnamefont {Weinstein}},\ }\href
  {\doibase 10.1103/RevModPhys.89.045002} {\bibfield  {journal} {\bibinfo
  {journal} {Rev. Mod. Phys.}\ }\textbf {\bibinfo {volume} {89}},\ \bibinfo
  {pages} {045002} (\bibinfo {year} {2017})}\BibitemShut {NoStop}%
\bibitem [{\citenamefont {Rohe}\ \emph {et~al.}(2004)\citenamefont {Rohe},
  \citenamefont {Armstrong}, \citenamefont {Asaturyan}, \citenamefont {Baker},
  \citenamefont {Bueltmann} \emph {et~al.}}]{PhysRevLett.93.182501}%
  \BibitemOpen
  \bibfield  {author} {\bibinfo {author} {\bibfnamefont {D.}~\bibnamefont
  {Rohe}}, \bibinfo {author} {\bibfnamefont {C.~S.}\ \bibnamefont {Armstrong}},
  \bibinfo {author} {\bibfnamefont {R.}~\bibnamefont {Asaturyan}}, \bibinfo
  {author} {\bibfnamefont {O.~K.}\ \bibnamefont {Baker}}, \bibinfo {author}
  {\bibfnamefont {S.}~\bibnamefont {Bueltmann}},  \emph {et~al.} (\bibinfo
  {collaboration} {E97-006 Collaboration}),\ }\href {\doibase
  10.1103/PhysRevLett.93.182501} {\bibfield  {journal} {\bibinfo  {journal}
  {Phys. Rev. Lett.}\ }\textbf {\bibinfo {volume} {93}},\ \bibinfo {pages}
  {182501} (\bibinfo {year} {2004})}\BibitemShut {NoStop}%
\bibitem [{\citenamefont {Onderwater}\ \emph {et~al.}(1998)\citenamefont
  {Onderwater}, \citenamefont {Allaart}, \citenamefont {Aschenauer} \emph
  {et~al.}}]{PhysRevLett.81.2213}%
  \BibitemOpen
  \bibfield  {author} {\bibinfo {author} {\bibfnamefont {C.~J.~G.}\
  \bibnamefont {Onderwater}}, \bibinfo {author} {\bibfnamefont
  {K.}~\bibnamefont {Allaart}}, \bibinfo {author} {\bibfnamefont {E.~C.}\
  \bibnamefont {Aschenauer}},  \emph {et~al.},\ }\href {\doibase
  10.1103/PhysRevLett.81.2213} {\bibfield  {journal} {\bibinfo  {journal}
  {Phys. Rev. Lett.}\ }\textbf {\bibinfo {volume} {81}},\ \bibinfo {pages}
  {2213} (\bibinfo {year} {1998})}\BibitemShut {NoStop}%
\bibitem [{\citenamefont {Starink}\ \emph {et~al.}(2000)\citenamefont
  {Starink}, \citenamefont {{Van Batenburg}}, \citenamefont {Cisbani},
  \citenamefont {Dickhoff}, \citenamefont {Frullani} \emph
  {et~al.}}]{STARINK200033}%
  \BibitemOpen
  \bibfield  {author} {\bibinfo {author} {\bibfnamefont {R.}~\bibnamefont
  {Starink}}, \bibinfo {author} {\bibfnamefont {M.}~\bibnamefont {{Van
  Batenburg}}}, \bibinfo {author} {\bibfnamefont {E.}~\bibnamefont {Cisbani}},
  \bibinfo {author} {\bibfnamefont {W.}~\bibnamefont {Dickhoff}}, \bibinfo
  {author} {\bibfnamefont {S.}~\bibnamefont {Frullani}},  \emph {et~al.},\
  }\href {\doibase https://doi.org/10.1016/S0370-2693(99)01510-5} {\bibfield
  {journal} {\bibinfo  {journal} {Physics Letters B}\ }\textbf {\bibinfo
  {volume} {474}},\ \bibinfo {pages} {33} (\bibinfo {year} {2000})}\BibitemShut
  {NoStop}%
\bibitem [{\citenamefont {Panda}\ \emph {et~al.}(2006)\citenamefont {Panda},
  \citenamefont {Provid\^encia},\ and\ \citenamefont
  {Provid\^encia}}]{PhysRevC.73.035805}%
  \BibitemOpen
  \bibfield  {author} {\bibinfo {author} {\bibfnamefont {P.~K.}\ \bibnamefont
  {Panda}}, \bibinfo {author} {\bibfnamefont {J.~a.~d.}\ \bibnamefont
  {Provid\^encia}}, \ and\ \bibinfo {author} {\bibfnamefont {C.~m.~c.}\
  \bibnamefont {Provid\^encia}},\ }\href {\doibase 10.1103/PhysRevC.73.035805}
  {\bibfield  {journal} {\bibinfo  {journal} {Phys. Rev. C}\ }\textbf {\bibinfo
  {volume} {73}},\ \bibinfo {pages} {035805} (\bibinfo {year}
  {2006})}\BibitemShut {NoStop}%
\bibitem [{\citenamefont {Sammarruca}(2014)}]{PhysRevC.90.064312}%
  \BibitemOpen
  \bibfield  {author} {\bibinfo {author} {\bibfnamefont {F.}~\bibnamefont
  {Sammarruca}},\ }\href {\doibase 10.1103/PhysRevC.90.064312} {\bibfield
  {journal} {\bibinfo  {journal} {Phys. Rev. C}\ }\textbf {\bibinfo {volume}
  {90}},\ \bibinfo {pages} {064312} (\bibinfo {year} {2014})}\BibitemShut
  {NoStop}%
\bibitem [{\citenamefont {Cai}\ and\ \citenamefont
  {Li}(2015)}]{PhysRevC.92.011601}%
  \BibitemOpen
  \bibfield  {author} {\bibinfo {author} {\bibfnamefont {B.-J.}\ \bibnamefont
  {Cai}}\ and\ \bibinfo {author} {\bibfnamefont {B.-A.}\ \bibnamefont {Li}},\
  }\href {\doibase 10.1103/PhysRevC.92.011601} {\bibfield  {journal} {\bibinfo
  {journal} {Phys. Rev. C}\ }\textbf {\bibinfo {volume} {92}},\ \bibinfo
  {pages} {011601} (\bibinfo {year} {2015})}\BibitemShut {NoStop}%
\bibitem [{\citenamefont {Hen}\ \emph {et~al.}(2015{\natexlab{a}})\citenamefont
  {Hen}, \citenamefont {Li}, \citenamefont {Guo}, \citenamefont {Weinstein},\
  and\ \citenamefont {Piasetzky}}]{PhysRevC.91.025803}%
  \BibitemOpen
  \bibfield  {author} {\bibinfo {author} {\bibfnamefont {O.}~\bibnamefont
  {Hen}}, \bibinfo {author} {\bibfnamefont {B.-A.}\ \bibnamefont {Li}},
  \bibinfo {author} {\bibfnamefont {W.-J.}\ \bibnamefont {Guo}}, \bibinfo
  {author} {\bibfnamefont {L.~B.}\ \bibnamefont {Weinstein}}, \ and\ \bibinfo
  {author} {\bibfnamefont {E.}~\bibnamefont {Piasetzky}},\ }\href {\doibase
  10.1103/PhysRevC.91.025803} {\bibfield  {journal} {\bibinfo  {journal} {Phys.
  Rev. C}\ }\textbf {\bibinfo {volume} {91}},\ \bibinfo {pages} {025803}
  (\bibinfo {year} {2015}{\natexlab{a}})}\BibitemShut {NoStop}%
\bibitem [{\citenamefont {Wang}\ \emph {et~al.}(2017)\citenamefont {Wang},
  \citenamefont {Xu}, \citenamefont {Ren},\ and\ \citenamefont
  {Gao}}]{PhysRevC.96.054603}%
  \BibitemOpen
  \bibfield  {author} {\bibinfo {author} {\bibfnamefont {Z.}~\bibnamefont
  {Wang}}, \bibinfo {author} {\bibfnamefont {C.}~\bibnamefont {Xu}}, \bibinfo
  {author} {\bibfnamefont {Z.}~\bibnamefont {Ren}}, \ and\ \bibinfo {author}
  {\bibfnamefont {C.}~\bibnamefont {Gao}},\ }\href {\doibase
  10.1103/PhysRevC.96.054603} {\bibfield  {journal} {\bibinfo  {journal} {Phys.
  Rev. C}\ }\textbf {\bibinfo {volume} {96}},\ \bibinfo {pages} {054603}
  (\bibinfo {year} {2017})}\BibitemShut {NoStop}%
\bibitem [{\citenamefont {Yang}\ \emph {et~al.}(2019)\citenamefont {Yang},
  \citenamefont {Shang}, \citenamefont {Yong}, \citenamefont {Zuo},\ and\
  \citenamefont {Gao}}]{PhysRevC.100.054325}%
  \BibitemOpen
  \bibfield  {author} {\bibinfo {author} {\bibfnamefont {Z.~X.}\ \bibnamefont
  {Yang}}, \bibinfo {author} {\bibfnamefont {X.~L.}\ \bibnamefont {Shang}},
  \bibinfo {author} {\bibfnamefont {G.~C.}\ \bibnamefont {Yong}}, \bibinfo
  {author} {\bibfnamefont {W.}~\bibnamefont {Zuo}}, \ and\ \bibinfo {author}
  {\bibfnamefont {Y.}~\bibnamefont {Gao}},\ }\href {\doibase
  10.1103/PhysRevC.100.054325} {\bibfield  {journal} {\bibinfo  {journal}
  {Phys. Rev. C}\ }\textbf {\bibinfo {volume} {100}},\ \bibinfo {pages}
  {054325} (\bibinfo {year} {2019})}\BibitemShut {NoStop}%
\bibitem [{\citenamefont {Weinstein}\ \emph {et~al.}(2011)\citenamefont
  {Weinstein}, \citenamefont {Piasetzky}, \citenamefont {Higinbotham},
  \citenamefont {Gomez}, \citenamefont {Hen},\ and\ \citenamefont
  {Shneor}}]{PhysRevLett.106.052301}%
  \BibitemOpen
  \bibfield  {author} {\bibinfo {author} {\bibfnamefont {L.~B.}\ \bibnamefont
  {Weinstein}}, \bibinfo {author} {\bibfnamefont {E.}~\bibnamefont
  {Piasetzky}}, \bibinfo {author} {\bibfnamefont {D.~W.}\ \bibnamefont
  {Higinbotham}}, \bibinfo {author} {\bibfnamefont {J.}~\bibnamefont {Gomez}},
  \bibinfo {author} {\bibfnamefont {O.}~\bibnamefont {Hen}}, \ and\ \bibinfo
  {author} {\bibfnamefont {R.}~\bibnamefont {Shneor}},\ }\href {\doibase
  10.1103/PhysRevLett.106.052301} {\bibfield  {journal} {\bibinfo  {journal}
  {Phys. Rev. Lett.}\ }\textbf {\bibinfo {volume} {106}},\ \bibinfo {pages}
  {052301} (\bibinfo {year} {2011})}\BibitemShut {NoStop}%
\bibitem [{\citenamefont {Hen}\ \emph {et~al.}(2012)\citenamefont {Hen},
  \citenamefont {Piasetzky},\ and\ \citenamefont
  {Weinstein}}]{PhysRevC.85.047301}%
  \BibitemOpen
  \bibfield  {author} {\bibinfo {author} {\bibfnamefont {O.}~\bibnamefont
  {Hen}}, \bibinfo {author} {\bibfnamefont {E.}~\bibnamefont {Piasetzky}}, \
  and\ \bibinfo {author} {\bibfnamefont {L.~B.}\ \bibnamefont {Weinstein}},\
  }\href {\doibase 10.1103/PhysRevC.85.047301} {\bibfield  {journal} {\bibinfo
  {journal} {Phys. Rev. C}\ }\textbf {\bibinfo {volume} {85}},\ \bibinfo
  {pages} {047301} (\bibinfo {year} {2012})}\BibitemShut {NoStop}%
\bibitem [{\citenamefont {Fields}\ \emph {et~al.}(2013)\citenamefont {Fields},
  \citenamefont {Chvojka}, \citenamefont {Aliaga} \emph
  {et~al.}}]{PhysRevLett.111.022501}%
  \BibitemOpen
  \bibfield  {author} {\bibinfo {author} {\bibfnamefont {L.}~\bibnamefont
  {Fields}}, \bibinfo {author} {\bibfnamefont {J.}~\bibnamefont {Chvojka}},
  \bibinfo {author} {\bibfnamefont {L.}~\bibnamefont {Aliaga}},  \emph {et~al.}
  (\bibinfo {collaboration} {MINERvA Collaboration}),\ }\href {\doibase
  10.1103/PhysRevLett.111.022501} {\bibfield  {journal} {\bibinfo  {journal}
  {Phys. Rev. Lett.}\ }\textbf {\bibinfo {volume} {111}},\ \bibinfo {pages}
  {022501} (\bibinfo {year} {2013})}\BibitemShut {NoStop}%
\bibitem [{\citenamefont {Fiorentini}\ \emph {et~al.}(2013)\citenamefont
  {Fiorentini}, \citenamefont {Schmitz}, \citenamefont {Rodrigues},\ and\
  \citenamefont {et~al.}}]{PhysRevLett.111.022502}%
  \BibitemOpen
  \bibfield  {author} {\bibinfo {author} {\bibfnamefont {G.~A.}\ \bibnamefont
  {Fiorentini}}, \bibinfo {author} {\bibfnamefont {D.~W.}\ \bibnamefont
  {Schmitz}}, \bibinfo {author} {\bibfnamefont {P.~A.}\ \bibnamefont
  {Rodrigues}}, \ and\ \bibinfo {author} {\bibnamefont {et~al.}} (\bibinfo
  {collaboration} {MINERvA Collaboration}),\ }\href {\doibase
  10.1103/PhysRevLett.111.022502} {\bibfield  {journal} {\bibinfo  {journal}
  {Phys. Rev. Lett.}\ }\textbf {\bibinfo {volume} {111}},\ \bibinfo {pages}
  {022502} (\bibinfo {year} {2013})}\BibitemShut {NoStop}%
\bibitem [{\citenamefont {Souza}\ \emph
  {et~al.}(2020{\natexlab{a}})\citenamefont {Souza}, \citenamefont {Dutra},
  \citenamefont {Lenzi},\ and\ \citenamefont {Louren\ifmmode~\mbox{\c{c}}\else
  \c{c}\fi{}o}}]{PhysRevC.101.065202}%
  \BibitemOpen
  \bibfield  {author} {\bibinfo {author} {\bibfnamefont {L.~A.}\ \bibnamefont
  {Souza}}, \bibinfo {author} {\bibfnamefont {M.}~\bibnamefont {Dutra}},
  \bibinfo {author} {\bibfnamefont {C.~H.}\ \bibnamefont {Lenzi}}, \ and\
  \bibinfo {author} {\bibfnamefont {O.}~\bibnamefont
  {Louren\ifmmode~\mbox{\c{c}}\else \c{c}\fi{}o}},\ }\href {\doibase
  10.1103/PhysRevC.101.065202} {\bibfield  {journal} {\bibinfo  {journal}
  {Phys. Rev. C}\ }\textbf {\bibinfo {volume} {101}},\ \bibinfo {pages}
  {065202} (\bibinfo {year} {2020}{\natexlab{a}})}\BibitemShut {NoStop}%
\bibitem [{\citenamefont {Hong}\ \emph {et~al.}(2022)\citenamefont {Hong},
  \citenamefont {Ren},\ and\ \citenamefont {Mu}}]{Hong_2022}%
  \BibitemOpen
  \bibfield  {author} {\bibinfo {author} {\bibfnamefont {B.}~\bibnamefont
  {Hong}}, \bibinfo {author} {\bibfnamefont {Z.}~\bibnamefont {Ren}}, \ and\
  \bibinfo {author} {\bibfnamefont {X.-L.}\ \bibnamefont {Mu}},\ }\href
  {\doibase 10.1088/1674-1137/ac588d} {\bibfield  {journal} {\bibinfo
  {journal} {Chinese Physics C}\ }\textbf {\bibinfo {volume} {46}},\ \bibinfo
  {pages} {065104} (\bibinfo {year} {2022})}\BibitemShut {NoStop}%
\bibitem [{\citenamefont {Li}\ \emph {et~al.}(2019)\citenamefont {Li},
  \citenamefont {Ren}, \citenamefont {Hong}, \citenamefont {Lu},\ and\
  \citenamefont {Bai}}]{LI2019118}%
  \BibitemOpen
  \bibfield  {author} {\bibinfo {author} {\bibfnamefont {Z.}~\bibnamefont
  {Li}}, \bibinfo {author} {\bibfnamefont {Z.}~\bibnamefont {Ren}}, \bibinfo
  {author} {\bibfnamefont {B.}~\bibnamefont {Hong}}, \bibinfo {author}
  {\bibfnamefont {H.}~\bibnamefont {Lu}}, \ and\ \bibinfo {author}
  {\bibfnamefont {D.}~\bibnamefont {Bai}},\ }\href {\doibase
  https://doi.org/10.1016/j.nuclphysa.2019.07.002} {\bibfield  {journal}
  {\bibinfo  {journal} {Nuclear Physics A}\ }\textbf {\bibinfo {volume}
  {990}},\ \bibinfo {pages} {118} (\bibinfo {year} {2019})}\BibitemShut
  {NoStop}%
\bibitem [{\citenamefont {Lu}\ \emph {et~al.}(2021)\citenamefont {Lu},
  \citenamefont {Ren},\ and\ \citenamefont {Bai}}]{LU2021122200}%
  \BibitemOpen
  \bibfield  {author} {\bibinfo {author} {\bibfnamefont {H.}~\bibnamefont
  {Lu}}, \bibinfo {author} {\bibfnamefont {Z.}~\bibnamefont {Ren}}, \ and\
  \bibinfo {author} {\bibfnamefont {D.}~\bibnamefont {Bai}},\ }\href {\doibase
  https://doi.org/10.1016/j.nuclphysa.2021.122200} {\bibfield  {journal}
  {\bibinfo  {journal} {Nuclear Physics A}\ }\textbf {\bibinfo {volume}
  {1011}},\ \bibinfo {pages} {122200} (\bibinfo {year} {2021})}\BibitemShut
  {NoStop}%
\bibitem [{\citenamefont {Louren\ifmmode~\mbox{\c{c}}\else \c{c}\fi{}o}\ \emph
  {et~al.}(2022)\citenamefont {Louren\ifmmode~\mbox{\c{c}}\else \c{c}\fi{}o},
  \citenamefont {Frederico},\ and\ \citenamefont
  {Dutra}}]{PhysRevD.105.023008}%
  \BibitemOpen
  \bibfield  {author} {\bibinfo {author} {\bibfnamefont {O.}~\bibnamefont
  {Louren\ifmmode~\mbox{\c{c}}\else \c{c}\fi{}o}}, \bibinfo {author}
  {\bibfnamefont {T.}~\bibnamefont {Frederico}}, \ and\ \bibinfo {author}
  {\bibfnamefont {M.}~\bibnamefont {Dutra}},\ }\href {\doibase
  10.1103/PhysRevD.105.023008} {\bibfield  {journal} {\bibinfo  {journal}
  {Phys. Rev. D}\ }\textbf {\bibinfo {volume} {105}},\ \bibinfo {pages}
  {023008} (\bibinfo {year} {2022})}\BibitemShut {NoStop}%
\bibitem [{\citenamefont {Souza}\ \emph
  {et~al.}(2020{\natexlab{b}})\citenamefont {Souza}, \citenamefont {Negreiros},
  \citenamefont {Dutra}, \citenamefont {Menezes},\ and\ \citenamefont
  {Lourenço}}]{souza2020shortrange}%
  \BibitemOpen
  \bibfield  {author} {\bibinfo {author} {\bibfnamefont {L.~A.}\ \bibnamefont
  {Souza}}, \bibinfo {author} {\bibfnamefont {R.}~\bibnamefont {Negreiros}},
  \bibinfo {author} {\bibfnamefont {M.}~\bibnamefont {Dutra}}, \bibinfo
  {author} {\bibfnamefont {D.~P.}\ \bibnamefont {Menezes}}, \ and\ \bibinfo
  {author} {\bibfnamefont {O.}~\bibnamefont {Lourenço}},\ }\href
  {https://doi.org/10.48550/arXiv.2004.10309} {\bibfield  {journal} {\bibinfo
  {journal} {arXiv}\ } (\bibinfo {year} {2020}{\natexlab{b}})}\BibitemShut
  {NoStop}%
\bibitem [{\citenamefont {Bertone}\ \emph {et~al.}(2005)\citenamefont
  {Bertone}, \citenamefont {Hooper},\ and\ \citenamefont
  {Silk}}]{BERTONE2005279}%
  \BibitemOpen
  \bibfield  {author} {\bibinfo {author} {\bibfnamefont {G.}~\bibnamefont
  {Bertone}}, \bibinfo {author} {\bibfnamefont {D.}~\bibnamefont {Hooper}}, \
  and\ \bibinfo {author} {\bibfnamefont {J.}~\bibnamefont {Silk}},\ }\href
  {\doibase https://doi.org/10.1016/j.physrep.2004.08.031} {\bibfield
  {journal} {\bibinfo  {journal} {Physics Reports}\ }\textbf {\bibinfo {volume}
  {405}},\ \bibinfo {pages} {279} (\bibinfo {year} {2005})}\BibitemShut
  {NoStop}%
\bibitem [{\citenamefont {Bertone}\ and\ \citenamefont
  {Hooper}(2018)}]{RevModPhys.90.045002}%
  \BibitemOpen
  \bibfield  {author} {\bibinfo {author} {\bibfnamefont {G.}~\bibnamefont
  {Bertone}}\ and\ \bibinfo {author} {\bibfnamefont {D.}~\bibnamefont
  {Hooper}},\ }\href {\doibase 10.1103/RevModPhys.90.045002} {\bibfield
  {journal} {\bibinfo  {journal} {Rev. Mod. Phys.}\ }\textbf {\bibinfo {volume}
  {90}},\ \bibinfo {pages} {045002} (\bibinfo {year} {2018})}\BibitemShut
  {NoStop}%
\bibitem [{\citenamefont {Buen-Abad}\ \emph {et~al.}(2022)\citenamefont
  {Buen-Abad}, \citenamefont {Essig}, \citenamefont {McKeen},\ and\
  \citenamefont {Zhong}}]{BUENABAD20221}%
  \BibitemOpen
  \bibfield  {author} {\bibinfo {author} {\bibfnamefont {M.~A.}\ \bibnamefont
  {Buen-Abad}}, \bibinfo {author} {\bibfnamefont {R.}~\bibnamefont {Essig}},
  \bibinfo {author} {\bibfnamefont {D.}~\bibnamefont {McKeen}}, \ and\ \bibinfo
  {author} {\bibfnamefont {Y.-M.}\ \bibnamefont {Zhong}},\ }\href {\doibase
  https://doi.org/10.1016/j.physrep.2022.02.006} {\bibfield  {journal}
  {\bibinfo  {journal} {Physics Reports}\ }\textbf {\bibinfo {volume} {961}},\
  \bibinfo {pages} {1} (\bibinfo {year} {2022})}\BibitemShut {NoStop}%
\bibitem [{\citenamefont {Kouvaris}\ and\ \citenamefont
  {Tinyakov}(2011)}]{PhysRevD.83.083512}%
  \BibitemOpen
  \bibfield  {author} {\bibinfo {author} {\bibfnamefont {C.}~\bibnamefont
  {Kouvaris}}\ and\ \bibinfo {author} {\bibfnamefont {P.}~\bibnamefont
  {Tinyakov}},\ }\href {\doibase 10.1103/PhysRevD.83.083512} {\bibfield
  {journal} {\bibinfo  {journal} {Phys. Rev. D}\ }\textbf {\bibinfo {volume}
  {83}},\ \bibinfo {pages} {083512} (\bibinfo {year} {2011})}\BibitemShut
  {NoStop}%
\bibitem [{\citenamefont {Quddus}\ \emph {et~al.}(2020)\citenamefont {Quddus},
  \citenamefont {Panotopoulos}, \citenamefont {Kumar}, \citenamefont {Ahmad},\
  and\ \citenamefont {Patra}}]{Quddus_2020}%
  \BibitemOpen
  \bibfield  {author} {\bibinfo {author} {\bibfnamefont {A.}~\bibnamefont
  {Quddus}}, \bibinfo {author} {\bibfnamefont {G.}~\bibnamefont
  {Panotopoulos}}, \bibinfo {author} {\bibfnamefont {B.}~\bibnamefont {Kumar}},
  \bibinfo {author} {\bibfnamefont {S.}~\bibnamefont {Ahmad}}, \ and\ \bibinfo
  {author} {\bibfnamefont {S.~K.}\ \bibnamefont {Patra}},\ }\href {\doibase
  10.1088/1361-6471/ab9d36} {\bibfield  {journal} {\bibinfo  {journal} {Journal
  of Physics G: Nuclear and Particle Physics}\ }\textbf {\bibinfo {volume}
  {47}},\ \bibinfo {pages} {095202} (\bibinfo {year} {2020})}\BibitemShut
  {NoStop}%
\bibitem [{\citenamefont {Das}\ \emph {et~al.}(2021)\citenamefont {Das},
  \citenamefont {Kumar},\ and\ \citenamefont {Patra}}]{PhysRevD.104.063028}%
  \BibitemOpen
  \bibfield  {author} {\bibinfo {author} {\bibfnamefont {H.~C.}\ \bibnamefont
  {Das}}, \bibinfo {author} {\bibfnamefont {A.}~\bibnamefont {Kumar}}, \ and\
  \bibinfo {author} {\bibfnamefont {S.~K.}\ \bibnamefont {Patra}},\ }\href
  {\doibase 10.1103/PhysRevD.104.063028} {\bibfield  {journal} {\bibinfo
  {journal} {Phys. Rev. D}\ }\textbf {\bibinfo {volume} {104}},\ \bibinfo
  {pages} {063028} (\bibinfo {year} {2021})}\BibitemShut {NoStop}%
\bibitem [{\citenamefont {Das}\ \emph {et~al.}(2020)\citenamefont {Das},
  \citenamefont {Kumar}, \citenamefont {Kumar}, \citenamefont {Biswal},
  \citenamefont {Nakatsukasa}, \citenamefont {Li},\ and\ \citenamefont
  {Patra}}]{10.1093/mnras/staa1435}%
  \BibitemOpen
  \bibfield  {author} {\bibinfo {author} {\bibfnamefont {H.~C.}\ \bibnamefont
  {Das}}, \bibinfo {author} {\bibfnamefont {A.}~\bibnamefont {Kumar}}, \bibinfo
  {author} {\bibfnamefont {B.}~\bibnamefont {Kumar}}, \bibinfo {author}
  {\bibfnamefont {S.~K.}\ \bibnamefont {Biswal}}, \bibinfo {author}
  {\bibfnamefont {T.}~\bibnamefont {Nakatsukasa}}, \bibinfo {author}
  {\bibfnamefont {A.}~\bibnamefont {Li}}, \ and\ \bibinfo {author}
  {\bibfnamefont {S.~K.}\ \bibnamefont {Patra}},\ }\href {\doibase
  10.1093/mnras/staa1435} {\bibfield  {journal} {\bibinfo  {journal} {Monthly
  Notices of the Royal Astronomical Society}\ }\textbf {\bibinfo {volume}
  {495}},\ \bibinfo {pages} {4893} (\bibinfo {year} {2020})}\BibitemShut
  {NoStop}%
\bibitem [{\citenamefont {Kong}\ \emph {et~al.}(2017)\citenamefont {Kong},
  \citenamefont {Xu}, \citenamefont {Chen}, \citenamefont {Li},\ and\
  \citenamefont {Ma}}]{PhysRevC.95.034324}%
  \BibitemOpen
  \bibfield  {author} {\bibinfo {author} {\bibfnamefont {H.-Y.}\ \bibnamefont
  {Kong}}, \bibinfo {author} {\bibfnamefont {J.}~\bibnamefont {Xu}}, \bibinfo
  {author} {\bibfnamefont {L.-W.}\ \bibnamefont {Chen}}, \bibinfo {author}
  {\bibfnamefont {B.-A.}\ \bibnamefont {Li}}, \ and\ \bibinfo {author}
  {\bibfnamefont {Y.-G.}\ \bibnamefont {Ma}},\ }\href {\doibase
  10.1103/PhysRevC.95.034324} {\bibfield  {journal} {\bibinfo  {journal} {Phys.
  Rev. C}\ }\textbf {\bibinfo {volume} {95}},\ \bibinfo {pages} {034324}
  (\bibinfo {year} {2017})}\BibitemShut {NoStop}%
\bibitem [{\citenamefont {Tang}\ \emph {et~al.}(2021)\citenamefont {Tang},
  \citenamefont {Jiang}, \citenamefont {Han}, \citenamefont {Fan},\ and\
  \citenamefont {Wei}}]{PhysRevD.104.063032}%
  \BibitemOpen
  \bibfield  {author} {\bibinfo {author} {\bibfnamefont {S.-P.}\ \bibnamefont
  {Tang}}, \bibinfo {author} {\bibfnamefont {J.-L.}\ \bibnamefont {Jiang}},
  \bibinfo {author} {\bibfnamefont {M.-Z.}\ \bibnamefont {Han}}, \bibinfo
  {author} {\bibfnamefont {Y.-Z.}\ \bibnamefont {Fan}}, \ and\ \bibinfo
  {author} {\bibfnamefont {D.-M.}\ \bibnamefont {Wei}},\ }\href {\doibase
  10.1103/PhysRevD.104.063032} {\bibfield  {journal} {\bibinfo  {journal}
  {Phys. Rev. D}\ }\textbf {\bibinfo {volume} {104}},\ \bibinfo {pages}
  {063032} (\bibinfo {year} {2021})}\BibitemShut {NoStop}%
\bibitem [{\citenamefont {Adhikari}\ \emph {et~al.}(2021)\citenamefont
  {Adhikari}, \citenamefont {Albataineh}, \citenamefont {Androic},\ and\
  \citenamefont {Aniol}}]{PhysRevLett.126.172502}%
  \BibitemOpen
  \bibfield  {author} {\bibinfo {author} {\bibfnamefont {D.}~\bibnamefont
  {Adhikari}}, \bibinfo {author} {\bibfnamefont {H.}~\bibnamefont
  {Albataineh}}, \bibinfo {author} {\bibfnamefont {D.}~\bibnamefont {Androic}},
  \ and\ \bibinfo {author} {\bibfnamefont {K.}~\bibnamefont {Aniol}} (\bibinfo
  {collaboration} {PREX Collaboration}),\ }\href {\doibase
  10.1103/PhysRevLett.126.172502} {\bibfield  {journal} {\bibinfo  {journal}
  {Phys. Rev. Lett.}\ }\textbf {\bibinfo {volume} {126}},\ \bibinfo {pages}
  {172502} (\bibinfo {year} {2021})}\BibitemShut {NoStop}%
\bibitem [{\citenamefont {Reed}\ \emph {et~al.}(2021)\citenamefont {Reed},
  \citenamefont {Fattoyev}, \citenamefont {Horowitz},\ and\ \citenamefont
  {Piekarewicz}}]{PhysRevLett.126.172503}%
  \BibitemOpen
  \bibfield  {author} {\bibinfo {author} {\bibfnamefont {B.~T.}\ \bibnamefont
  {Reed}}, \bibinfo {author} {\bibfnamefont {F.~J.}\ \bibnamefont {Fattoyev}},
  \bibinfo {author} {\bibfnamefont {C.~J.}\ \bibnamefont {Horowitz}}, \ and\
  \bibinfo {author} {\bibfnamefont {J.}~\bibnamefont {Piekarewicz}},\ }\href
  {\doibase 10.1103/PhysRevLett.126.172503} {\bibfield  {journal} {\bibinfo
  {journal} {Phys. Rev. Lett.}\ }\textbf {\bibinfo {volume} {126}},\ \bibinfo
  {pages} {172503} (\bibinfo {year} {2021})}\BibitemShut {NoStop}%
\bibitem [{\citenamefont {Jones}(2002)}]{Jones}%
  \BibitemOpen
  \bibfield  {author} {\bibinfo {author} {\bibfnamefont {D.~I.}\ \bibnamefont
  {Jones}},\ }\href {\doibase 10.1088/0264-9381/19/7/304} {\bibfield  {journal}
  {\bibinfo  {journal} {Classical and Quantum Gravity}\ }\textbf {\bibinfo
  {volume} {19}},\ \bibinfo {pages} {1255} (\bibinfo {year}
  {2002})}\BibitemShut {NoStop}%
\bibitem [{\citenamefont {Kokkotas}\ and\ \citenamefont
  {Schutz}(1992)}]{10.1093/mnras/255.1.119}%
  \BibitemOpen
  \bibfield  {author} {\bibinfo {author} {\bibfnamefont {K.~D.}\ \bibnamefont
  {Kokkotas}}\ and\ \bibinfo {author} {\bibfnamefont {B.~F.}\ \bibnamefont
  {Schutz}},\ }\href {\doibase 10.1093/mnras/255.1.119} {\bibfield  {journal}
  {\bibinfo  {journal} {Monthly Notices of the Royal Astronomical Society}\
  }\textbf {\bibinfo {volume} {255}},\ \bibinfo {pages} {119} (\bibinfo {year}
  {1992})}\BibitemShut {NoStop}%
\bibitem [{\citenamefont {Lau}\ \emph {et~al.}(2010)\citenamefont {Lau},
  \citenamefont {Leung},\ and\ \citenamefont {Lin}}]{Lau_2010}%
  \BibitemOpen
  \bibfield  {author} {\bibinfo {author} {\bibfnamefont {H.~K.}\ \bibnamefont
  {Lau}}, \bibinfo {author} {\bibfnamefont {P.~T.}\ \bibnamefont {Leung}}, \
  and\ \bibinfo {author} {\bibfnamefont {L.~M.}\ \bibnamefont {Lin}},\ }\href
  {\doibase 10.1088/0004-637x/714/2/1234} {\bibfield  {journal} {\bibinfo
  {journal} {The Astrophysical Journal}\ }\textbf {\bibinfo {volume} {714}},\
  \bibinfo {pages} {1234} (\bibinfo {year} {2010})}\BibitemShut {NoStop}%
\bibitem [{\citenamefont {Kokkotas}\ and\ \citenamefont
  {Schmidt}(1999)}]{RN84}%
  \BibitemOpen
  \bibfield  {author} {\bibinfo {author} {\bibfnamefont {K.~D.}\ \bibnamefont
  {Kokkotas}}\ and\ \bibinfo {author} {\bibfnamefont {B.~G.}\ \bibnamefont
  {Schmidt}},\ }\href {\doibase 10.12942/lrr-1999-2} {\bibfield  {journal}
  {\bibinfo  {journal} {Living Reviews in Relativity}\ }\textbf {\bibinfo
  {volume} {2}},\ \bibinfo {pages} {2} (\bibinfo {year} {1999})}\BibitemShut
  {NoStop}%
\bibitem [{\citenamefont {Passamonti}\ \emph {et~al.}(2007)\citenamefont
  {Passamonti}, \citenamefont {Stergioulas},\ and\ \citenamefont
  {Nagar}}]{PhysRevD.75.084038}%
  \BibitemOpen
  \bibfield  {author} {\bibinfo {author} {\bibfnamefont {A.}~\bibnamefont
  {Passamonti}}, \bibinfo {author} {\bibfnamefont {N.}~\bibnamefont
  {Stergioulas}}, \ and\ \bibinfo {author} {\bibfnamefont {A.}~\bibnamefont
  {Nagar}},\ }\href {\doibase 10.1103/PhysRevD.75.084038} {\bibfield  {journal}
  {\bibinfo  {journal} {Phys. Rev. D}\ }\textbf {\bibinfo {volume} {75}},\
  \bibinfo {pages} {084038} (\bibinfo {year} {2007})}\BibitemShut {NoStop}%
\bibitem [{\citenamefont {Passamonti}\ \emph {et~al.}(2006)\citenamefont
  {Passamonti}, \citenamefont {Bruni}, \citenamefont {Gualtieri}, \citenamefont
  {Nagar},\ and\ \citenamefont {Sopuerta}}]{PhysRevD.73.084010}%
  \BibitemOpen
  \bibfield  {author} {\bibinfo {author} {\bibfnamefont {A.}~\bibnamefont
  {Passamonti}}, \bibinfo {author} {\bibfnamefont {M.}~\bibnamefont {Bruni}},
  \bibinfo {author} {\bibfnamefont {L.}~\bibnamefont {Gualtieri}}, \bibinfo
  {author} {\bibfnamefont {A.}~\bibnamefont {Nagar}}, \ and\ \bibinfo {author}
  {\bibfnamefont {C.~F.}\ \bibnamefont {Sopuerta}},\ }\href {\doibase
  10.1103/PhysRevD.73.084010} {\bibfield  {journal} {\bibinfo  {journal} {Phys.
  Rev. D}\ }\textbf {\bibinfo {volume} {73}},\ \bibinfo {pages} {084010}
  (\bibinfo {year} {2006})}\BibitemShut {NoStop}%
\bibitem [{\citenamefont {{Vaeth}}\ and\ \citenamefont
  {{Chanmugam}}(1992)}]{1992AA}%
  \BibitemOpen
  \bibfield  {author} {\bibinfo {author} {\bibfnamefont {H.~M.}\ \bibnamefont
  {{Vaeth}}}\ and\ \bibinfo {author} {\bibfnamefont {G.}~\bibnamefont
  {{Chanmugam}}},\ }\href {\doibase
  https://ui.adsabs.harvard.edu/abs/1992A&A...260..250V} {\bibfield  {journal}
  {\bibinfo  {journal} {A\&A}\ }\textbf {\bibinfo {volume} {260}},\ \bibinfo
  {pages} {250} (\bibinfo {year} {1992})}\BibitemShut {NoStop}%
\bibitem [{\citenamefont {V\'asquez~Flores}\ and\ \citenamefont
  {Lugones}(2010)}]{PhysRevD.82.063006}%
  \BibitemOpen
  \bibfield  {author} {\bibinfo {author} {\bibfnamefont {C.}~\bibnamefont
  {V\'asquez~Flores}}\ and\ \bibinfo {author} {\bibfnamefont {G.}~\bibnamefont
  {Lugones}},\ }\href {\doibase 10.1103/PhysRevD.82.063006} {\bibfield
  {journal} {\bibinfo  {journal} {Phys. Rev. D}\ }\textbf {\bibinfo {volume}
  {82}},\ \bibinfo {pages} {063006} (\bibinfo {year} {2010})}\BibitemShut
  {NoStop}%
\bibitem [{\citenamefont {Flores}\ \emph {et~al.}(2017)\citenamefont {Flores},
  \citenamefont {Hall},\ and\ \citenamefont {Jaikumar}}]{PhysRevC.96.065803}%
  \BibitemOpen
  \bibfield  {author} {\bibinfo {author} {\bibfnamefont {C.~V.}\ \bibnamefont
  {Flores}}, \bibinfo {author} {\bibfnamefont {Z.~B.}\ \bibnamefont {Hall}}, \
  and\ \bibinfo {author} {\bibfnamefont {P.}~\bibnamefont {Jaikumar}},\ }\href
  {\doibase 10.1103/PhysRevC.96.065803} {\bibfield  {journal} {\bibinfo
  {journal} {Phys. Rev. C}\ }\textbf {\bibinfo {volume} {96}},\ \bibinfo
  {pages} {065803} (\bibinfo {year} {2017})}\BibitemShut {NoStop}%
\bibitem [{\citenamefont {Ranea-Sandoval}\ \emph {et~al.}(2018)\citenamefont
  {Ranea-Sandoval}, \citenamefont {Guilera}, \citenamefont {Mariani},\ and\
  \citenamefont {Orsaria}}]{Ranea_Sandoval_2018}%
  \BibitemOpen
  \bibfield  {author} {\bibinfo {author} {\bibfnamefont {I.~F.}\ \bibnamefont
  {Ranea-Sandoval}}, \bibinfo {author} {\bibfnamefont {O.~M.}\ \bibnamefont
  {Guilera}}, \bibinfo {author} {\bibfnamefont {M.}~\bibnamefont {Mariani}}, \
  and\ \bibinfo {author} {\bibfnamefont {M.~G.}\ \bibnamefont {Orsaria}},\
  }\href {\doibase 10.1088/1475-7516/2018/12/031} {\bibfield  {journal}
  {\bibinfo  {journal} {Journal of Cosmology and Astroparticle Physics}\
  }\textbf {\bibinfo {volume} {2018}},\ \bibinfo {pages} {031} (\bibinfo {year}
  {2018})}\BibitemShut {NoStop}%
\bibitem [{\citenamefont {Pereira}\ \emph {et~al.}(2018)\citenamefont
  {Pereira}, \citenamefont {Flores},\ and\ \citenamefont
  {Lugones}}]{Pereira_2018}%
  \BibitemOpen
  \bibfield  {author} {\bibinfo {author} {\bibfnamefont {J.~P.}\ \bibnamefont
  {Pereira}}, \bibinfo {author} {\bibfnamefont {C.~V.}\ \bibnamefont {Flores}},
  \ and\ \bibinfo {author} {\bibfnamefont {G.}~\bibnamefont {Lugones}},\ }\href
  {\doibase 10.3847/1538-4357/aabfbf} {\bibfield  {journal} {\bibinfo
  {journal} {The Astrophysical Journal}\ }\textbf {\bibinfo {volume} {860}},\
  \bibinfo {pages} {12} (\bibinfo {year} {2018})}\BibitemShut {NoStop}%
\bibitem [{\citenamefont {Sun}\ \emph {et~al.}(2021)\citenamefont {Sun},
  \citenamefont {Zheng}, \citenamefont {Chen}, \citenamefont {Burgio},\ and\
  \citenamefont {Schulze}}]{PhysRevD.103.103003}%
  \BibitemOpen
  \bibfield  {author} {\bibinfo {author} {\bibfnamefont {T.-T.}\ \bibnamefont
  {Sun}}, \bibinfo {author} {\bibfnamefont {Z.-Y.}\ \bibnamefont {Zheng}},
  \bibinfo {author} {\bibfnamefont {H.}~\bibnamefont {Chen}}, \bibinfo {author}
  {\bibfnamefont {G.~F.}\ \bibnamefont {Burgio}}, \ and\ \bibinfo {author}
  {\bibfnamefont {H.-J.}\ \bibnamefont {Schulze}},\ }\href {\doibase
  10.1103/PhysRevD.103.103003} {\bibfield  {journal} {\bibinfo  {journal}
  {Phys. Rev. D}\ }\textbf {\bibinfo {volume} {103}},\ \bibinfo {pages}
  {103003} (\bibinfo {year} {2021})}\BibitemShut {NoStop}%
\bibitem [{\citenamefont {Jaikumar}\ \emph {et~al.}(2021)\citenamefont
  {Jaikumar}, \citenamefont {Semposki}, \citenamefont {Prakash},\ and\
  \citenamefont {Constantinou}}]{PhysRevD.103.123009}%
  \BibitemOpen
  \bibfield  {author} {\bibinfo {author} {\bibfnamefont {P.}~\bibnamefont
  {Jaikumar}}, \bibinfo {author} {\bibfnamefont {A.}~\bibnamefont {Semposki}},
  \bibinfo {author} {\bibfnamefont {M.}~\bibnamefont {Prakash}}, \ and\
  \bibinfo {author} {\bibfnamefont {C.}~\bibnamefont {Constantinou}},\ }\href
  {\doibase 10.1103/PhysRevD.103.123009} {\bibfield  {journal} {\bibinfo
  {journal} {Phys. Rev. D}\ }\textbf {\bibinfo {volume} {103}},\ \bibinfo
  {pages} {123009} (\bibinfo {year} {2021})}\BibitemShut {NoStop}%
\bibitem [{\citenamefont {Lau}\ and\ \citenamefont
  {Yagi}(2021)}]{PhysRevD.103.063015}%
  \BibitemOpen
  \bibfield  {author} {\bibinfo {author} {\bibfnamefont {S.~Y.}\ \bibnamefont
  {Lau}}\ and\ \bibinfo {author} {\bibfnamefont {K.}~\bibnamefont {Yagi}},\
  }\href {\doibase 10.1103/PhysRevD.103.063015} {\bibfield  {journal} {\bibinfo
   {journal} {Phys. Rev. D}\ }\textbf {\bibinfo {volume} {103}},\ \bibinfo
  {pages} {063015} (\bibinfo {year} {2021})}\BibitemShut {NoStop}%
\bibitem [{\citenamefont {Bauswein}\ \emph {et~al.}(2014)\citenamefont
  {Bauswein}, \citenamefont {Stergioulas},\ and\ \citenamefont
  {Janka}}]{PhysRevD.90.023002}%
  \BibitemOpen
  \bibfield  {author} {\bibinfo {author} {\bibfnamefont {A.}~\bibnamefont
  {Bauswein}}, \bibinfo {author} {\bibfnamefont {N.}~\bibnamefont
  {Stergioulas}}, \ and\ \bibinfo {author} {\bibfnamefont {H.-T.}\ \bibnamefont
  {Janka}},\ }\href {\doibase 10.1103/PhysRevD.90.023002} {\bibfield  {journal}
  {\bibinfo  {journal} {Phys. Rev. D}\ }\textbf {\bibinfo {volume} {90}},\
  \bibinfo {pages} {023002} (\bibinfo {year} {2014})}\BibitemShut {NoStop}%
\bibitem [{\citenamefont {Takami}\ \emph {et~al.}(2014)\citenamefont {Takami},
  \citenamefont {Rezzolla},\ and\ \citenamefont
  {Baiotti}}]{PhysRevLett.113.091104}%
  \BibitemOpen
  \bibfield  {author} {\bibinfo {author} {\bibfnamefont {K.}~\bibnamefont
  {Takami}}, \bibinfo {author} {\bibfnamefont {L.}~\bibnamefont {Rezzolla}}, \
  and\ \bibinfo {author} {\bibfnamefont {L.}~\bibnamefont {Baiotti}},\ }\href
  {\doibase 10.1103/PhysRevLett.113.091104} {\bibfield  {journal} {\bibinfo
  {journal} {Phys. Rev. Lett.}\ }\textbf {\bibinfo {volume} {113}},\ \bibinfo
  {pages} {091104} (\bibinfo {year} {2014})}\BibitemShut {NoStop}%
\bibitem [{\citenamefont {Sotani}\ \emph
  {et~al.}(2012{\natexlab{a}})\citenamefont {Sotani}, \citenamefont {Nakazato},
  \citenamefont {Iida},\ and\ \citenamefont
  {Oyamatsu}}]{PhysRevLett.108.201101}%
  \BibitemOpen
  \bibfield  {author} {\bibinfo {author} {\bibfnamefont {H.}~\bibnamefont
  {Sotani}}, \bibinfo {author} {\bibfnamefont {K.}~\bibnamefont {Nakazato}},
  \bibinfo {author} {\bibfnamefont {K.}~\bibnamefont {Iida}}, \ and\ \bibinfo
  {author} {\bibfnamefont {K.}~\bibnamefont {Oyamatsu}},\ }\href {\doibase
  10.1103/PhysRevLett.108.201101} {\bibfield  {journal} {\bibinfo  {journal}
  {Phys. Rev. Lett.}\ }\textbf {\bibinfo {volume} {108}},\ \bibinfo {pages}
  {201101} (\bibinfo {year} {2012}{\natexlab{a}})}\BibitemShut {NoStop}%
\bibitem [{\citenamefont {Sotani}\ \emph
  {et~al.}(2012{\natexlab{b}})\citenamefont {Sotani}, \citenamefont {Nakazato},
  \citenamefont {Iida},\ and\ \citenamefont
  {Oyamatsu}}]{10.1093/mnrasl/sls006}%
  \BibitemOpen
  \bibfield  {author} {\bibinfo {author} {\bibfnamefont {H.}~\bibnamefont
  {Sotani}}, \bibinfo {author} {\bibfnamefont {K.}~\bibnamefont {Nakazato}},
  \bibinfo {author} {\bibfnamefont {K.}~\bibnamefont {Iida}}, \ and\ \bibinfo
  {author} {\bibfnamefont {K.}~\bibnamefont {Oyamatsu}},\ }\href {\doibase
  10.1093/mnrasl/sls006} {\bibfield  {journal} {\bibinfo  {journal} {Monthly
  Notices of the Royal Astronomical Society: Letters}\ }\textbf {\bibinfo
  {volume} {428}},\ \bibinfo {pages} {L21} (\bibinfo {year}
  {2012}{\natexlab{b}})}\BibitemShut {NoStop}%
\bibitem [{\citenamefont {Sotani}\ \emph {et~al.}(2018)\citenamefont {Sotani},
  \citenamefont {Iida},\ and\ \citenamefont
  {Oyamatsu}}]{10.1093/mnras/sty1755}%
  \BibitemOpen
  \bibfield  {author} {\bibinfo {author} {\bibfnamefont {H.}~\bibnamefont
  {Sotani}}, \bibinfo {author} {\bibfnamefont {K.}~\bibnamefont {Iida}}, \ and\
  \bibinfo {author} {\bibfnamefont {K.}~\bibnamefont {Oyamatsu}},\ }\href
  {\doibase 10.1093/mnras/sty1755} {\bibfield  {journal} {\bibinfo  {journal}
  {Monthly Notices of the Royal Astronomical Society}\ }\textbf {\bibinfo
  {volume} {479}},\ \bibinfo {pages} {4735} (\bibinfo {year}
  {2018})}\BibitemShut {NoStop}%
\bibitem [{\citenamefont {Sotani}\ \emph {et~al.}(2019)\citenamefont {Sotani},
  \citenamefont {Iida},\ and\ \citenamefont
  {Oyamatsu}}]{10.1093/mnras/stz2385}%
  \BibitemOpen
  \bibfield  {author} {\bibinfo {author} {\bibfnamefont {H.}~\bibnamefont
  {Sotani}}, \bibinfo {author} {\bibfnamefont {K.}~\bibnamefont {Iida}}, \ and\
  \bibinfo {author} {\bibfnamefont {K.}~\bibnamefont {Oyamatsu}},\ }\href
  {\doibase 10.1093/mnras/stz2385} {\bibfield  {journal} {\bibinfo  {journal}
  {Monthly Notices of the Royal Astronomical Society}\ }\textbf {\bibinfo
  {volume} {489}},\ \bibinfo {pages} {3022} (\bibinfo {year}
  {2019})}\BibitemShut {NoStop}%
\bibitem [{\citenamefont {Perkins}\ \emph {et~al.}(2021)\citenamefont
  {Perkins}, \citenamefont {Yunes},\ and\ \citenamefont
  {Berti}}]{PhysRevD.103.044024}%
  \BibitemOpen
  \bibfield  {author} {\bibinfo {author} {\bibfnamefont {S.~E.}\ \bibnamefont
  {Perkins}}, \bibinfo {author} {\bibfnamefont {N.}~\bibnamefont {Yunes}}, \
  and\ \bibinfo {author} {\bibfnamefont {E.}~\bibnamefont {Berti}},\ }\href
  {\doibase 10.1103/PhysRevD.103.044024} {\bibfield  {journal} {\bibinfo
  {journal} {Phys. Rev. D}\ }\textbf {\bibinfo {volume} {103}},\ \bibinfo
  {pages} {044024} (\bibinfo {year} {2021})}\BibitemShut {NoStop}%
\bibitem [{\citenamefont {Punturo}\ \emph {et~al.}(2010)\citenamefont
  {Punturo}, \citenamefont {Abernathy}, \citenamefont {Acernese},\ and\
  \citenamefont {et~al.}}]{Punturo_2010}%
  \BibitemOpen
  \bibfield  {author} {\bibinfo {author} {\bibfnamefont {M.}~\bibnamefont
  {Punturo}}, \bibinfo {author} {\bibfnamefont {M.}~\bibnamefont {Abernathy}},
  \bibinfo {author} {\bibfnamefont {F.}~\bibnamefont {Acernese}}, \ and\
  \bibinfo {author} {\bibnamefont {et~al.}},\ }\href {\doibase
  10.1088/0264-9381/27/8/084007} {\bibfield  {journal} {\bibinfo  {journal}
  {Classical and Quantum Gravity}\ }\textbf {\bibinfo {volume} {27}},\ \bibinfo
  {pages} {084007} (\bibinfo {year} {2010})}\BibitemShut {NoStop}%
\bibitem [{\citenamefont {Dwyer}\ \emph {et~al.}(2015)\citenamefont {Dwyer},
  \citenamefont {Sigg}, \citenamefont {Ballmer}, \citenamefont {Barsotti},
  \citenamefont {Mavalvala},\ and\ \citenamefont {Evans}}]{PhysRevD.91.082001}%
  \BibitemOpen
  \bibfield  {author} {\bibinfo {author} {\bibfnamefont {S.}~\bibnamefont
  {Dwyer}}, \bibinfo {author} {\bibfnamefont {D.}~\bibnamefont {Sigg}},
  \bibinfo {author} {\bibfnamefont {S.~W.}\ \bibnamefont {Ballmer}}, \bibinfo
  {author} {\bibfnamefont {L.}~\bibnamefont {Barsotti}}, \bibinfo {author}
  {\bibfnamefont {N.}~\bibnamefont {Mavalvala}}, \ and\ \bibinfo {author}
  {\bibfnamefont {M.}~\bibnamefont {Evans}},\ }\href {\doibase
  10.1103/PhysRevD.91.082001} {\bibfield  {journal} {\bibinfo  {journal} {Phys.
  Rev. D}\ }\textbf {\bibinfo {volume} {91}},\ \bibinfo {pages} {082001}
  (\bibinfo {year} {2015})}\BibitemShut {NoStop}%
\bibitem [{\citenamefont {Panotopoulos}\ and\ \citenamefont
  {Lopes}(2017{\natexlab{a}})}]{PhysRevD.96.083004}%
  \BibitemOpen
  \bibfield  {author} {\bibinfo {author} {\bibfnamefont {G.}~\bibnamefont
  {Panotopoulos}}\ and\ \bibinfo {author} {\bibfnamefont {I.}~\bibnamefont
  {Lopes}},\ }\href {\doibase 10.1103/PhysRevD.96.083004} {\bibfield  {journal}
  {\bibinfo  {journal} {Phys. Rev. D}\ }\textbf {\bibinfo {volume} {96}},\
  \bibinfo {pages} {083004} (\bibinfo {year} {2017}{\natexlab{a}})}\BibitemShut
  {NoStop}%
\bibitem [{\citenamefont {Murakami}\ and\ \citenamefont
  {Wells}(2001)}]{PhysRevD.64.015001}%
  \BibitemOpen
  \bibfield  {author} {\bibinfo {author} {\bibfnamefont {B.}~\bibnamefont
  {Murakami}}\ and\ \bibinfo {author} {\bibfnamefont {J.~D.}\ \bibnamefont
  {Wells}},\ }\href {\doibase 10.1103/PhysRevD.64.015001} {\bibfield  {journal}
  {\bibinfo  {journal} {Phys. Rev. D}\ }\textbf {\bibinfo {volume} {64}},\
  \bibinfo {pages} {015001} (\bibinfo {year} {2001})}\BibitemShut {NoStop}%
\bibitem [{\citenamefont {Cline}\ \emph {et~al.}(2013)\citenamefont {Cline},
  \citenamefont {Scott}, \citenamefont {Kainulainen},\ and\ \citenamefont
  {Weniger}}]{PhysRevD.88.055025}%
  \BibitemOpen
  \bibfield  {author} {\bibinfo {author} {\bibfnamefont {J.~M.}\ \bibnamefont
  {Cline}}, \bibinfo {author} {\bibfnamefont {P.}~\bibnamefont {Scott}},
  \bibinfo {author} {\bibfnamefont {K.}~\bibnamefont {Kainulainen}}, \ and\
  \bibinfo {author} {\bibfnamefont {C.}~\bibnamefont {Weniger}},\ }\href
  {\doibase 10.1103/PhysRevD.88.055025} {\bibfield  {journal} {\bibinfo
  {journal} {Phys. Rev. D}\ }\textbf {\bibinfo {volume} {88}},\ \bibinfo
  {pages} {055025} (\bibinfo {year} {2013})}\BibitemShut {NoStop}%
\bibitem [{\citenamefont {Tan}\ \emph {et~al.}(2016)\citenamefont {Tan} \emph
  {et~al.}}]{PhysRevLett.117.121303}%
  \BibitemOpen
  \bibfield  {author} {\bibinfo {author} {\bibfnamefont {A.}~\bibnamefont
  {Tan}} \emph {et~al.} (\bibinfo {collaboration} {PandaX-II Collaboration}),\
  }\href {\doibase 10.1103/PhysRevLett.117.121303} {\bibfield  {journal}
  {\bibinfo  {journal} {Phys. Rev. Lett.}\ }\textbf {\bibinfo {volume} {117}},\
  \bibinfo {pages} {121303} (\bibinfo {year} {2016})}\BibitemShut {NoStop}%
\bibitem [{\citenamefont {Meng}\ \emph {et~al.}(2021)\citenamefont {Meng} \emph
  {et~al.}}]{PhysRevLett.127.261802}%
  \BibitemOpen
  \bibfield  {author} {\bibinfo {author} {\bibfnamefont {Y.}~\bibnamefont
  {Meng}} \emph {et~al.} (\bibinfo {collaboration} {PandaX-4T Collaboration}),\
  }\href {\doibase 10.1103/PhysRevLett.127.261802} {\bibfield  {journal}
  {\bibinfo  {journal} {Phys. Rev. Lett.}\ }\textbf {\bibinfo {volume} {127}},\
  \bibinfo {pages} {261802} (\bibinfo {year} {2021})}\BibitemShut {NoStop}%
\bibitem [{\citenamefont {Dickhoff}\ and\ \citenamefont
  {Van~Neck}(2008)}]{doi:10.1142/6821}%
  \BibitemOpen
  \bibfield  {author} {\bibinfo {author} {\bibfnamefont {W.~H.}\ \bibnamefont
  {Dickhoff}}\ and\ \bibinfo {author} {\bibfnamefont {D.}~\bibnamefont
  {Van~Neck}},\ }\href {\doibase 10.1142/6821} {\emph {\bibinfo {title}
  {Many-Body Theory Exposed!}}},\ \bibinfo {edition} {2nd}\ ed.\ (\bibinfo
  {publisher} {WORLD SCIENTIFIC},\ \bibinfo {year} {2008})\BibitemShut
  {NoStop}%
\bibitem [{\citenamefont {Rios}\ \emph {et~al.}(2014)\citenamefont {Rios},
  \citenamefont {Polls},\ and\ \citenamefont {Dickhoff}}]{PhysRevC.89.044303}%
  \BibitemOpen
  \bibfield  {author} {\bibinfo {author} {\bibfnamefont {A.}~\bibnamefont
  {Rios}}, \bibinfo {author} {\bibfnamefont {A.}~\bibnamefont {Polls}}, \ and\
  \bibinfo {author} {\bibfnamefont {W.~H.}\ \bibnamefont {Dickhoff}},\ }\href
  {\doibase 10.1103/PhysRevC.89.044303} {\bibfield  {journal} {\bibinfo
  {journal} {Phys. Rev. C}\ }\textbf {\bibinfo {volume} {89}},\ \bibinfo
  {pages} {044303} (\bibinfo {year} {2014})}\BibitemShut {NoStop}%
\bibitem [{\citenamefont {Rios}(2020)}]{Rios_2020}%
  \BibitemOpen
  \bibfield  {author} {\bibinfo {author} {\bibfnamefont {A.}~\bibnamefont
  {Rios}},\ }\href {\doibase 10.1088/1742-6596/1643/1/012164} {\bibfield
  {journal} {\bibinfo  {journal} {Journal of Physics: Conference Series}\
  }\textbf {\bibinfo {volume} {1643}},\ \bibinfo {pages} {012164} (\bibinfo
  {year} {2020})}\BibitemShut {NoStop}%
\bibitem [{\citenamefont {Guo}\ \emph {et~al.}(2021)\citenamefont {Guo},
  \citenamefont {Li},\ and\ \citenamefont {Yong}}]{PhysRevC.104.034603}%
  \BibitemOpen
  \bibfield  {author} {\bibinfo {author} {\bibfnamefont {W.-M.}\ \bibnamefont
  {Guo}}, \bibinfo {author} {\bibfnamefont {B.-A.}\ \bibnamefont {Li}}, \ and\
  \bibinfo {author} {\bibfnamefont {G.-C.}\ \bibnamefont {Yong}},\ }\href
  {\doibase 10.1103/PhysRevC.104.034603} {\bibfield  {journal} {\bibinfo
  {journal} {Phys. Rev. C}\ }\textbf {\bibinfo {volume} {104}},\ \bibinfo
  {pages} {034603} (\bibinfo {year} {2021})}\BibitemShut {NoStop}%
\bibitem [{\citenamefont {Subedi}\ \emph {et~al.}(2008)\citenamefont {Subedi},
  \citenamefont {Shneor}, \citenamefont {Monaghan}, \citenamefont {Anderson},
  \citenamefont {Aniol}, \citenamefont {Annand}, \citenamefont {Arrington},
  \citenamefont {Benaoum}, \citenamefont {Benmokhtar}, \citenamefont
  {Boeglin},\ and\ \citenamefont {Chen}}]{R.SUBEDIR}%
  \BibitemOpen
  \bibfield  {author} {\bibinfo {author} {\bibfnamefont {R.}~\bibnamefont
  {Subedi}}, \bibinfo {author} {\bibfnamefont {R.}~\bibnamefont {Shneor}},
  \bibinfo {author} {\bibfnamefont {P.}~\bibnamefont {Monaghan}}, \bibinfo
  {author} {\bibfnamefont {B.~D.}\ \bibnamefont {Anderson}}, \bibinfo {author}
  {\bibfnamefont {K.}~\bibnamefont {Aniol}}, \bibinfo {author} {\bibfnamefont
  {J.}~\bibnamefont {Annand}}, \bibinfo {author} {\bibfnamefont
  {J.}~\bibnamefont {Arrington}}, \bibinfo {author} {\bibfnamefont
  {H.}~\bibnamefont {Benaoum}}, \bibinfo {author} {\bibfnamefont
  {F.}~\bibnamefont {Benmokhtar}}, \bibinfo {author} {\bibfnamefont
  {W.}~\bibnamefont {Boeglin}}, \ and\ \bibinfo {author} {\bibfnamefont
  {J.-P.}\ \bibnamefont {Chen}},\ }\href {\doibase 10.1126/science.1156675}
  {\bibfield  {journal} {\bibinfo  {journal} {Science}\ }\textbf {\bibinfo
  {volume} {320}},\ \bibinfo {pages} {1476} (\bibinfo {year}
  {2008})}\BibitemShut {NoStop}%
\bibitem [{\citenamefont {Hen}\ \emph {et~al.}(2015{\natexlab{b}})\citenamefont
  {Hen}, \citenamefont {Weinstein}, \citenamefont {Piasetzky}, \citenamefont
  {Miller}, \citenamefont {Sargsian},\ and\ \citenamefont
  {Sagi}}]{PhysRevC.92.045205}%
  \BibitemOpen
  \bibfield  {author} {\bibinfo {author} {\bibfnamefont {O.}~\bibnamefont
  {Hen}}, \bibinfo {author} {\bibfnamefont {L.~B.}\ \bibnamefont {Weinstein}},
  \bibinfo {author} {\bibfnamefont {E.}~\bibnamefont {Piasetzky}}, \bibinfo
  {author} {\bibfnamefont {G.~A.}\ \bibnamefont {Miller}}, \bibinfo {author}
  {\bibfnamefont {M.~M.}\ \bibnamefont {Sargsian}}, \ and\ \bibinfo {author}
  {\bibfnamefont {Y.}~\bibnamefont {Sagi}},\ }\href {\doibase
  10.1103/PhysRevC.92.045205} {\bibfield  {journal} {\bibinfo  {journal} {Phys.
  Rev. C}\ }\textbf {\bibinfo {volume} {92}},\ \bibinfo {pages} {045205}
  (\bibinfo {year} {2015}{\natexlab{b}})}\BibitemShut {NoStop}%
\bibitem [{\citenamefont {{Baym}}\ \emph {et~al.}(1971)\citenamefont {{Baym}},
  \citenamefont {{Pethick}},\ and\ \citenamefont
  {{Sutherland}}}]{1971ApJ...170..299B}%
  \BibitemOpen
  \bibfield  {author} {\bibinfo {author} {\bibfnamefont {G.}~\bibnamefont
  {{Baym}}}, \bibinfo {author} {\bibfnamefont {C.}~\bibnamefont {{Pethick}}}, \
  and\ \bibinfo {author} {\bibfnamefont {P.}~\bibnamefont {{Sutherland}}},\
  }\href {\doibase 10.1086/151216} {\bibfield  {journal} {\bibinfo  {journal}
  {\apj}\ }\textbf {\bibinfo {volume} {170}},\ \bibinfo {pages} {299} (\bibinfo
  {year} {1971})}\BibitemShut {NoStop}%
\bibitem [{\citenamefont {Cai}\ and\ \citenamefont
  {Li}(2016)}]{PhysRevC.93.014619}%
  \BibitemOpen
  \bibfield  {author} {\bibinfo {author} {\bibfnamefont {B.-J.}\ \bibnamefont
  {Cai}}\ and\ \bibinfo {author} {\bibfnamefont {B.-A.}\ \bibnamefont {Li}},\
  }\href {\doibase 10.1103/PhysRevC.93.014619} {\bibfield  {journal} {\bibinfo
  {journal} {Phys. Rev. C}\ }\textbf {\bibinfo {volume} {93}},\ \bibinfo
  {pages} {014619} (\bibinfo {year} {2016})}\BibitemShut {NoStop}%
\bibitem [{\citenamefont {Xu}\ \emph {et~al.}(2009)\citenamefont {Xu},
  \citenamefont {Chen}, \citenamefont {Li},\ and\ \citenamefont
  {Ma}}]{PhysRevC.79.035802}%
  \BibitemOpen
  \bibfield  {author} {\bibinfo {author} {\bibfnamefont {J.}~\bibnamefont
  {Xu}}, \bibinfo {author} {\bibfnamefont {L.-W.}\ \bibnamefont {Chen}},
  \bibinfo {author} {\bibfnamefont {B.-A.}\ \bibnamefont {Li}}, \ and\ \bibinfo
  {author} {\bibfnamefont {H.-R.}\ \bibnamefont {Ma}},\ }\href {\doibase
  10.1103/PhysRevC.79.035802} {\bibfield  {journal} {\bibinfo  {journal} {Phys.
  Rev. C}\ }\textbf {\bibinfo {volume} {79}},\ \bibinfo {pages} {035802}
  (\bibinfo {year} {2009})}\BibitemShut {NoStop}%
\bibitem [{\citenamefont {Carriere}\ \emph {et~al.}(2003)\citenamefont
  {Carriere}, \citenamefont {Horowitz},\ and\ \citenamefont
  {Piekarewicz}}]{Carriere_2003}%
  \BibitemOpen
  \bibfield  {author} {\bibinfo {author} {\bibfnamefont {J.}~\bibnamefont
  {Carriere}}, \bibinfo {author} {\bibfnamefont {C.~J.}\ \bibnamefont
  {Horowitz}}, \ and\ \bibinfo {author} {\bibfnamefont {J.}~\bibnamefont
  {Piekarewicz}},\ }\href {\doibase 10.1086/376515} {\bibfield  {journal}
  {\bibinfo  {journal} {The Astrophysical Journal}\ }\textbf {\bibinfo {volume}
  {593}},\ \bibinfo {pages} {463} (\bibinfo {year} {2003})}\BibitemShut
  {NoStop}%
\bibitem [{\citenamefont {Oppenheimer}\ and\ \citenamefont
  {Volkoff}(1939)}]{Oppenheimer1939374}%
  \BibitemOpen
  \bibfield  {author} {\bibinfo {author} {\bibfnamefont {J.}~\bibnamefont
  {Oppenheimer}}\ and\ \bibinfo {author} {\bibfnamefont {G.}~\bibnamefont
  {Volkoff}},\ }\href {\doibase 10.1103/PhysRev.55.374} {\bibfield  {journal}
  {\bibinfo  {journal} {Physical Review}\ }\textbf {\bibinfo {volume} {55}},\
  \bibinfo {pages} {374} (\bibinfo {year} {1939})},\ \bibinfo {note} {cited By
  1756}\BibitemShut {NoStop}%
\bibitem [{\citenamefont {Lalazissis}\ \emph {et~al.}(1997)\citenamefont
  {Lalazissis}, \citenamefont {K\"onig},\ and\ \citenamefont
  {Ring}}]{PhysRevC.55.540}%
  \BibitemOpen
  \bibfield  {author} {\bibinfo {author} {\bibfnamefont {G.~A.}\ \bibnamefont
  {Lalazissis}}, \bibinfo {author} {\bibfnamefont {J.}~\bibnamefont {K\"onig}},
  \ and\ \bibinfo {author} {\bibfnamefont {P.}~\bibnamefont {Ring}},\ }\href
  {\doibase 10.1103/PhysRevC.55.540} {\bibfield  {journal} {\bibinfo  {journal}
  {Phys. Rev. C}\ }\textbf {\bibinfo {volume} {55}},\ \bibinfo {pages} {540}
  (\bibinfo {year} {1997})}\BibitemShut {NoStop}%
\bibitem [{\citenamefont {Todd-Rutel}\ and\ \citenamefont
  {Piekarewicz}(2005)}]{PhysRevLett.95.122501}%
  \BibitemOpen
  \bibfield  {author} {\bibinfo {author} {\bibfnamefont {B.~G.}\ \bibnamefont
  {Todd-Rutel}}\ and\ \bibinfo {author} {\bibfnamefont {J.}~\bibnamefont
  {Piekarewicz}},\ }\href {\doibase 10.1103/PhysRevLett.95.122501} {\bibfield
  {journal} {\bibinfo  {journal} {Phys. Rev. Lett.}\ }\textbf {\bibinfo
  {volume} {95}},\ \bibinfo {pages} {122501} (\bibinfo {year}
  {2005})}\BibitemShut {NoStop}%
\bibitem [{\citenamefont {Glendenning}\ and\ \citenamefont
  {Moszkowski}(1991)}]{PhysRevLett.67.2414}%
  \BibitemOpen
  \bibfield  {author} {\bibinfo {author} {\bibfnamefont {N.~K.}\ \bibnamefont
  {Glendenning}}\ and\ \bibinfo {author} {\bibfnamefont {S.~A.}\ \bibnamefont
  {Moszkowski}},\ }\href {\doibase 10.1103/PhysRevLett.67.2414} {\bibfield
  {journal} {\bibinfo  {journal} {Phys. Rev. Lett.}\ }\textbf {\bibinfo
  {volume} {67}},\ \bibinfo {pages} {2414} (\bibinfo {year}
  {1991})}\BibitemShut {NoStop}%
\bibitem [{\citenamefont {Liu}\ \emph {et~al.}(2002)\citenamefont {Liu},
  \citenamefont {Greco}, \citenamefont {Baran}, \citenamefont {Colonna},\ and\
  \citenamefont {Di~Toro}}]{PhysRevC.65.045201}%
  \BibitemOpen
  \bibfield  {author} {\bibinfo {author} {\bibfnamefont {B.}~\bibnamefont
  {Liu}}, \bibinfo {author} {\bibfnamefont {V.}~\bibnamefont {Greco}}, \bibinfo
  {author} {\bibfnamefont {V.}~\bibnamefont {Baran}}, \bibinfo {author}
  {\bibfnamefont {M.}~\bibnamefont {Colonna}}, \ and\ \bibinfo {author}
  {\bibfnamefont {M.}~\bibnamefont {Di~Toro}},\ }\href {\doibase
  10.1103/PhysRevC.65.045201} {\bibfield  {journal} {\bibinfo  {journal} {Phys.
  Rev. C}\ }\textbf {\bibinfo {volume} {65}},\ \bibinfo {pages} {045201}
  (\bibinfo {year} {2002})}\BibitemShut {NoStop}%
\bibitem [{\citenamefont {Abbott}\ \emph {et~al.}(2017)\citenamefont {Abbott},
  \citenamefont {Abbott}, \citenamefont {Abbott} \emph
  {et~al.}}]{PhysRevLett.119.161101}%
  \BibitemOpen
  \bibfield  {author} {\bibinfo {author} {\bibfnamefont {B.~P.}\ \bibnamefont
  {Abbott}}, \bibinfo {author} {\bibfnamefont {R.}~\bibnamefont {Abbott}},
  \bibinfo {author} {\bibfnamefont {T.~D.}\ \bibnamefont {Abbott}},  \emph
  {et~al.} (\bibinfo {collaboration} {LIGO Scientific Collaboration and Virgo
  Collaboration}),\ }\href {\doibase 10.1103/PhysRevLett.119.161101} {\bibfield
   {journal} {\bibinfo  {journal} {Phys. Rev. Lett.}\ }\textbf {\bibinfo
  {volume} {119}},\ \bibinfo {pages} {161101} (\bibinfo {year}
  {2017})}\BibitemShut {NoStop}%
\bibitem [{\citenamefont {Abbott}\ \emph {et~al.}(2018)\citenamefont {Abbott},
  \citenamefont {Abbott}, \citenamefont {Abbott} \emph
  {et~al.}}]{PhysRevLett.121.161101}%
  \BibitemOpen
  \bibfield  {author} {\bibinfo {author} {\bibfnamefont {B.~P.}\ \bibnamefont
  {Abbott}}, \bibinfo {author} {\bibfnamefont {R.}~\bibnamefont {Abbott}},
  \bibinfo {author} {\bibfnamefont {T.~D.}\ \bibnamefont {Abbott}},  \emph
  {et~al.} (\bibinfo {collaboration} {The LIGO Scientific Collaboration and the
  Virgo Collaboration}),\ }\href {\doibase 10.1103/PhysRevLett.121.161101}
  {\bibfield  {journal} {\bibinfo  {journal} {Phys. Rev. Lett.}\ }\textbf
  {\bibinfo {volume} {121}},\ \bibinfo {pages} {161101} (\bibinfo {year}
  {2018})}\BibitemShut {NoStop}%
\bibitem [{\citenamefont {Annala}\ \emph {et~al.}(2018)\citenamefont {Annala},
  \citenamefont {Gorda}, \citenamefont {Kurkela},\ and\ \citenamefont
  {Vuorinen}}]{PhysRevLett.120.172703}%
  \BibitemOpen
  \bibfield  {author} {\bibinfo {author} {\bibfnamefont {E.}~\bibnamefont
  {Annala}}, \bibinfo {author} {\bibfnamefont {T.}~\bibnamefont {Gorda}},
  \bibinfo {author} {\bibfnamefont {A.}~\bibnamefont {Kurkela}}, \ and\
  \bibinfo {author} {\bibfnamefont {A.}~\bibnamefont {Vuorinen}},\ }\href
  {\doibase 10.1103/PhysRevLett.120.172703} {\bibfield  {journal} {\bibinfo
  {journal} {Phys. Rev. Lett.}\ }\textbf {\bibinfo {volume} {120}},\ \bibinfo
  {pages} {172703} (\bibinfo {year} {2018})}\BibitemShut {NoStop}%
\bibitem [{\citenamefont {Most}\ \emph {et~al.}(2018)\citenamefont {Most},
  \citenamefont {Weih}, \citenamefont {Rezzolla},\ and\ \citenamefont
  {Schaffner-Bielich}}]{PhysRevLett.120.261103}%
  \BibitemOpen
  \bibfield  {author} {\bibinfo {author} {\bibfnamefont {E.~R.}\ \bibnamefont
  {Most}}, \bibinfo {author} {\bibfnamefont {L.~R.}\ \bibnamefont {Weih}},
  \bibinfo {author} {\bibfnamefont {L.}~\bibnamefont {Rezzolla}}, \ and\
  \bibinfo {author} {\bibfnamefont {J.}~\bibnamefont {Schaffner-Bielich}},\
  }\href {\doibase 10.1103/PhysRevLett.120.261103} {\bibfield  {journal}
  {\bibinfo  {journal} {Phys. Rev. Lett.}\ }\textbf {\bibinfo {volume} {120}},\
  \bibinfo {pages} {261103} (\bibinfo {year} {2018})}\BibitemShut {NoStop}%
\bibitem [{\citenamefont {De}\ \emph {et~al.}(2018)\citenamefont {De},
  \citenamefont {Finstad}, \citenamefont {Lattimer}, \citenamefont {Brown},
  \citenamefont {Berger},\ and\ \citenamefont
  {Biwer}}]{PhysRevLett.121.091102}%
  \BibitemOpen
  \bibfield  {author} {\bibinfo {author} {\bibfnamefont {S.}~\bibnamefont
  {De}}, \bibinfo {author} {\bibfnamefont {D.}~\bibnamefont {Finstad}},
  \bibinfo {author} {\bibfnamefont {J.~M.}\ \bibnamefont {Lattimer}}, \bibinfo
  {author} {\bibfnamefont {D.~A.}\ \bibnamefont {Brown}}, \bibinfo {author}
  {\bibfnamefont {E.}~\bibnamefont {Berger}}, \ and\ \bibinfo {author}
  {\bibfnamefont {C.~M.}\ \bibnamefont {Biwer}},\ }\href {\doibase
  10.1103/PhysRevLett.121.091102} {\bibfield  {journal} {\bibinfo  {journal}
  {Phys. Rev. Lett.}\ }\textbf {\bibinfo {volume} {121}},\ \bibinfo {pages}
  {091102} (\bibinfo {year} {2018})}\BibitemShut {NoStop}%
\bibitem [{\citenamefont {Miller}\ \emph {et~al.}(2021)\citenamefont {Miller},
  \citenamefont {Lamb}, \citenamefont {Dittmann} \emph {et~al.}}]{Miller_2021}%
  \BibitemOpen
  \bibfield  {author} {\bibinfo {author} {\bibfnamefont {M.~C.}\ \bibnamefont
  {Miller}}, \bibinfo {author} {\bibfnamefont {F.~K.}\ \bibnamefont {Lamb}},
  \bibinfo {author} {\bibfnamefont {A.~J.}\ \bibnamefont {Dittmann}},  \emph
  {et~al.},\ }\href {\doibase 10.3847/2041-8213/ac089b} {\bibfield  {journal}
  {\bibinfo  {journal} {The Astrophysical Journal Letters}\ }\textbf {\bibinfo
  {volume} {918}},\ \bibinfo {pages} {L28} (\bibinfo {year}
  {2021})}\BibitemShut {NoStop}%
\bibitem [{\citenamefont {Riley}\ \emph {et~al.}(2021)\citenamefont {Riley}
  \emph {et~al.}}]{Riley_2021}%
  \BibitemOpen
  \bibfield  {author} {\bibinfo {author} {\bibfnamefont {T.~E.}\ \bibnamefont
  {Riley}} \emph {et~al.},\ }\href {\doibase 10.3847/2041-8213/ac0a81}
  {\bibfield  {journal} {\bibinfo  {journal} {The Astrophysical Journal
  Letters}\ }\textbf {\bibinfo {volume} {918}},\ \bibinfo {pages} {L27}
  (\bibinfo {year} {2021})}\BibitemShut {NoStop}%
\bibitem [{\citenamefont {Hornick}\ \emph {et~al.}(2018)\citenamefont
  {Hornick}, \citenamefont {Tolos}, \citenamefont {Zacchi}, \citenamefont
  {Christian},\ and\ \citenamefont {Schaffner-Bielich}}]{PhysRevC.98.065804}%
  \BibitemOpen
  \bibfield  {author} {\bibinfo {author} {\bibfnamefont {N.}~\bibnamefont
  {Hornick}}, \bibinfo {author} {\bibfnamefont {L.}~\bibnamefont {Tolos}},
  \bibinfo {author} {\bibfnamefont {A.}~\bibnamefont {Zacchi}}, \bibinfo
  {author} {\bibfnamefont {J.-E.}\ \bibnamefont {Christian}}, \ and\ \bibinfo
  {author} {\bibfnamefont {J.}~\bibnamefont {Schaffner-Bielich}},\ }\href
  {\doibase 10.1103/PhysRevC.98.065804} {\bibfield  {journal} {\bibinfo
  {journal} {Phys. Rev. C}\ }\textbf {\bibinfo {volume} {98}},\ \bibinfo
  {pages} {065804} (\bibinfo {year} {2018})}\BibitemShut {NoStop}%
\bibitem [{\citenamefont {Wu}\ \emph {et~al.}(2021)\citenamefont {Wu},
  \citenamefont {Bao}, \citenamefont {Shen},\ and\ \citenamefont
  {Xu}}]{PhysRevC.104.015802}%
  \BibitemOpen
  \bibfield  {author} {\bibinfo {author} {\bibfnamefont {X.}~\bibnamefont
  {Wu}}, \bibinfo {author} {\bibfnamefont {S.}~\bibnamefont {Bao}}, \bibinfo
  {author} {\bibfnamefont {H.}~\bibnamefont {Shen}}, \ and\ \bibinfo {author}
  {\bibfnamefont {R.}~\bibnamefont {Xu}},\ }\href {\doibase
  10.1103/PhysRevC.104.015802} {\bibfield  {journal} {\bibinfo  {journal}
  {Phys. Rev. C}\ }\textbf {\bibinfo {volume} {104}},\ \bibinfo {pages}
  {015802} (\bibinfo {year} {2021})}\BibitemShut {NoStop}%
\bibitem [{\citenamefont {Hu}\ \emph {et~al.}(2020)\citenamefont {Hu},
  \citenamefont {Bao}, \citenamefont {Zhang}, \citenamefont {Nakazato},
  \citenamefont {Sumiyoshi},\ and\ \citenamefont
  {Shen}}]{10.1093/ptep/ptaa016}%
  \BibitemOpen
  \bibfield  {author} {\bibinfo {author} {\bibfnamefont {J.}~\bibnamefont
  {Hu}}, \bibinfo {author} {\bibfnamefont {S.}~\bibnamefont {Bao}}, \bibinfo
  {author} {\bibfnamefont {Y.}~\bibnamefont {Zhang}}, \bibinfo {author}
  {\bibfnamefont {K.}~\bibnamefont {Nakazato}}, \bibinfo {author}
  {\bibfnamefont {K.}~\bibnamefont {Sumiyoshi}}, \ and\ \bibinfo {author}
  {\bibfnamefont {H.}~\bibnamefont {Shen}},\ }\href
  {https://doi.org/10.1093/ptep/ptaa016} {\bibfield  {journal} {\bibinfo
  {journal} {Progress of Theoretical and Experimental Physics}\ }\textbf
  {\bibinfo {volume} {2020}} (\bibinfo {year} {2020})}\BibitemShut {NoStop}%
\bibitem [{\citenamefont {Huang}\ \emph {et~al.}(2020)\citenamefont {Huang},
  \citenamefont {Hu}, \citenamefont {Zhang},\ and\ \citenamefont
  {Shen}}]{Huang_2020}%
  \BibitemOpen
  \bibfield  {author} {\bibinfo {author} {\bibfnamefont {K.}~\bibnamefont
  {Huang}}, \bibinfo {author} {\bibfnamefont {J.}~\bibnamefont {Hu}}, \bibinfo
  {author} {\bibfnamefont {Y.}~\bibnamefont {Zhang}}, \ and\ \bibinfo {author}
  {\bibfnamefont {H.}~\bibnamefont {Shen}},\ }\href {\doibase
  10.3847/1538-4357/abbb37} {\bibfield  {journal} {\bibinfo  {journal} {The
  Astrophysical Journal}\ }\textbf {\bibinfo {volume} {904}},\ \bibinfo {pages}
  {39} (\bibinfo {year} {2020})}\BibitemShut {NoStop}%
\bibitem [{\citenamefont {Cavagnoli}\ \emph {et~al.}(2011)\citenamefont
  {Cavagnoli}, \citenamefont {Menezes},\ and\ \citenamefont
  {Provid\^encia}}]{PhysRevC.84.065810}%
  \BibitemOpen
  \bibfield  {author} {\bibinfo {author} {\bibfnamefont {R.}~\bibnamefont
  {Cavagnoli}}, \bibinfo {author} {\bibfnamefont {D.~P.}\ \bibnamefont
  {Menezes}}, \ and\ \bibinfo {author} {\bibfnamefont {C.~m.~c.}\ \bibnamefont
  {Provid\^encia}},\ }\href {\doibase 10.1103/PhysRevC.84.065810} {\bibfield
  {journal} {\bibinfo  {journal} {Phys. Rev. C}\ }\textbf {\bibinfo {volume}
  {84}},\ \bibinfo {pages} {065810} (\bibinfo {year} {2011})}\BibitemShut
  {NoStop}%
\bibitem [{\citenamefont {Danielewicz}\ \emph {et~al.}(2002)\citenamefont
  {Danielewicz}, \citenamefont {Lacey},\ and\ \citenamefont
  {Lynch}}]{doi:10.1126/science.1078070}%
  \BibitemOpen
  \bibfield  {author} {\bibinfo {author} {\bibfnamefont {P.}~\bibnamefont
  {Danielewicz}}, \bibinfo {author} {\bibfnamefont {R.}~\bibnamefont {Lacey}},
  \ and\ \bibinfo {author} {\bibfnamefont {W.~G.}\ \bibnamefont {Lynch}},\
  }\href {\doibase 10.1126/science.1078070} {\bibfield  {journal} {\bibinfo
  {journal} {Science}\ }\textbf {\bibinfo {volume} {298}},\ \bibinfo {pages}
  {1592} (\bibinfo {year} {2002})}\BibitemShut {NoStop}%
\bibitem [{\citenamefont {{Thorne}}\ and\ \citenamefont
  {{Campolattaro}}(1967)}]{1967ApJ...149..591T}%
  \BibitemOpen
  \bibfield  {author} {\bibinfo {author} {\bibfnamefont {K.~S.}\ \bibnamefont
  {{Thorne}}}\ and\ \bibinfo {author} {\bibfnamefont {A.}~\bibnamefont
  {{Campolattaro}}},\ }\href {\doibase 10.1086/149288} {\bibfield  {journal}
  {\bibinfo  {journal} {The Astrophysical Journal}\ }\textbf {\bibinfo {volume}
  {149}},\ \bibinfo {pages} {591} (\bibinfo {year} {1967})}\BibitemShut
  {NoStop}%
\bibitem [{\citenamefont {Cowling}(1941)}]{10.1093/mnras/101.8.367}%
  \BibitemOpen
  \bibfield  {author} {\bibinfo {author} {\bibfnamefont {T.~G.}\ \bibnamefont
  {Cowling}},\ }\href {\doibase 10.1093/mnras/101.8.367} {\bibfield  {journal}
  {\bibinfo  {journal} {Monthly Notices of the Royal Astronomical Society}\
  }\textbf {\bibinfo {volume} {101}},\ \bibinfo {pages} {367} (\bibinfo {year}
  {1941})}\BibitemShut {NoStop}%
\bibitem [{\citenamefont {{McDermott}}\ \emph {et~al.}(1983)\citenamefont
  {{McDermott}}, \citenamefont {{van Horn}},\ and\ \citenamefont
  {{Scholl}}}]{1983ApJ...268..837M}%
  \BibitemOpen
  \bibfield  {author} {\bibinfo {author} {\bibfnamefont {P.~N.}\ \bibnamefont
  {{McDermott}}}, \bibinfo {author} {\bibfnamefont {H.~M.}\ \bibnamefont {{van
  Horn}}}, \ and\ \bibinfo {author} {\bibfnamefont {J.~F.}\ \bibnamefont
  {{Scholl}}},\ }\href {\doibase 10.1086/161006} {\bibfield  {journal}
  {\bibinfo  {journal} {The Astrophysical Journal}\ }\textbf {\bibinfo {volume}
  {268}},\ \bibinfo {pages} {837} (\bibinfo {year} {1983})}\BibitemShut
  {NoStop}%
\bibitem [{\citenamefont {Torres-Forné}\ \emph {et~al.}(2017)\citenamefont
  {Torres-Forné}, \citenamefont {Cerdá-Durán}, \citenamefont {Passamonti},\
  and\ \citenamefont {Font}}]{10.1093/mnras/stx3067}%
  \BibitemOpen
  \bibfield  {author} {\bibinfo {author} {\bibfnamefont {A.}~\bibnamefont
  {Torres-Forné}}, \bibinfo {author} {\bibfnamefont {P.}~\bibnamefont
  {Cerdá-Durán}}, \bibinfo {author} {\bibfnamefont {A.}~\bibnamefont
  {Passamonti}}, \ and\ \bibinfo {author} {\bibfnamefont {J.~A.}\ \bibnamefont
  {Font}},\ }\href {\doibase 10.1093/mnras/stx3067} {\bibfield  {journal}
  {\bibinfo  {journal} {Monthly Notices of the Royal Astronomical Society}\
  }\textbf {\bibinfo {volume} {474}},\ \bibinfo {pages} {5272} (\bibinfo {year}
  {2017})}\BibitemShut {NoStop}%
\bibitem [{\citenamefont {Yoshida}\ and\ \citenamefont
  {Kojima}(1997)}]{10.1093/mnras/289.1.117}%
  \BibitemOpen
  \bibfield  {author} {\bibinfo {author} {\bibfnamefont {S.}~\bibnamefont
  {Yoshida}}\ and\ \bibinfo {author} {\bibfnamefont {Y.}~\bibnamefont
  {Kojima}},\ }\href {\doibase 10.1093/mnras/289.1.117} {\bibfield  {journal}
  {\bibinfo  {journal} {Monthly Notices of the Royal Astronomical Society}\
  }\textbf {\bibinfo {volume} {289}},\ \bibinfo {pages} {117} (\bibinfo {year}
  {1997})}\BibitemShut {NoStop}%
\bibitem [{\citenamefont {Sotani}\ \emph {et~al.}(2001)\citenamefont {Sotani},
  \citenamefont {Tominaga},\ and\ \citenamefont {Maeda}}]{PhysRevD.65.024010}%
  \BibitemOpen
  \bibfield  {author} {\bibinfo {author} {\bibfnamefont {H.}~\bibnamefont
  {Sotani}}, \bibinfo {author} {\bibfnamefont {K.}~\bibnamefont {Tominaga}}, \
  and\ \bibinfo {author} {\bibfnamefont {K.-i.}\ \bibnamefont {Maeda}},\ }\href
  {\doibase 10.1103/PhysRevD.65.024010} {\bibfield  {journal} {\bibinfo
  {journal} {Phys. Rev. D}\ }\textbf {\bibinfo {volume} {65}},\ \bibinfo
  {pages} {024010} (\bibinfo {year} {2001})}\BibitemShut {NoStop}%
\bibitem [{\citenamefont {Flores}\ and\ \citenamefont
  {Lugones}(2014)}]{V_squez_Flores_2014}%
  \BibitemOpen
  \bibfield  {author} {\bibinfo {author} {\bibfnamefont {C.~V.}\ \bibnamefont
  {Flores}}\ and\ \bibinfo {author} {\bibfnamefont {G.}~\bibnamefont
  {Lugones}},\ }\href {\doibase 10.1088/0264-9381/31/15/155002} {\bibfield
  {journal} {\bibinfo  {journal} {Classical and Quantum Gravity}\ }\textbf
  {\bibinfo {volume} {31}},\ \bibinfo {pages} {155002} (\bibinfo {year}
  {2014})}\BibitemShut {NoStop}%
\bibitem [{\citenamefont {Passamonti}\ \emph {et~al.}(2013)\citenamefont
  {Passamonti}, \citenamefont {Gaertig}, \citenamefont {Kokkotas},\ and\
  \citenamefont {Doneva}}]{PhysRevD.87.084010}%
  \BibitemOpen
  \bibfield  {author} {\bibinfo {author} {\bibfnamefont {A.}~\bibnamefont
  {Passamonti}}, \bibinfo {author} {\bibfnamefont {E.}~\bibnamefont {Gaertig}},
  \bibinfo {author} {\bibfnamefont {K.~D.}\ \bibnamefont {Kokkotas}}, \ and\
  \bibinfo {author} {\bibfnamefont {D.}~\bibnamefont {Doneva}},\ }\href
  {\doibase 10.1103/PhysRevD.87.084010} {\bibfield  {journal} {\bibinfo
  {journal} {Phys. Rev. D}\ }\textbf {\bibinfo {volume} {87}},\ \bibinfo
  {pages} {084010} (\bibinfo {year} {2013})}\BibitemShut {NoStop}%
\bibitem [{\citenamefont {Doneva}\ \emph {et~al.}(2013)\citenamefont {Doneva},
  \citenamefont {Gaertig}, \citenamefont {Kokkotas},\ and\ \citenamefont
  {Kr\"uger}}]{PhysRevD.88.044052}%
  \BibitemOpen
  \bibfield  {author} {\bibinfo {author} {\bibfnamefont {D.~D.}\ \bibnamefont
  {Doneva}}, \bibinfo {author} {\bibfnamefont {E.}~\bibnamefont {Gaertig}},
  \bibinfo {author} {\bibfnamefont {K.~D.}\ \bibnamefont {Kokkotas}}, \ and\
  \bibinfo {author} {\bibfnamefont {C.}~\bibnamefont {Kr\"uger}},\ }\href
  {\doibase 10.1103/PhysRevD.88.044052} {\bibfield  {journal} {\bibinfo
  {journal} {Phys. Rev. D}\ }\textbf {\bibinfo {volume} {88}},\ \bibinfo
  {pages} {044052} (\bibinfo {year} {2013})}\BibitemShut {NoStop}%
\bibitem [{\citenamefont {Benhar}\ \emph {et~al.}(2004)\citenamefont {Benhar},
  \citenamefont {Ferrari},\ and\ \citenamefont
  {Gualtieri}}]{PhysRevD.70.124015}%
  \BibitemOpen
  \bibfield  {author} {\bibinfo {author} {\bibfnamefont {O.}~\bibnamefont
  {Benhar}}, \bibinfo {author} {\bibfnamefont {V.}~\bibnamefont {Ferrari}}, \
  and\ \bibinfo {author} {\bibfnamefont {L.}~\bibnamefont {Gualtieri}},\ }\href
  {\doibase 10.1103/PhysRevD.70.124015} {\bibfield  {journal} {\bibinfo
  {journal} {Phys. Rev. D}\ }\textbf {\bibinfo {volume} {70}},\ \bibinfo
  {pages} {124015} (\bibinfo {year} {2004})}\BibitemShut {NoStop}%
\bibitem [{\citenamefont {Gaertig}\ \emph {et~al.}(2011)\citenamefont
  {Gaertig}, \citenamefont {Glampedakis}, \citenamefont {Kokkotas},\ and\
  \citenamefont {Zink}}]{PhysRevLett.107.101102}%
  \BibitemOpen
  \bibfield  {author} {\bibinfo {author} {\bibfnamefont {E.}~\bibnamefont
  {Gaertig}}, \bibinfo {author} {\bibfnamefont {K.}~\bibnamefont
  {Glampedakis}}, \bibinfo {author} {\bibfnamefont {K.~D.}\ \bibnamefont
  {Kokkotas}}, \ and\ \bibinfo {author} {\bibfnamefont {B.}~\bibnamefont
  {Zink}},\ }\href {\doibase 10.1103/PhysRevLett.107.101102} {\bibfield
  {journal} {\bibinfo  {journal} {Phys. Rev. Lett.}\ }\textbf {\bibinfo
  {volume} {107}},\ \bibinfo {pages} {101102} (\bibinfo {year}
  {2011})}\BibitemShut {NoStop}%
\bibitem [{\citenamefont {Pradhan}\ and\ \citenamefont
  {Chatterjee}(2021)}]{PhysRevC.103.035810}%
  \BibitemOpen
  \bibfield  {author} {\bibinfo {author} {\bibfnamefont {B.~K.}\ \bibnamefont
  {Pradhan}}\ and\ \bibinfo {author} {\bibfnamefont {D.}~\bibnamefont
  {Chatterjee}},\ }\href {\doibase 10.1103/PhysRevC.103.035810} {\bibfield
  {journal} {\bibinfo  {journal} {Phys. Rev. C}\ }\textbf {\bibinfo {volume}
  {103}},\ \bibinfo {pages} {035810} (\bibinfo {year} {2021})}\BibitemShut
  {NoStop}%
\bibitem [{\citenamefont {Wen}\ \emph {et~al.}(2019)\citenamefont {Wen},
  \citenamefont {Li}, \citenamefont {Chen},\ and\ \citenamefont
  {Zhang}}]{PhysRevC.99.045806}%
  \BibitemOpen
  \bibfield  {author} {\bibinfo {author} {\bibfnamefont {D.-H.}\ \bibnamefont
  {Wen}}, \bibinfo {author} {\bibfnamefont {B.-A.}\ \bibnamefont {Li}},
  \bibinfo {author} {\bibfnamefont {H.-Y.}\ \bibnamefont {Chen}}, \ and\
  \bibinfo {author} {\bibfnamefont {N.-B.}\ \bibnamefont {Zhang}},\ }\href
  {\doibase 10.1103/PhysRevC.99.045806} {\bibfield  {journal} {\bibinfo
  {journal} {Phys. Rev. C}\ }\textbf {\bibinfo {volume} {99}},\ \bibinfo
  {pages} {045806} (\bibinfo {year} {2019})}\BibitemShut {NoStop}%
\bibitem [{\citenamefont {Andersson}\ and\ \citenamefont
  {Kokkotas}(1996)}]{PhysRevLett.77.4134}%
  \BibitemOpen
  \bibfield  {author} {\bibinfo {author} {\bibfnamefont {N.}~\bibnamefont
  {Andersson}}\ and\ \bibinfo {author} {\bibfnamefont {K.~D.}\ \bibnamefont
  {Kokkotas}},\ }\href {\doibase 10.1103/PhysRevLett.77.4134} {\bibfield
  {journal} {\bibinfo  {journal} {Phys. Rev. Lett.}\ }\textbf {\bibinfo
  {volume} {77}},\ \bibinfo {pages} {4134} (\bibinfo {year}
  {1996})}\BibitemShut {NoStop}%
\bibitem [{\citenamefont {Andersson}\ and\ \citenamefont
  {Kokkotas}(1998)}]{10.1046/j.1365-8711.1998.01840.x}%
  \BibitemOpen
  \bibfield  {author} {\bibinfo {author} {\bibfnamefont {N.}~\bibnamefont
  {Andersson}}\ and\ \bibinfo {author} {\bibfnamefont {K.~D.}\ \bibnamefont
  {Kokkotas}},\ }\href {\doibase 10.1046/j.1365-8711.1998.01840.x} {\bibfield
  {journal} {\bibinfo  {journal} {Monthly Notices of the Royal Astronomical
  Society}\ }\textbf {\bibinfo {volume} {299}},\ \bibinfo {pages} {1059}
  (\bibinfo {year} {1998})}\BibitemShut {NoStop}%
\bibitem [{\citenamefont {Chandrasekhar}(1964)}]{PhysRevLett.12.114}%
  \BibitemOpen
  \bibfield  {author} {\bibinfo {author} {\bibfnamefont {S.}~\bibnamefont
  {Chandrasekhar}},\ }\href {\doibase 10.1103/PhysRevLett.12.114} {\bibfield
  {journal} {\bibinfo  {journal} {Phys. Rev. Lett.}\ }\textbf {\bibinfo
  {volume} {12}},\ \bibinfo {pages} {114} (\bibinfo {year} {1964})}\BibitemShut
  {NoStop}%
\bibitem [{\citenamefont {{Glass}}\ and\ \citenamefont
  {{Lindblom}}(1983)}]{1983ApJS...53...93G}%
  \BibitemOpen
  \bibfield  {author} {\bibinfo {author} {\bibfnamefont {E.~N.}\ \bibnamefont
  {{Glass}}}\ and\ \bibinfo {author} {\bibfnamefont {L.}~\bibnamefont
  {{Lindblom}}},\ }\href {\doibase 10.1086/190885} {\bibfield  {journal}
  {\bibinfo  {journal} {The Astrophysical Journal Letters}\ }\textbf {\bibinfo
  {volume} {53}},\ \bibinfo {pages} {93} (\bibinfo {year} {1983})}\BibitemShut
  {NoStop}%
\bibitem [{\citenamefont {{Chanmugam}}(1977)}]{1977ApJ...217..799C}%
  \BibitemOpen
  \bibfield  {author} {\bibinfo {author} {\bibfnamefont {G.}~\bibnamefont
  {{Chanmugam}}},\ }\href {\doibase 10.1086/155627} {\bibfield  {journal}
  {\bibinfo  {journal} {The Astrophysical Journal}\ }\textbf {\bibinfo {volume}
  {217}},\ \bibinfo {pages} {799} (\bibinfo {year} {1977})}\BibitemShut
  {NoStop}%
\bibitem [{\citenamefont {Di~Clemente}\ \emph {et~al.}(2020)\citenamefont
  {Di~Clemente}, \citenamefont {Mannarelli},\ and\ \citenamefont
  {Tonelli}}]{PhysRevD.101.103003}%
  \BibitemOpen
  \bibfield  {author} {\bibinfo {author} {\bibfnamefont {F.}~\bibnamefont
  {Di~Clemente}}, \bibinfo {author} {\bibfnamefont {M.}~\bibnamefont
  {Mannarelli}}, \ and\ \bibinfo {author} {\bibfnamefont {F.}~\bibnamefont
  {Tonelli}},\ }\href {\doibase 10.1103/PhysRevD.101.103003} {\bibfield
  {journal} {\bibinfo  {journal} {Phys. Rev. D}\ }\textbf {\bibinfo {volume}
  {101}},\ \bibinfo {pages} {103003} (\bibinfo {year} {2020})}\BibitemShut
  {NoStop}%
\bibitem [{\citenamefont {Sagun}\ \emph {et~al.}(2020)\citenamefont {Sagun},
  \citenamefont {Panotopoulos},\ and\ \citenamefont
  {Lopes}}]{PhysRevD.101.063025}%
  \BibitemOpen
  \bibfield  {author} {\bibinfo {author} {\bibfnamefont {V.}~\bibnamefont
  {Sagun}}, \bibinfo {author} {\bibfnamefont {G.}~\bibnamefont {Panotopoulos}},
  \ and\ \bibinfo {author} {\bibfnamefont {I.}~\bibnamefont {Lopes}},\ }\href
  {\doibase 10.1103/PhysRevD.101.063025} {\bibfield  {journal} {\bibinfo
  {journal} {Phys. Rev. D}\ }\textbf {\bibinfo {volume} {101}},\ \bibinfo
  {pages} {063025} (\bibinfo {year} {2020})}\BibitemShut {NoStop}%
\bibitem [{\citenamefont {{Gondek}}\ \emph {et~al.}(1997)\citenamefont
  {{Gondek}}, \citenamefont {{Haensel}},\ and\ \citenamefont
  {{Zdunik}}}]{1997A&A...325..217G}%
  \BibitemOpen
  \bibfield  {author} {\bibinfo {author} {\bibfnamefont {D.}~\bibnamefont
  {{Gondek}}}, \bibinfo {author} {\bibfnamefont {P.}~\bibnamefont {{Haensel}}},
  \ and\ \bibinfo {author} {\bibfnamefont {J.~L.}\ \bibnamefont {{Zdunik}}},\
  }\href {\doibase https://ui.adsabs.harvard.edu/abs/1997A&A...325..217G}
  {\bibfield  {journal} {\bibinfo  {journal} {A\&A}\ }\textbf {\bibinfo
  {volume} {325}},\ \bibinfo {pages} {217} (\bibinfo {year}
  {1997})}\BibitemShut {NoStop}%
\bibitem [{\citenamefont {{Lopes, I. P.}}(2001)}]{refId0}%
  \BibitemOpen
  \bibfield  {author} {\bibinfo {author} {\bibnamefont {{Lopes, I. P.}}},\
  }\href {\doibase 10.1051/0004-6361:20010130} {\bibfield  {journal} {\bibinfo
  {journal} {A\&A}\ }\textbf {\bibinfo {volume} {373}},\ \bibinfo {pages} {916}
  (\bibinfo {year} {2001})}\BibitemShut {NoStop}%
\bibitem [{\citenamefont {Panotopoulos}\ and\ \citenamefont
  {Lopes}(2017{\natexlab{b}})}]{PhysRevD.96.083013}%
  \BibitemOpen
  \bibfield  {author} {\bibinfo {author} {\bibfnamefont {G.}~\bibnamefont
  {Panotopoulos}}\ and\ \bibinfo {author} {\bibfnamefont {I.}~\bibnamefont
  {Lopes}},\ }\href {\doibase 10.1103/PhysRevD.96.083013} {\bibfield  {journal}
  {\bibinfo  {journal} {Phys. Rev. D}\ }\textbf {\bibinfo {volume} {96}},\
  \bibinfo {pages} {083013} (\bibinfo {year} {2017}{\natexlab{b}})}\BibitemShut
  {NoStop}%
\end{thebibliography}%

\end{document}